\documentclass[preprintnumbers,secnumarabic,amsmath,amssymb,nofootinbib,floatfix]{revtex4}
\usepackage{varioref,exscale,latexsym,amsmath,amssymb}
\usepackage{graphicx}

\usepackage{graphicx}
\usepackage{slashed}
\usepackage{dcolumn}
\usepackage{bm}
\usepackage{hyperref}
\usepackage{color}
\usepackage{xcolor}

\newcommand{\beq}{\begin{equation}}
\newcommand{\eeq}{\end{equation}}
\newcommand{\bea}{\begin{eqnarray}}
\newcommand{\eea}{\end{eqnarray}}
\newcommand{\nn}{\nonumber}

\def\lsi{\raise0.3ex\hbox{$<$\kern-0.75em\raise-1.1ex\hbox{$\sim$}}}
\def\gsi{\raise0.3ex\hbox{$>$\kern-0.75em\raise-1.1ex\hbox{$\sim$}}}

\def\beq{\begin{equation}}

\def\eeq{\end{equation}}

\def\cB{{\cal B}}

\def\cL{{\cal L}}

\begin{document}
\preprint{ACFI-T23-07}

\title{{\bf Higher Derivative Sigma Models}}

\medskip\

\medskip\

\author{John F. Donoghue${}^{1}$}
\email{donoghue@physics.umass.edu}
\author{Gabriel Menezes${}^{2}$~\footnote{On leave of absence from Departamento de F\'{i}sica, Universidade Federal Rural do Rio de Janeiro.}}
\email{gabrielmenezes@ufrrj.br}
\affiliation{
${}^1$Department of Physics,
University of Massachusetts,
Amherst, MA  01003, USA\\
${}^2$Instituto de F\'isica Te\'orica, Universidade Estadual Paulista,
Rua Dr.~Bento Teobaldo Ferraz, 271 - Bloco II, 01140-070 S\~ao Paulo, S\~ao Paulo, Brazil}

\begin{abstract}
We explore the nature of running couplings in the higher derivative linear and nonlinear sigma models and show that the results in dimensional regularization for the physical running couplings do not always match the values quoted in the literature. Heat kernel methods identify divergences correctly, but not all of these divergences are related to physical running couplings. Likewise the running found using the Functional Renormalization Group does not always appear as running couplings in physical processes, even for the case of logarithmic running. The basic coupling of the higher derivative SU(N) nonlinear sigma model does not run at all at one loop,  in contrast to published claims for asymptotic freedom. At one loop we describe how to properly identify the physical running couplings in these theories, and provide revised numbers for the higher derivative nonlinear sigma model.
\end{abstract}

\maketitle

\section{Introduction}

We use running coupling constants routinely in quantum field theory. By defining a scale dependent coupling constant one can sum up a set of quantum corrections which appear at that scale. The use of the running coupling constant in physical reactions yields a better perturbative expansion at that scale than does using a coupling defined at very distant scales. 

{ Despite dependence on the renormalization scheme, the running coupling constants appear in all reactions at a given scale because in standard four-dimensional theories the running is tied to the renormalization of the bare couplings.} In standard renormalizable theories the running is  logarithmic in the energy scales. There are a set of techniques which are used to calculate the beta functions for the couplings which exploit this connection to the divergences of the theory. 

The goal of the present paper is to illustrate how some of these techniques no longer yield physical running couplings when applied to theories which involve higher derivatives, and to propose a solution. {By physical running couplings, we mean the couplings which appear in the physical on-shell amplitudes of the theory. In this case, the running has to involve the kinematic variables of the physical amplitudes. } This problem has been identified and studied in detail in a simple model in recent work by Buccio, Percacci and one of present authors \cite{Buccio:2023lzo}, and we also review that example in Section \ref{Shiftinvariant}. We have also employed these techniques in the study of the two dimensional $CP(1)$ model \cite{Buccio:2024vaf} and in quadratic gravity \cite{Buccio:2024hys}, in collaboration with Buccio and Percacci. We motivate such theories in Section \ref{theories}.  However, first we describe the general reasons why the calculations in the literature for such theories do not yield physical scale dependent running couplings.

 \subsection{Identifying running couplings}
 
\begin{figure}
\begin{center}
\includegraphics[height=40mm,width=100mm]{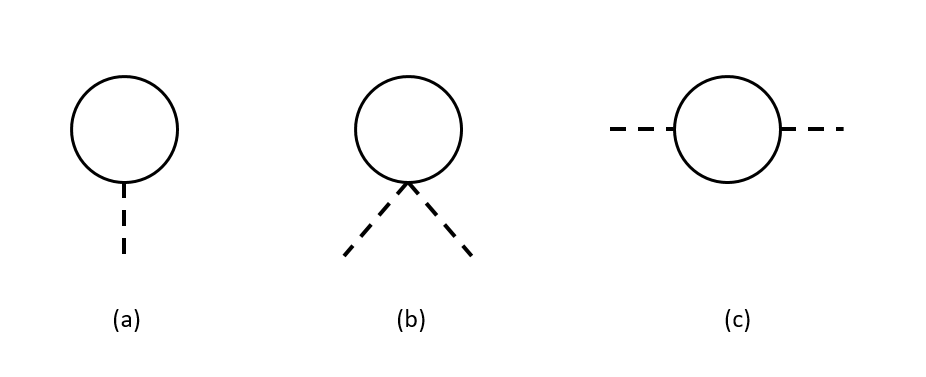}
\caption{Diagrams (a) and (b) are referred to as tadpole diagrams, with any number of external legs. Diagram (c) is a bubble diagram. The important distinction is whether the external momentum flows through the loop or not. }
\label{diagrams}
\end{center}
\end{figure}

 Let us start off with what appears to be a curiosity. In theories which include four derivatives in the kinetic energy, the propagator falls as $1/p^4$. In calculations, one often encounters the tadpole diagram of Fig \ref{diagrams}. This integral is logarithmically divergent, with 
 \beq\label{cutoffint}
I_{\text{tad}} =-i \int d^4 p \frac1{p^4} \sim \log \frac{\Lambda^2}{k^2}    ~~~~~~~~~~{\rm (UV/IR ~cutoffs)}
\eeq
where $\Lambda$ is a UV cutoff and $k$ is an IR cutoff. However, in dimensional regularization this integral is scaleless and vanishes
\beq\label{difint}
I_{\text{tad}} =-i \int d^d p \frac1{p^4} =0 ~~~~~~~~~~~~~~~~~{\rm (dimensional ~regularization)}\ \ .
\eeq
This is a well-known oddity of dimensional regularization, here applied to quartic propagators.  Despite this difference there is no disagreement in physical processes, which are all independent of the regularization scheme. The $ \log \Lambda/k$ factor disappears in the renormalization process and does not lead to any physical effects.

However, this curiosity does have important consequences for the calculation of beta functions. With cutoff regularization, we often calculate the beta function using $\Lambda d/d\Lambda$ on the relevant quantum corrections. The Functional Renormalization Group, as {generally} applied in the Asymptotic Safety program, uses the infrared cutoff dependence $k d/dk$.  However, the vanishing of the purely quartic integral, Eq. \ref{difint}, already tells us that this must be wrong - the tadpole integral cannot lead to running couplings in physical reactions if the integral vanishes.  This can be understood physically because no external momentum flows through the tadpole integral. The whole integral is just a constant which is absorbed into the renormalization of the parameter and there is no residual dependence on the energy scale. Identifying $\log \Lambda^2$, $\log k^2$ (or $\log \mu^2$ in dimensional regularization) does not always tell us about $\log E^2$, where $E$ represents the energy scale of whatever reaction which we are studying. We will show that this distinction has created errors in the calculation of the physical  beta functions which exist in the literature.

More generally, identifying the divergences is not sufficient in cases where there are other dimensionful parameters. This issue is not just the presence of tadpole diagrams themselves, as the scalar tadpole integral also arises in the Passarino-Veltman reduction of general bubble, triangle, box, etc., diagrams. One must separate the $\log E^2/\mu^2$ factors from the $\log m^2/\mu^2$ ones. Heat kernel methods are good at identifying the divergences but do not tell us the form of the logarithms. There needs to be an extra step to identify which divergences are associated with kinematic logarithms and we will provide this separation in Sect. \ref{physical}. This issue surfaces in various ways in the theories described in this paper, and also more widely in the literature. 

 \subsection{Theories with higher derivatives} {\label{theories}}

In the space of possible quantum field theories, we normally limit ourselves to the sector with only two derivatives in the kinetic energy terms. However, there can be reasons for going beyond this limitation. 

Lee and Wick explored higher derivative theories in order to have finite quantum field theories, without the usual divergences~\cite{Lee:1970iw}. The higher derivative kinetic energies improves the high energy behavior of propagators and perturbatively gives finite loop corrections for theories such as QED. For a similar reason, extra derivative kinetic energies can also turn non-renormalizable theories into renormalizable ones. Here the example with the greatest physical interest is gravity. General relativity by itself is non-renormalizable but when terms proportional to the curvature squared are included -- bringing in four derivatives -- the resulting theory of Quadratic Gravity is renormalizable~\cite{Stelle:1976gc, Salvio:2018crh, Donoghue:2021cza}.

Higher derivative operators can also be generated by quantum corrections. When treated as simple perturbations using effective field theory techniques the propagators can remain the same as in the original theory. However, in some settings all the higher order operators are on the same footing. For example, with Functional Renormalization Group (FRG) techniques \cite{Dupuis:2020fhh, Saueressig:2023irs} one in principle includes all local operators, with scale transformations yielding a renormalization group flow of the couplings of these operators. Such techniques will always face the situation with higher derivative kinetic energies. The present practice of Asymptotic Safety most often treats gravity using FRG techniques \cite{Niedermaier:2006wt, Reuter:2019byg}. {Our discussion and comparison of physical running vs the running of couplings in the FRG builds from the initial work of \cite{Buccio:2023lzo}, where an explicit amplitude is compared to FRG methods. }

Sigma models are an ideal testing ground for this class of theories. They are simple enough that one can focus on the essential new physics without too much complication. The SU(N) higher derivative nonlinear sigma model is the closest analogy to Quadratic Gravity as it shares all of the higher derivative features, but does not have the subtle feature of general coordinate invariance. The running of the couplings has been studied by Hasenfratz \cite{Hasenfratz:1988rf}, and it has also been treated using FRG techniques by Percacci and Zanusso \cite{Percacci:2009fh}, {with results which agree within the appropriate limit. However those results differ from the physical running that we find below. Moreover higher derivative sigma models are potentially useful because that can be simulated using lattice methods. }

Higher derivative theories have features which differ from conventional QFT, and these are much debated in the literature. We have reviewed our own studies in \cite{Donoghue:2021cza}. In the present paper we do not address these other aspects, but focus uniquely on the issue of renormalization group flow. 

\subsection{Renormalization group techniques and physical running}

We are interested in the running of couplings with the energy scale of the process in physical amplitudes. Renormalization group techniques do not always track this directly, but follow the logarithmic divergences such as $log \Lambda^2 $ when using a cutoff $\Lambda$ or $\log \mu^2$ when using dimensional regularization. Often this can be sufficient because, in theories with no other significant dimensionful factors, the logarithms must also involve the kinematic factors for dimensional reasons, i.e. $\log (\Lambda^2/E^2)$ or $\log (\mu^2/E^2) $, where $E$ represents the energy scale of the process. However, when there is another scale in the problem, $m$, there can be also factors of $\log (\Lambda^2/m^2)$ or $\log (\mu^2/m^2)$. In this case, following $\log \Lambda^2$ or $\log \mu^2$ does not track the true dependence of the amplitudes with the energy scale. All the theories discussed in this paper have such an extra factor of $m$, which comes from the relative scale of the 2-derivative and 4-derivative contributions to the kinetic energy.

There is an additional feature that the form of the running couplings changes between the low energy and high energy regions. In theories with higher derivatives, the mass scale $m$ is related to the relative size of the two derivative and  four derivative kinetic energy terms. In this case, low energy refers to $E<m$ and high energy to $E>m$. At low energy the running couplings can be found by Effective Field Theory (EFT) methods. The high energy region requires the full theory.

In order to find the physical running parameters one needs to identify the diagrams which contain the kinematic variables of the quantum corrections. This is aided by the Passarino-Veltman reduction which allows all one loop Feynman integrals to be written in terms of scalar tadpole, bubble, triangle and box diagrams times overall momentum factors. Only the scalar tadpole and bubble diagrams will be relevant for the renormalization of the couplings. The scalar tadpole diagram does not contain any factors of the external momentum and can never lead to factors of $\log E^2$. The bubble diagram does contain the external momentum. Massless bubble diagrams always involve $\log ( \Lambda^2/E^2)$ or $\log (\mu^2/E^2) $ and this leads to physical running. For bubble diagrams with massive internal lines, the $\log E^2$ factor emerges at high energy, when $E>m$. The techniques used throughout this paper depend on this separation of scalar bubble diagrams from tadpoles.

{In applying running couplings, there is residual scheme dependence coming from the procedure used to define the measured couplings. This extra scheme dependence is present in the theories which we discuss here also. However, that is distinct from the issue of whether the coupling runs or does not run. We are addressing the latter issue in this paper.}

There is also a second feature to our calculation which deserves some explanation. This is that some relevant logarithms appear in infared-sensitive parts of the Feynman integrals. This was found in the study of the two dimensional $CP(1)$ model \cite{Buccio:2024vaf} where the scalar bubble diagram is UV finite in two dimensions but which neverthless produces the $\log E^2$ factor which leads to physical running. These logarithms were also important in the treatment of quadratic gravity \cite{Buccio:2024hys}, where they were needed to obtain general covariance of the final result. We can see the effect in a simple example. In calculating bubble diagrams, one could find a particular contribution of the form
\beq
M_1 = T_{\mu\nu\alpha\beta} \int \frac{d^4k}{(2\pi)^4} \frac{k^\mu k^\nu k^\alpha k^\beta}{k^4 (k-q)^4}
\eeq
where $T_{\mu\nu\alpha\beta}$ is some tensor consisting of external momenta and metric factors. A related case could be
\beq
M_2 = T_{\mu\nu\alpha\beta} \int \frac{d^4k}{(2\pi)^4} \frac{k^\mu k^\nu (k-q)^\alpha (k-q)^\beta}{k^4 (k-q)^4}
\eeq
with the same tensor. Power counting shows that the two Feynman integrals in these examples have the same logarithmic UV divergence. However they have differing IR sensitivities. This can make a difference. Consider the case where $T_{\mu\nu\alpha\beta} = g_{\mu\nu}g_{\alpha\beta}$. In this case, these integrals turn into
\beq
M_1  = \int  \frac{d^4k}{(2\pi)^4} \frac{1}{ (k-q)^4}\eeq
and
\beq
M_2  = \int  \frac{d^4k}{(2\pi)^4} \frac{1}{k^2 (k-q)^2}\eeq
After a shift in the momentum integration, the first of these is a pure tadpole integral with a result that does not depend on $q$. In contrast, $M_2$ is exactly the scalar bubble diagram, which contains a factor of $\log q^2$. The difference has come from the infrared sensitive portion of the integral. In evaluating the impact of these Feynman integrals, one needs to keep the full tensor evaluation of the Feynman integral combined with the external factors, not just the divergent pieces. This was first identified  in \cite{Buccio:2024hys}.  We explore this topic, and evaluate these integrals in more detail, in Section \ref{effectiveaction}.

\subsection{Structure for the paper}

{Given these introductory comments, we study sigma models of increasing complexity. In the first example, Sect. \ref{linear} we study the higher derivative linear sigma model. In this case we encounter a parameter, the mass, which carries cutoff or $\mu$ dependence but which does not run in physical process. We also see a parameter, the coupling $\lambda$ which runs at low energy (although not at high energy) despite not carrying any cutoff or $\mu$ dependence. Then in Sect. \ref{Shiftinvariant} we encounter a coupling which does not run at low energy, but does at high energy. Effective field theory reasoning is useful in sorting out these behaviors. Finally in Sect. \ref{nonlinear}, we find that the primary coupling does not run at all at one loop in physical reactions even though standard techniques in the literature have indicated an asymptotically free running behavior.  The remaining couplings have a variety of different behaviors differing from that in the literature. Finally we describe the overarching lessons of these investigations in Sect. \ref{physical} and in the conclusions.}

\section{The higher derivative linear sigma model}\label{linear}

The higher derivative $O(N)$ linear sigma model is defined by the Lagrangian
\beq
{\cal L}= \frac12 \partial_\mu \boldsymbol{\phi}\cdot \partial^\mu \boldsymbol{\phi} -\frac1{2m^2} \Box \boldsymbol{\phi} \cdot \Box \boldsymbol{\phi}+\frac{\mu_0^2}{2} \boldsymbol{\phi}\cdot \boldsymbol{\phi} -\frac{\lambda}{4} (\boldsymbol{\phi} \cdot \boldsymbol{\phi})^2  \ \ .
\eeq
Here the field $\boldsymbol{\phi} = \{ \phi_1,  ... ,\phi_N\}$ is conventionally normalized and the higher derivative kinetic energy is parameterized by the mass $m$. Symmetry breaking takes place as normal, with the vacuum expectation value $v=\sqrt{\mu_0^2/\lambda}$, a heavy scalar with mass $m_\sigma^2 =2\mu_0^2$ and $N-1$ Goldstone bosons. The higher derivative term implies extra massive degrees of freedom, even for the Goldstones whose propagator is
\beq
D_F(p) =\frac1{p^2 -\frac{p^4}{m^2 } }= \frac1{p^2} -\frac1{p^2-m^2}
\eeq
The negative norm of the massive state is a well-known feature of higher derivative theories. For us the main feature is that it causes cancellations within loop integrals.

This theory is renormalizable. Because the higher derivative term improves the UV convergence, and the normal sigma model is already renormalizable, one might expect that the higher derivative version would actually be finite. However, the mass term in the theory, $\mu_0^2$, rather famously has a quadratic divergence, and the higher derivative modification merely reduces that to a logarithmic divergence. There are no divergences related to the quartic coupling $\lambda$. 

Jansen, Kuti and Liu have explored a very similar model model both analytically and numerically as a probe of naturalness in the Higgs sector \cite{Jansen:1992xv, Jansen:1993ji}. Their model involves a yet higher derivative kinetic energy, $\phi \Box^3 \phi$, such that the theory is finite rather than renormalizable. Nevertheless, the results described below for the running of $\lambda$ is present in their work. Our discussion highlights those features relevant for the remaining sections of this paper.

\subsection{Renormalization without running}

The divergence in the mass term comes from the tadpole diagram in Figure \ref{diagrams}. The full evaluation of this tadpole reveals more about the physics that was not evident in our motivation section of Eq. \ref{cutoffint} and Eq. \ref{difint}. That is, even in cases where the tadpole diagram does not vanish, it still does not lead to running couplings. 

A quadratic term in the propagator acts as an infrared cutoff in the tadpole integral. In contrast with the pure quartic integral of Eq. \ref{difint} the result now does not vanish. 
One finds  
\beq
\delta m_\sigma^2 = {(N +2)\lambda } \tilde{I}_{tad}
\eeq
with 
\bea\label{taddim}
\tilde{I}_{tad} =-i \int \frac{d^dp}{(2\pi)^d} \frac1{p^2-\frac{p^4}{m^2 } } &=&   -i\int \frac{d^dp}{(2\pi)^d}  \left[ \frac1{p^2} -\frac1{p^2-m^2} \right]   \nonumber \\
&=&-\frac{m^2}{16\pi^2} \left[\frac1{\epsilon} + \log 4\pi - \gamma -\log \left(\frac{m^2}{\mu^2} \right) +1 \right]   \ \ .
\eea
with $\epsilon = 2/(4-d)$. (For the rest of the paper, unless noted we display only the divergences and the logarithms and suppress the remaining constants.)

Often when using dimensional regularization one follows the $1/\epsilon$ factors or equivalently uses $\mu \frac{d}{d\mu}$. Despite the divergence and the factor of $\log \mu^2$, this does not lead to a running mass. This is seen from the fact that the logarithm involves $\log m^2/\mu^2$ and not any kinematic quantity.  After renormalization,
\beq
m_{\sigma, {\rm ren}}^2 = m_{\sigma, {\rm bare}} + \delta m_\sigma^2
\eeq
there is no residual dependence on any external scale. Measurement of the mass term at any scale will yield the same value. Using the $\log \mu$ dependence to define a running coupling would be incorrect.

\subsection{Running without renormalization}

The quantum corrections to the quartic coupling $\lambda$ do not involve any divergences. However, as first noted by Jansen, Kuti and Liu, the coupling has the interesting feature of running at low energies, and then the beta function vanishes at high energies. Our discussion here is appropriate for the unbroken phase at energies above the scale of symmetry breaking. For an analysis of the broken phase, see Section \ref{Shiftinvariant}.

The one loop correction to this coupling involves the scalar bubble diagram. Using the partial fraction decomposition of the propagator one readily finds that the scalar bubble integrals involve
\beq\label{deltalambda}
{\hat I} (q) = \left[I_2(0,0,q) -2 I_2(0,m,q) + I_2(m,m,q) \right]
\eeq
where the bubble integral is
\beq\label{bubble}
I_2(m_1,m_2,q) = \frac1{16\pi^2}\left[ \frac1{\epsilon} - \int dx \log\left(\frac{x m_1^2+(1-x)m_2^2 -q^2  x(1-x)}{\mu^2}\right) \right] \ \ .
\eeq
One sees that the divergences and $\mu$ dependences cancel in ${\hat I} (q)$. Nevertheless, the result carries a logarithmic momentum dependence at low energy
\beq
{\hat I}(q) =- \frac1{16\pi^2}\log (-q^2/m^2 ) +...      ~~~~~~,~(q^2\ll m^2)
\eeq
that leads to a running coupling. One finds 
the scattering amplitude for $\phi_1+\phi_1\to \phi_1+\phi_1$ to be
\beq
{\cal M}= 6\lambda\left[ 1 - \lambda \frac{ (N+8)}{3} \left( {\hat I}(s) +{\hat I}(t) +{\hat I}(u) \right)\right]
\eeq
When measuring the coupling in the scattering amplitude using the renormalization point $s=t=u=\mu_R^2 \ll m^2$,  the physical beta function
\beq
\beta_\lambda = \mu_R \frac{\partial }{\partial \mu_R} \lambda (\mu_R) = \frac{(N+8) \lambda^2}{8 \pi^2} ~~~~~~~~~~,~{\rm (\mu_R^2 \ll m^2)}  \ \ .
\eeq

In contrast, at high energy it is easy to see that the mass $m$ becomes irrelevant and the energy dependence also cancels out between the terms in ${\hat I}$
\beq
{\hat I} (q) \sim 0   ~~~~~~~~~~~~,~{q^2 \gg m^2}
\eeq
with the result that 
\beq
\beta_\lambda =0   ~~~~~~~~~~,~{\rm (\mu_R^2 \gg m^2)} \ \ .
\eeq

\subsection{The EFT divide}

The result for the running of $\lambda$ illustrate a universal feature in these higher derivative theories. There will be two energy regions with a running behavior that will in general be different. { The logic involves the ideas of effective field theory (EFT) \cite{Donoghue:1992dd, Donoghue:1994dn,Burgess:2020tbq, Petrov:2016azi}. The full theory is able to describe the results at all energy scales. However, at low energy the heavy particles are not dynamically active. They can be integrated out of the theory. By the uncertainty principle, this leaves residual effects which appear local when working only at low energy. This gives the effective field theory. Because the symmetry of the system is not changed in these theories the basic interaction will appear in the effective field theory, as well as higher derivative interactions. In particular one can identify the renormalized coupling already by measuring it at low energy. At high energy, the heavy particles are dynamically active, and the form of the predictions change. But one can continue to describe these predictions using the renormalized coupling measured at low energy. So we are led to describe the theory predictions somewhat differently in the low and high energy regions. }

{Because the higher derivative theories always involve the heavy ghost field with mass $m$, the EFT divide happens at the energy $m$.  This leaves an effective field theory (EFT) at low energy  .  For energies below $m$, the couplings run like described by the EFT. At energies above $m$ the full theory is required. Beta functions then generally have to be given separately for the two regions. In the present theory, the EFT is just the usual linear sigma model, so that the coupling $\lambda$ runs in the usual way at low energy. }

\section{The U(1) non-linear sigma model}
\label{Shiftinvariant}

Next consider the Lagrangian
\beq
\label{notation}
\cL =-\frac{1}{2} \partial_\mu \phi \partial^\mu\phi - \frac{1}{2m^2}\Box\phi \Box \phi - \frac{g}{4M^4}( \partial_\mu \phi \partial^\mu\phi )( \partial_\nu  \phi \partial^\nu \phi) .
\eeq
Without the higher derivative kinetic energy, this is a standard example of the low energy limit of the U(1) linear sigma model in the symmetry broken phase. With the extra kinetic energy it is similarly the low energy limit of the {symmetry broken phase of the } higher derivative linear sigma model studied in the previous section. In Appendix A we provide this demonstration, with the identification 
\beq
\frac{g}{M^4} = \frac{\lambda}{m^2 m_\sigma^2} +\frac{\lambda}{m_\sigma^4}  \ \ .
\eeq 
We will consider the parameter $M$ as fixed and  will use $g$ as a coupling which potentially may be a running parameter. We will also here only describe the case where $m\lesssim M$ so that the physics associated with the higher derivative term is active in the symmetry broken phase.

Despite this connection to the linear model, this theory is renormalizeable and can be treated on its own as a complete QFT. It has been the focus of several recent studies \cite{Buccio:2023lzo, Buccio:2022egr, Tseytlin:2022flu, Holdom:2023usn}. In particular Ref \cite{Buccio:2023lzo}
has studied the renormalization and running behavior of this theory in great detail, including the calculation of the scattering amplitudes in all energy regions, {and compares the result to the FRG analysis of Ref. \cite{Buccio:2022egr}}. Here we recall and recast their results in order to compare and contrast with our other results. 

At low energy, the coupling $g$ is renormalized, but does not run,
 \beq
\beta_g =0 ~~~~~~~~~~(E<<m)   \ \ .
\eeq
This can be understood because at low energy the heavy mass particle can be integrated out leaving a normal effective field theory with the same interaction term. Treated as a massless effective field theory, tadpole corrections vanish in dimensional regularization and bubble diagrams are of order $g^2E^8/M^8$ because of the need for two interaction terms. This implies that in the EFT treatment there is no loop correction to $g$ and hence it cannot run. This is reproduced by the full theory because the one loop correction is of the form $g^2(m^8/M^8)[1/ \epsilon ~- \log m^2/\mu^2 ]$.  Once measured at low energy, the coupling does not run because there is no kinematic dependence.

However, here the interesting feature is that although there is no further renormalization needed, the coupling does start running at high energy. This is nontrivial and does not generalize to all related theories. In this case, it was seen by calculating the scattering amplitude, which reveals that a coupling defined at $s=t=u=\mu_R^2$ of 
\beq\label{runningg}
 \bar{g}(\mu_R) =  g +\frac{5g^2m^4}{32\pi^2 M^4}  \left[  \log \left(\frac{\mu_R^2}{m^2} \right) {-\frac{17}{30}}\right]  \  \  .
 \eeq
 removes all large logarithms from the amplitude. While there remain other finite logarithms, one can see in the amplitude
 \bea
{\cal M} &=&-\frac{\bar{g}(\mu_R)}{2M^4} (s^2+t^2+u^2) \nonumber \\
&& -\frac{\bar g^2m^4}{192\pi^2 M^8}\left[ \log \left(\frac{-s}{\mu_R^2}\right)(13s^2 +t^2 +u^2)+  \log \left(\frac{-t}{\mu_R^2}\right)(s^2 + 13t^2 +u^2)+ \log \left(\frac{-u}{\mu_R^2}\right)(s^2 +t^2 +13u^2)  \right] \,
\label{heamp}
\eea
 that the use of a running coupling is appropriate for this physical amplitude. 
 The coupling obeys an renormalization group equation with
\beq
\beta_{\bar{g}} =  \frac{5\bar{g}^2 m^4}{16\pi^2 M^4}  ~~~~~~~~~~~~~(E>>m)\ \ . 
\label{betabg}
\eeq

The reason that this this new kinematic logarithm emerges was described in Ref. \cite{Buccio:2023lzo} and we will rephrase it using the background field method in order to use the same result in the next section.  If we expand the field $\phi$ around a background field via
\beq
\phi = \bar{\phi} + \eta
\eeq
with 
\beq
\cL (\phi) = \cL (\bar{\phi}) + \cL_2 (\bar{\phi} ,\eta)  \ \ 
\eeq
 we find
 \beq\label{Bexpansion}
  \cL_2 (\bar{\phi} ,\eta) = \frac12 \partial_\mu \eta \partial^\mu \eta - \frac1{2m^2} \Box\eta \Box\eta - \cB_{\mu\nu} \partial^\mu \eta \partial^\nu \eta
 \eeq
 with
 \beq
 \cB_{\mu\nu} =2g \partial_\lambda \bar{\phi} \partial^\lambda \bar{\phi} \eta_{\mu\nu}+ 4g \partial_\mu \bar{\phi }\partial_\nu \bar{\phi}
 \eeq
 At one loop there are two divergent diagrams. The tadpole diagram is linear in $\cB_{\mu\nu}$, and does not involve any kinematic quantity.  It contribute only to wavefunction renormalization, which then does not show any physical running. The bubble diagram contains two factors of $\cB_{\mu\nu}$. 
 The loop integral for this is proportional to 
 \bea
 I_{\mu\nu\alpha\beta} &=&m^4\int \frac{d^dp}{(2\pi)^d} \frac{p_\mu p_\nu ( p-q)_\alpha (p-q)_\beta}{[m^2 p^2 - p^4][(m^2 (p-q)^2 - (p-q)^4]}  \nonumber \\
 &=& F(q^2) (\eta_{\mu\nu}\eta_{\alpha\beta} +\eta_{\mu\alpha}\eta_{\nu\beta} + \eta_{\mu\beta}\eta_{\nu\alpha}) 
 \nn\\
 &&+ \, {\cal O}(q) {\rm ~terms} .
 \eea
 The only divergence appears in the first term $F$. We can simply evaluate this divergence by taking the trace of this integral
 \bea
 \eta^{\mu\nu}\eta^{\alpha\beta} I_{\mu\nu\alpha\beta} &=& m^4 \int\frac{d^dp}{(2\pi)^d} \frac1{[m^2 - p^2][(m^2  - (p-q)^2]}   \nonumber \\
  &=& m^4 I_2(m,m,q) =  d(d+2) F+....
  \eea
where $I_2$ is given in Eq. \ref{bubble}.  This is just the scalar bubble diagram, which has the behavior
\bea
I_2 (m,m,q) &=& \frac1{16\pi^2} \left[ \frac1{\epsilon} - \log \frac{m^2}{\mu^2}\right]  ~~~~~~~(q^2<<m^2)   \nonumber \\
&=& \frac1{16\pi^2} \left[ \frac1{\epsilon} - \log \frac{q^2}{\mu^2}\right]  ~~~~~~~(q^2>>m^2) \ \ .
\eea
From this we can see that below the scale $m$ we assign a $\log m^2$ along with the divergence, which explains the lack of physical running at low energy. At higher energies the heavy particle is dynamically active and it is appropriate to assign the physical running associated with $\log q^2$ along with the divergence. This was the result demonstrated in the full calculation of \cite{Buccio:2023lzo}.
   
 We can extract a more general lesson from this calculation. When the background field expansion has the form of Eq. \ref{Bexpansion}, for any $\cB_{\mu\nu}$, {the calculation of the one-loop diagrams reveals that the divergences have the form 
 \beq
\delta \cL = \frac1{16\pi^2} \frac1{\epsilon} \left[ \frac14 m^2 \cB^\lambda_\lambda + \frac1{24}\cB_{\mu\nu}\cB^{\mu\nu} + \frac1{48}\cB^\lambda_\lambda \cB^\sigma_\sigma \right]  \ \ .
\eeq
While this result was obtained by Feynman diagrams, we can anticipate the heat kernel language used in the next section and identify} the {Seeley-DeWitt coefficients} $a_1$  in the first term, and  $a_2$  in the last two terms. However,  we know from the direct calculation that the $a_1$ term arises from the tadpole diagram and does not correspond to physical running at any scale. In contrast, the terms of order $\cB^2$ do not indicate running at scales below $m$ but do have real physical effects when the scales are above $m$. {The heat kernel is adept at identifying the divergences, but does not always identify the momentum dependence, such that further techniques are required. } This result will be generalized further in Sec. \ref{physical}.

\section{The SU(N) non-linear sigma model}\label{nonlinear}

Let us now discuss a more intricate model that resembles more closely the aspects and challenges one expects to meet in Quadratic Gravity. This is the higher-derivative nonlinear SU(N) sigma model (HDNLSM), whose one-loop renormalization, to the best of our knowledge, was first discussed by Hasenfratz~\cite{Hasenfratz:1988rf}. Here we aim to give a modern perspective of the problem, emphasizing some key features and discussing explicitly the issues concerning the evaluation of the associated beta functions. Ultimately we will compare our results to those derived by Hasenfratz and also to Functxional Renormaization Group thechniques \cite{Percacci:2009fh}. 

The action for the HDNLSM reads~\cite{Hasenfratz:1988rf} ($N \geq 4$)
\beq
S = \frac{1}{c_0} S_0 + \frac{1}{f^2} S_1 + \sum_{i=2}^{5} \alpha_i^2 S_i 
\eeq
where
\bea
S_0 &=& \int d^4 x \, \textrm{Tr}\Bigl( A_{\mu}(x) A^{\mu}(x)  \Bigr)
=  - \int d^4 x \, \textrm{Tr}\Bigl( \partial_{\mu}U^{-1}\partial^{\mu} U \Bigr)
\nn\\
S_1 &=& \frac{1}{2} \int d^4 x \, \textrm{Tr}\Bigl( \partial_{\mu} A^{\mu}(x) \partial_{\nu} A^{\nu}(x) 
+ \partial_{\mu} A_{\nu}(x) \partial^{\mu} A^{\nu}(x)  \Bigr)
\nn\\
S_2 &=& - \frac{1}{2} \int d^4 x \, \textrm{Tr}\Bigl( \partial_{\mu} A^{\mu}(x) \partial_{\nu} A^{\nu}(x) 
- \partial_{\mu} A_{\nu}(x) \partial^{\mu} A^{\nu}(x)  \Bigr)
\nn\\
S_3 &=& - \frac{1}{2} \int d^4 x \, \textrm{Tr}\Bigl( A_{\mu}(x) A^{\mu}(x) A_{\nu}(x) A^{\nu}(x) 
+ A_{\mu}(x) A_{\nu}(x) A^{\mu}(x) A^{\nu}(x)  \Bigr)
\nn\\
S_4 &=& - \int d^4 x \, \textrm{Tr}\Bigl( A_{\mu}(x) A^{\mu}(x)  \Bigr)
 \textrm{Tr}\Bigl( A_{\nu}(x) A^{\nu}(x)  \Bigr)
\nn\\
&=&  \int d^4 x \, \textrm{Tr}\Bigl( A_{\mu}(x) A^{\mu}(x)  \Bigr)
 \textrm{Tr}\Bigl( \partial_{\nu}U^{-1}\partial^{\nu} U  \Bigr) 
=  - \int d^4 x \, \textrm{Tr}\Bigl( \partial_{\mu}U^{-1}\partial^{\mu} U  \Bigr)
 \textrm{Tr}\Bigl( \partial_{\nu}U^{-1}\partial^{\nu} U  \Bigr)
  \nn\\
S_5 &=& - \int d^4 x \, \textrm{Tr}\Bigl( A_{\mu}(x) A_{\nu}(x)  \Bigr)
 \textrm{Tr}\Bigl( A^{\mu}(x) A^{\nu}(x)  \Bigr)
\nn\\
&=&  \int d^4 x \, \textrm{Tr}\Bigl( A_{\mu}(x) A_{\nu}(x)  \Bigr)
 \textrm{Tr}\Bigl( \partial^{\mu}U^{-1}\partial^{\nu} U  \Bigr) 
=  - \int d^4 x \, \textrm{Tr}\Bigl( \partial_{\mu}U^{-1}\partial_{\nu} U  \Bigr)
 \textrm{Tr}\Bigl( \partial^{\mu}U^{-1}\partial^{\nu} U  \Bigr)
\eea
and we have defined
\beq
A_{\mu}(x) = U^{-1}(x) \partial_{\mu} U(x).
\eeq
In $S_0, S_4, S_5$ we have used the property 
$U^{-1}(x) \partial_{\mu} U(x) = - \partial_{\mu} U^{-1}(x)  U(x)$ which is a consequence of the unitarity of $U$. Moreover, despite the position of the indices, we work in Euclidean space, with the metric 
$\delta_{\mu\nu}$. The SU(N) matrix field $U(x)$ reads
\beq
U(x) = e^{i f \pi^{a}(x) t^a}
\eeq
where $t^a$, $a = 1, 2, \ldots, N^2-1$ are the SU(N) generators and the fields $\pi^a(x)$ are dimensionless (we consider the fundamental representation). In the context of chiral perturbation theory, the latter are identified as Goldstone fields. The model enjoys a set of symmetries fully discussed in Ref.~\cite{Hasenfratz:1988rf}, among them a global chiral symmetry. For standard chiral perturbation theory, see also Ref.~\cite{Gasser:1984gg}. 

In the present case, it is more useful to employ the following parametrization~\cite{Gasser:1984gg}
\bea
U(x) &=&  u(x) e^{i \pi(x)} u(x) = u(x) \left( 1 + i \pi - \frac{1}{2} \pi^2 + \cdots \right) u(x)
\nn\\
\tilde{U}(x) &=& u^2
\eea
so that
\beq
\tilde{A}_{\mu}(x) = \tilde{U}^{-1}(x) \partial_{\mu} \tilde{U}(x).
\eeq
Here $\pi(x) = \pi^{a}(x) t^a$ is the fluctuation over the classical background described by the classical solution $\tilde{U}(x)$ (or $\tilde{A}_{\mu}(x)$). The corresponding Euclidean path integral reads
\beq
e^{W[\tilde{U}]} = \int d\mu(\pi) \, e^{S}.
\eeq
We are interested in the quadratic part in $\pi$. Again following Ref.~\cite{Gasser:1984gg}, we introduce the following anti-Hermitian matrices
\bea
\Gamma_{\mu} &=& \frac{1}{2} \left[ u^{-1}, \partial_{\mu} u \right]
\nn\\
\Delta_{\mu} &=& \frac{1}{2} \left\{ u^{-1} , \partial_{\mu}u \right\}
= - \frac{1}{2} \left\{ u, \partial_{\mu} u^{-1} \right\} 
\eea
which allows us to define an appropriate covariant derivative of $\pi$ as follows:
\beq
d_{\mu} \pi = \partial_{\mu} \pi + \left[ \Gamma_{\mu}, \pi \right] 
\eeq
or, in components:
\beq
( d_{\mu} \pi )^a t^a =  \partial_{\mu} \pi^a t^a + \pi^a \left[ \Gamma_{\mu}, t^a \right] .
\eeq
We can rewrite the background action in terms of $\Delta$ and its derivatives; we find that
\bea
\tilde{S} &=& \int d^4 x \,
\Biggl\{ \frac{2}{c_0} \delta^{cd} \delta^{\alpha\beta} \Delta^c_{\alpha} \Delta^{d}_{\beta}
+ \frac{1}{f^2} {\cal P}^{\alpha\beta\gamma\delta} \delta^{cd}
( d_{\alpha} \Delta_{\beta} )^c ( d_{\gamma} \Delta_{\delta} )^d
\nn\\
&+& 4 \left\{  \left[ \alpha_6^2 \textrm{Tr}\bigl( t^e t^f t^c t^d \bigr) 
- \alpha_5^2 \delta^{ef} \delta^{cd} \right]  \delta^{\alpha\gamma} \delta^{\beta\delta}
- \left[ \alpha_7^2 \textrm{Tr}\bigl( t^e t^f t^c t^d \bigr)
+ \alpha_4^2 \delta^{ef} \delta^{cd} \right] \delta^{\alpha\beta} \delta^{\gamma\delta} \right\}
\Delta^{c}_ {\alpha} \Delta^{d}_{\beta} \Delta^e_{\gamma } \Delta^f_{\delta}   
\Biggr\} 
\eea
where we have employed the following definitions:
\bea
{\cal P}^{\alpha\beta\gamma\delta} &=& 
\left(1 - f^2\alpha_2^2 \right) \delta^{\alpha\beta} \delta^{\gamma\delta}
+ \left(1 + f^2\alpha_2^2 \right) \delta^{\alpha\gamma} \delta^{\beta\delta}
\nn\\
\alpha_6^2 &=& \frac{1}{f^2} + \alpha_2^2 - 2\alpha_3^2
\nn\\
\alpha_7^2 &=& \frac{1}{f^2} + \alpha_2^2 + 2\alpha_3^2 .
\eea
In this representation, we should understand that $\alpha_2 = \alpha_2(\alpha_6, \alpha_7, f)$ and 
$\alpha_3 = \alpha_3(\alpha_6, \alpha_7)$ in the subsequent expressions.

Concerning the quadratic part in $\pi$, the expansion in the fluctuation produces (in terms of the components $\pi^{a}$)
\bea
S_{\textrm{quad}} &=& \frac{1}{2} \int d^4 x \, \pi^a {\cal D}^{ab} \pi^b
\nn\\
{\cal D}^{ab} &=& - \frac{1}{f^2} \delta^{ab}  D_{\mu} D^{\mu} D_{\nu} D^{\nu}
+ {\cal B}_{(\mu\nu)}^{ab} D^{\nu} D^{\mu} 
+ {\cal C}_{\mu}^{ab} D^{\mu}
+ {\cal E}^{ab}
\nn\\
{\cal C}_{\mu}^{ab} &=& \frac{1}{f^2} Z_{\mu}^{ab (1)}
+ \sum_{i=2}^{5} \frac{1}{\alpha_i^2} Z_{\mu}^{ab (i)}
\nn\\
{\cal E}^{ab} &=& \frac{1}{c_0} \hat{\sigma}^{ab} + \frac{1}{f^2} Q^{ab (1)}
+ \sum_{i=2}^{5} \alpha_i^2 Q^{ab (i)}  
+  \frac{1}{2} {\cal B}_{\nu\mu}^{ac} {\cal R}^{\mu\nu, cb}
\label{14}
\eea
where
\bea
d_{\mu} \pi  &=& ( d_{\mu} \pi )^a t^a = ( D_{\mu} \pi^a ) t^a
\nn\\
D_{\mu} \pi^a &=& \delta^{ab} \partial_{\mu} \pi^b  + \hat{\Gamma}^{ab}_{\mu} \pi^b
\nn\\
\hat{\Gamma}^{ab}_{\mu} &=& - 2  \textrm{Tr}\left( [ t^a,t^b ] \Gamma_{\mu} \right)
\nn\\
\hat{\sigma}^{ab} &=& 2 \textrm{Tr}\left( [ \Delta_{\mu},t^a ] [ \Delta^{\mu},t^b ] \right).
\eea
In addition, as usual the curvature (or field strength) ${\cal R}_{\alpha\beta}$ arises as the commutator of the covariant derivatives
\beq
\bigl( D_{\nu} D_{\mu} - D_{\mu} D_{\nu} \bigr) \pi^a =  {\cal R}^{ab}_{\nu\mu} \pi^b,
\eeq
but it can be also calculated from $\hat{\Gamma}_{\mu}$ (which acts naturally as a connection)
\beq
{\cal R}_{\mu\nu} = \partial_{\mu} \hat{\Gamma}_{\nu} - \partial_{\nu} \hat{\Gamma}_{\mu}
+ [ \hat{\Gamma}_{\mu}, \hat{\Gamma}_{\nu} ]
\eeq
or, in components
\beq
{\cal R}^{ab}_{\mu\nu} = - 2  \textrm{Tr}\left( [ t^a,t^b ] f_{\mu\nu} \right)
= \partial_{\mu} \hat{\Gamma}^{ab}_{\nu} - \partial_{\nu} \hat{\Gamma}^{ab}_{\mu}
+  \hat{\Gamma}^{ac}_{\mu}  \hat{\Gamma}^{cb}_{\nu}
- \hat{\Gamma}^{ac}_{\nu}  \hat{\Gamma}^{cb}_{\mu} 
\eeq
where
\beq
\left[ d_{\nu}, d_{\mu} \right] \pi =
\left[ \partial_{\nu}\Gamma_{\mu} - \partial_{\mu}\Gamma_{\nu} + [\Gamma_{\nu},\Gamma_{\mu}], \pi \right]
 = [ f_{\nu\mu}, \pi ] .
\eeq
Also:
\beq
f_{\nu\mu} = - [ \Delta_{\nu}, \Delta_{\mu}] .
\eeq
In view of very lengthy terms concerning ${\cal B}_{\nu\mu}^{ab}$ and $Q^{ab (i)}$, we gather their complete contribution in Appendix B. Moreover, even though we will not need the explicit expression for 
${\cal C}_{\mu}^{ab}$ for the heat-kernel computation, we also present the full expressions for 
$Z_{\mu}^{ab (i)}$ in Appendix B.

\section{Identifying physical running}\label{physical}

The ultimate goal of this section is to provide a prescription for going from the generic heat kernel to the beta functions. We will do this explicitly with the HDNLSM presented above, but the general reasoning should also work to other higher-derivative theories, including the other ones discussed in this paper. We also use dimensional regularization.

We now move to a full discussion of the one-loop renormalization of the HDNLSM. We are particularly interested in the application of the Schwinger-DeWitt technique -- for this we need the minimal fourth-order operator in the standard form. This is obtained by multiplying ${\cal D}^{ab}$ by $- f^2$. It is obvious that this multiplication does not affect the divergences. Now we can use that
\beq
\textrm{Tr} \ln {\cal D} = \int d^d x \textrm{Tr} \langle x| \ln {\cal D} |x\rangle
= - \int d^d x \int_{0}^{\infty} \frac{d\tau}{\tau} \textrm{Tr} \langle x|  e^{-\tau{\cal D}} |x\rangle
\eeq
and evaluate the action using the heat kernel $\langle x|  e^{-\tau{\cal D}} |x\rangle$. The expansion in terms of the Seeley-DeWitt coefficients yields
\beq
\langle x|  e^{-\tau{\cal D}} |x\rangle = \frac{i}{(4\pi)^{d/2}} \frac{e^{-\tau^{1/2} m^2}}{\tau^{d/4}}
\sum_{n=0}^{\infty} \tau^{n/2} a_n(x)
\eeq
where $m$ is an infrared regulator. The above evaluation can be done through a simple Mellin transform and the result is
\beq
\textrm{Tr} \langle x| \ln {\cal D} |x\rangle = - \frac{2 i}{(4\pi)^{d/2}}
\sum_{n=0}^{\infty} m^{d-2 n} \Gamma \left(n-\frac{d}{2}\right)
\textrm{Tr} \, a_n(x)
\eeq
plus a divergent constant having no physical consequences. We see that as $d \to 4$ the UV divergent part resides in the first coefficients in the expansion; in particular, for $m \to 0$, the contributions for $n=0$ and $n=1$ vanish. For a detailed discussion of Seeley-DeWitt coefficients, see Refs.~\cite{bos,bar,Donoghue:2017fvm,Donoghue:1992dd,Barvinsky:2021ijq}. 

We briefly discuss the calculation of the coefficients $a_0$, $a_1$ and $a_2$ in Appendix C. We find
\beq
a_0 = \frac{\Gamma(d/4)}{2 \Gamma(d/2)} {\bf 1}
\eeq
\beq
a_1 = - \frac{\Gamma \left(\frac{d/2-1}{2}\right)}{2 \Gamma \left(\frac{d}{2}-1\right)}
\frac{f^2 {\cal B}}{2d}
\eeq 
and
\beq
a_2 = \frac{\Gamma(d/4)}{4\Gamma(d/2)}
\left[ \frac{(d-2)}{6} {\cal R}_{\mu\nu}  {\cal R}^{\mu\nu}
+ \frac{1}{(d + 2)} \left(  \frac{f^4}{2} {\cal B}_{(\mu\nu)}  {\cal B}^{\mu\nu} 
+ \frac{f^4}{4}  {\cal B}^2  \right) + 2 f^2 {\cal E} 
\right]
\eeq
where ${\cal B} = \delta_{\mu\nu} {\cal B}^{\mu\nu}$. A simple dimensional analysis shows us that the divergence associated with $a_2$ comes from the bubble diagram -- the ${\cal B}^2$ terms -- and the tadpole -- last term in the expression for $a_2$; triangles and boxes do not contribute to the divergences. However, as we will discuss below, in the ${\cal B}^2$ and ${\cal B}_{\mu\nu}^2$ terms we have the presence of {\it hidden} tadpoles which will have an important impact on the calculation of associated beta functions.

Let us now address the one-loop divergences encoded in the coefficient $a_2$ -- the evaluation of the associated traces can be found in Appendix B. The final result reads
\bea
\Gamma^{(1)}_{\textrm{Div}} &=& - \frac{\mu^{d-4}}{16 \pi ^2 (d-4)} \int d^d x
\Biggl\{ \frac{f^4}{2} \frac{ (N^2-1)}{c_0^2} 
+ \frac{b_0}{c_0} \delta ^{c d} \delta^{\alpha\beta} (\Delta_{\alpha})^c (\Delta_{\beta})^d  
\nn\\
&-& \frac{N}{2} \left[ \left(1 - {f^2}{\alpha_2^2}  \right) \delta^{\alpha\beta} \delta^{\gamma\delta} 
+ \left(1 + {f^2}{\alpha_2^2}  \right) \delta^{\alpha\gamma} \delta^{\beta\delta} \right]
\delta^{c d} (d_{\alpha} \Delta_{\beta})^c (d_{\gamma} \Delta_{\delta})^d 
\nn\\
&+& \biggl[ \Bigl( \beta_1 \delta ^{e f} \delta ^{c d} 
+ \beta_2 \text{Tr}(t^e t^f t^c t^d ) \Bigr)
\delta^{\alpha\beta} \delta^{\gamma\delta}
 + \left( \beta_3 \delta ^{e f} \delta ^{c d} 
+ \beta_4 \text{Tr}(t^e t^f t^c t^d) 
+ \frac{2 N}{6} \textrm{Tr}\bigl( [t^e, t^f] [t^c, t^d] \bigr)
\right)
\delta^{\alpha\gamma} \delta^{\beta\delta}
 \biggr] 
 \nn\\
 &\times& (\Delta_{\alpha})^c (\Delta_{\beta})^d (\Delta_{\gamma})^e (\Delta_{\delta})^f
 \Biggr\}
\nn\\
&=& - \frac{\mu^{d-4}}{16 \pi ^2 (d-4)} \int d^d x
\Biggl\{ \frac{f^4}{2} \frac{ (N^2-1)}{c_0^2} 
+ \frac{b_0}{c_0} \delta ^{c d} \delta^{\alpha\beta} (\Delta_{\alpha})^c (\Delta_{\beta})^d  
- \frac{N}{2} {\cal P}^{\alpha\beta\gamma\delta}
\delta^{c d} (d_{\alpha} \Delta_{\beta})^c (d_{\gamma} \Delta_{\delta})^d 
\nn\\
&+& \biggl[ \Bigl( \beta_1 \delta ^{e f} \delta ^{c d} 
+ \left( \beta_2 - \frac{4 N}{6} \right) \text{Tr}(t^e t^f t^c t^d ) \Bigr)
\delta^{\alpha\beta} \delta^{\gamma\delta}
 + \left( \beta_3 \delta ^{e f} \delta ^{c d} 
+ \left( \beta_4 + \frac{4 N}{6} \right) \text{Tr}(t^e t^f t^c t^d) 
\right)
\delta^{\alpha\gamma} \delta^{\beta\delta}
 \biggr] 
 \nn\\
 &\times& (\Delta_{\alpha})^c (\Delta_{\beta})^d (\Delta_{\gamma})^e (\Delta_{\delta})^f
 \Biggr\}
\eea
where
\bea
b_0 &=& - f^4 \left( \frac{N}{2 f^2} + \frac{3 N}{2} \alpha _2^2 
+ \frac{7 N^2 - 12}{2 N} \alpha _3^2
+ {\left(4 N^2-2\right)}{\alpha _4^2} 
+ {\left(N^2+4\right)}{\alpha _5^2} \right)
\nn\\
\beta_1 &=& \frac{f^4}{24} b_1 + \frac{f^4}{48} b_5 - \frac{1}{2}  + {4 N f^2}{\alpha_4^2}
\nn\\
\beta_2 &=& \frac{f^4}{24} b_2 + \frac{f^4}{48} b_6 
-  2 N  + {10 N f^2}{\alpha_3^2} - {16 f^2}{\alpha_4^2} + {8 f^2}{\alpha_5^2}
\nn\\
\beta_3 &=& \frac{f^4}{24} b_3 + \frac{f^4}{48} b_7 
- 1 + {4 N f^2}{\alpha_5^2} 
\nn\\
\beta_4 &=& \frac{f^4}{24} b_4 + \frac{f^4}{48} b_8 
+ {6 N f^2}{\alpha_3^2} + {16 f^2}{\alpha_4^2} - {8 f^2}{\alpha_5^2} .
\label{36}
\eea
The definition of the functions $b_k$, $k=1, \ldots 8$, can also be found in Appendix B. 

{When one tracks the dependence on $\mu$,} the rate of change of the renormalized couplings at the scale $\mu$ would then be given by
\bea
 \mu \frac{d c_0}{d\mu} &=& \frac{b_0}{32 \pi^2} c_0 = \beta_{c_0}
\nn\\
\mu \frac{d f}{d\mu} &=& -  \frac{N}{64 \pi ^2} f^3 = \beta_{f}
\nn\\
 \mu \frac{d \alpha_7^2}{d\mu} &=& \frac{1}{64 \pi ^2 } \left( \beta_2 - \frac{4 N}{6} \right) 
= \beta_{\alpha_7}
\nn\\
\mu \frac{d \alpha_4^2}{d\mu} &=& \frac{1}{64 \pi ^2} \beta_1  = \beta_{\alpha_4}
\nn\\
\mu \frac{d \alpha_5^2}{d\mu} &=& \frac{1}{64 \pi ^2} \beta_3   = \beta_{\alpha_5}
\nn\\
\mu \frac{d \alpha_6^2}{d\mu} &=& - \frac{1}{64 \pi ^2} \left( \beta_4 + \frac{4 N}{6} \right)  
= \beta_{\alpha_6}
\eea
where we have identified the associated beta functions. Observe that only tadpole terms in the $a_2$ coefficient contribute to $\beta_{f}$. The expression for $\beta_{f}$ above and the coefficient $b_0$ both agree with the ones calculated in Ref.~\cite{Hasenfratz:1988rf}.

Given that $\textrm{Tr} \bigl( {\cal E} \bigr)$ is a tadpole, we expect it to renormalize couplings but not contribute to the beta functions. Hence let us remove its contribution to the aforementioned beta functions. We find that
\bea
\beta_1 &=& \frac{f^4}{24} b_1 + \frac{f^4}{48} b_5 
\nn\\
\beta_2 &=& \frac{f^4}{24} b_2 + \frac{f^4}{48} b_6 
\nn\\
\beta_3 &=& \frac{f^4}{24} b_3 + \frac{f^4}{48} b_7 
\nn\\
\beta_4 &=& \frac{f^4}{24} b_4 + \frac{f^4}{48} b_8 .
\eea
In particular, the only contribution to $\beta_f$ comes from $\textrm{Tr} \bigl( {\cal E} \bigr)$, and hence we also should set
\beq
\beta_f = 0.
\eeq
In other words, despite the interpretation given in Ref.~\cite{Hasenfratz:1988rf}, the coupling $f$ does not run. In turn, as the $1/c_0$ term only modifies the two point function and the only renormalization for this quantity at one loop is expected to be a tadpole diagram, this means that $1/c_0$ does not run either, which implies that
\beq
\beta_{c_0} = 0.
\eeq
Hence the factor $b_0$ is {\it not} to be regarded as a beta function -- it is just a factor that renormalizes the coupling $1/c_0$. Everything that contributes to $b_0$ should be considered as a tadpole since no momenta is running through the loop and therefore should be discarded from the beta functions. Observe that all contributions to $b_0$ come from the first term of the expression for $\textrm{Tr} \bigl( {\cal E} \bigr)$ as well as the second term of the traces of ${\cal B}^2$ terms -- that is why we termed such contributions as tadpoles. 

In order to recover the results of Gasser and Leutwyler from Ref.~\cite{Gasser:1984gg} for the SU(N) case, one should consider the limits $f \to \infty$ and $\alpha_i \to 0$ at the level of the differential operator given in Eq.~(\ref{14}) -- it can be shown that ${\cal C}_{\mu}^{ab}$ vanishes in these limits. We obtain
\bea
S_{\textrm{quad}} &=& \frac{1}{2 c_0} \int d^4 x \, \pi^a {\cal D}^{ab} \pi^b
\nn\\
{\cal D}^{ab} &=& \delta^{ab} D^{2} + \hat{\sigma}^{ab} .
\eea
The background action in terms of $\Delta$ now reads
\bea
\tilde{S} &=& \frac{2}{c_0} \int d^4 x \, \delta^{cd} \delta^{\alpha\beta} \Delta^c_{\alpha} \Delta^{d}_{\beta}
\eea
The coefficient $a_2$ for this quadratic operator reads
\beq
a_2 = \frac{1}{12} {\cal R}_{\mu\nu}  {\cal R}^{\mu\nu} + \frac{1}{2} \hat{\sigma}^2 .
\eeq
Hence the one-loop divergences are given by
\bea
\Gamma^{(1)}_{\textrm{Div}} &=& - \frac{\mu^{d-4}}{16 \pi ^2 (d-4)} \int d^d x
\biggl[ \Bigl( \bar{\beta}_1 \delta ^{e f} \delta ^{c d} 
+ \bar{\beta}_2 \text{Tr}(t^e t^f t^c t^d ) \Bigr)
\delta^{\alpha\beta} \delta^{\gamma\delta}
\nn\\
&+& \left( \bar{\beta}_3 \delta^{e f} \delta^{c d} 
+ \bar{\beta}_4 \text{Tr}(t^e t^f t^c t^d) 
+ \frac{2 N}{12} \textrm{Tr}\bigl( [t^e, t^f] [t^c, t^d] \bigr)
\right)
\delta^{\alpha\gamma} \delta^{\beta\delta}
 \biggr] 
 (\Delta_{\alpha})^c (\Delta_{\beta})^d (\Delta_{\gamma})^e (\Delta_{\delta})^f
\nn\\
&=& - \frac{\mu^{d-4}}{16 \pi ^2 (d-4)} \int d^d x
 \biggl[ \Bigl( \bar{\beta}_1 \delta^{e f} \delta^{c d} 
+ \left( \bar{\beta}_2 - \frac{4 N}{12} \right) \text{Tr}(t^e t^f t^c t^d ) \Bigr)
\delta^{\alpha\beta} \delta^{\gamma\delta}
\nn\\
&+& \left( \bar{\beta}_3 \delta^{e f} \delta^{c d} 
+ \left( \bar{\beta}_4 + \frac{4 N}{12} \right) \text{Tr}(t^e t^f t^c t^d) 
\right)
\delta^{\alpha\gamma} \delta^{\beta\delta}
 \biggr] 
  (\Delta_{\alpha})^c (\Delta_{\beta})^d (\Delta_{\gamma})^e (\Delta_{\delta})^f
\eea
where
\bea
\bar{\beta}_1 &=& \frac{1}{4} = - \frac{1}{2} \beta_1(f=0)
\nn\\
\bar{\beta}_2 &=& N = - \frac{1}{2} \beta_2(f=0)
\nn\\
\bar{\beta}_3 &=& \frac{1}{2} = - \frac{1}{2} \beta_3(f=0)
\nn\\
\bar{\beta}_4 &=& 0 = - \frac{1}{2} \beta_4(f=0)
\eea
where the $\beta_i$s on the right-hand sides refer to the ones in Eq.~(\ref{36}), i.e., with the tadpole contributions. So unsurprisingly we see that higher-derivative terms are needed for renormalization -- in particular, terms associated with the couplings $\alpha_i$, $i=4,5,6,7$, get renormalized. Therefore:
\bea
\mu \frac{d \alpha_7^2}{d\mu} &=& \frac{1}{64 \pi ^2 } \left( \bar{\beta}_2 - \frac{4 N}{12} \right) 
= \bar{\beta}_{\alpha_7}
\nn\\
\mu \frac{d \alpha_4^2}{d\mu} &=&  \frac{1}{64 \pi ^2 } \bar{\beta}_1  = \bar{\beta}_{\alpha_4}
\nn\\
\mu \frac{d \alpha_5^2}{d\mu} &=&  \frac{1}{64 \pi ^2 } \bar{\beta}_3  = \bar{\beta}_{\alpha_5}
\nn\\
\mu \frac{d \alpha_6^2}{d\mu} &=& - \frac{1}{64 \pi ^2 } \left( \bar{\beta}_4 + \frac{4 N}{12} \right)  
= \bar{\beta}_{\alpha_6} .
\eea
Observe that such couplings also run at low energies (below the scale set by $1/c_0$), but with different values. Furthermore, note that the renormalization of the coupling $c_0$ is not incorporated in the coefficient $a_2$, in contrast with the higher-derivative result.

For theories with only two derivatives in the kinetic energy, the separation of tadpoles and bubbles in the heat-kernel formalism is evident -- the former appears in the coefficient $a_1$ whereas the latter emerges in the coefficient $a_2$. {We have seen that this }clear separation is no longer true for higher-derivative theories; in this case one can also identify tadpole terms in the corresponding coefficient $a_2$. This entails non-trivial consequences for the beta functions of theory. The origin for this feature can be traced back to the fact that such theories carry an intrinsic mass scale with them.

In summary, the higher derivative full nonlinear sigma model has many couplings and perhaps the lessons of the calculation can be lost in the multiplicity of couplings. The most important lesson here is that the beta function of the fundamental coupling $f$ vanishes, in contrast with previous claims in the literature  \cite{Hasenfratz:1988rf, Percacci:2009fh} .

\section{One-loop effective action for the SU(N) non-linear sigma model}\label{effectiveaction}

In this section we calculate the full one loop effective action and compare it with the result obtained using heat-kernel methods. It is here that the logarithmic terms appear in infrared sensitive portions of the Feynman integrals, which are not directly tied to divergences.

As above, our background fields will be $\Delta_{\mu}$ and 
$\Gamma_{\mu}$ (of course, they are not independent). The associated background action is given by
\bea
\tilde{S} &=& \int d^4 x \,
\Biggl\{ \frac{1}{4N} \textrm{Tr} \bigl( {\cal R}_{\mu\nu}  {\cal R}^{\mu\nu} \bigr) 
+ \frac{1}{f^2} {\cal P}^{\alpha\beta\gamma\delta} \delta^{cd}
( d_{\alpha} \Delta_{\beta} )^c ( d_{\gamma} \Delta_{\delta} )^d
+ m^2 \delta^{cd} \delta^{\alpha\beta} \Delta^c_{\alpha} \Delta^{d}_{\beta}
\nn\\
&+& 4 \left\{  \left[ \left(\frac{1}{\alpha_6^2} - \frac14 \right) \textrm{Tr}\bigl( t^e t^f t^c t^d \bigr) 
- \frac{1}{\alpha_5^2} \delta^{ef} \delta^{cd} \right]  \delta^{\alpha\gamma} \delta^{\beta\delta}
\right.
\nn\\
&-& \left. \left[ \left( \frac{1}{\alpha_7^2} - \frac14 \right) \textrm{Tr}\bigl( t^e t^f t^c t^d \bigr)
+ \frac{1}{\alpha_4^2} \delta^{ef} \delta^{cd} \right] \delta^{\alpha\beta} \delta^{\gamma\delta} \right\}
\Delta^{c}_ {\alpha} \Delta^{d}_{\beta} \Delta^e_{\gamma } \Delta^f_{\delta}   
\Biggr\} 
\eea
where $m^2 = 2/c_0$. For convenience we rescale the pion field by $\pi \to i f \pi$. The differential operator is then
$$
{\cal D}^{ab} =  \delta^{ab}  D_{\mu} D^{\mu} D_{\nu} D^{\nu}
- f^2 {\cal B}_{(\mu\nu)}^{ab} D^{\nu} D^{\mu} -f^2 {\cal C}_{\mu}^{ab} D^{\mu}
-f^2 {\cal E}^{ab} .
$$
After integrating out the pions, we obtain the one-loop effective action:
\bea
S_{\textrm{eff}} &=& \frac{1}{2} \textrm{Tr} \ln {\cal D} 
= \frac{1}{2} \textrm{Tr} \ln( \Box^{2} - M^2 \Box )
\nn\\
&+& \frac{1}{2} \textrm{Tr}\left[\frac{1}{\Box^{2} - M^2 \Box} \partial_X^{(k)}
- \frac{1}{2}\frac{1}{\Box^{2} - M^2 \Box} \partial_X^{(k)}
 \frac{1}{\Box^{2} - M^2 \Box} \partial_X^{(k)} \right]
+ {\cal O}(\pi^3)
\eea
where $M^2 = f^2 m^2/2$ and the notation $\partial_X^{(k)}$ is schematically taking into account $k$ powers of $X=\Gamma$ and $X=\Delta$. 

The first term in the expression of the one-loop effective action gets canceled by a suitable normalization factor defined in the path integral. {Next come the tadpole and bubble diagrams.}  Considering states that are normalized such that
$$
{}^a\langle x|x' \rangle^b = \delta(x-x') \delta^{ab}
$$
one can define
\beq
{}^a\left\langle x \left| \frac{1}{\Box^{2} - M^2 \Box} \right|x' \right\rangle^b 
= \frac{1}{\Box_x^{2} - M^2 \Box_x} \delta^{ab} \delta(x-x')
= \delta^{ab} \int \frac{d^4 k}{(2\pi)^4} \frac{e^{-ik \cdot (x-x')}}{k^4 + M^2 k^2}
= D^{ab}(x-x')
\eeq
where $D^{ab}(x-x')$ is the Feynman propagator for the pion field. In turn, the trace of an operator $M$, which acts in this space, is defined by
$$
\textrm{Tr} M = \int d^{d}x  M_{xx}
= \int d^{d}x  \langle x|M|x \rangle.
$$
Therefore
\bea
\textrm{Tr}\left[\frac{1}{\Box^{2} - M^2 \Box} \partial_X^{(k)}  \right]
&=&  \int d^d x \int d^d x' \,{}^a\left\langle x \left|\frac{1}{\Box^{2} - M^2 \Box} \right|x' \right\rangle^b
{}^b\left\langle x' \left| \partial_X^{(k)} \right|x \right\rangle^a 
\nn\\
&=& \int d^{d}x \int d^d x' D^{ab}(x-x') \big( \partial_{X,x}^{(k)} \big)^{ba} \delta(x'-x).
\eea
This will produce tadpole integrals. Notice that in the massless (high-energy) limit, these are just scaleless integrals and therefore vanish within dimensional regularization. In any case, as a sane check, we will calculate the tadpoles explicitly keeping the full form of the propagator as we wish to reproduce the coefficient of the term $(d \Delta)^2$ as derived above using the heat kernel method. For this calculation we are going to need vertices with two pions and two and four $\Delta$s, as well as mixed vertices with one and two $\Gamma$s. The explicit computation of such vertices can be found in the Appendix D. {In addition, we relegate the computation of the full tadpole result to Appendix E, so that we can focus on the more relevant bubble diagrams.}

{The bubble diagrams are the ones which encode the running with the momentum.}  We need to work out two bubbles, namely
\beq
\Pi^{c,\alpha;c^{\prime},\alpha^{\prime}}(q) = 
\mu^{4-d} \int \frac{d^d k}{(2\pi)^d}
{\cal V}_{\Gamma}^{abc,\alpha}(k,q) 
D^{aa^{\prime}}(k)
{\cal V}_{\Gamma}^{a^{\prime}b^{\prime}c^{\prime},\alpha^{\prime}}(k,q) 
D^{bb^{\prime}}(k-q) 
\eeq
(two-point function for the $\Gamma$s) and the one associated with the four-point function for the 
$\Delta$s. First, consider $\Pi^{c,\alpha;c^{\prime},\alpha^{\prime}}(q)$. We will neglect any massive tadpoles generated in the tensor reduction. We find that
\bea
\Pi^{c,\alpha;e,\beta}(q) &=& \frac{\mu^{4-d}}{M^4} \int \frac{d^d k}{(2\pi)^d} 
\left( \frac{1}{k^2+M^2}\frac{1}{(k-q)^2+M^2}
+ \frac{1}{k^2}\frac{1}{(k-q)^2}
- \frac{1}{k^2+M^2}\frac{1}{(k-q)^2}
- \frac{1}{k^2}\frac{1}{(k-q)^2+M^2} \right)
\nn\\
&\times& 
f^{a b c} \big( k^2 + (k-q)^2 + M^2 \big) \big( - k^{\alpha} - (k^{\alpha}-q^{\alpha})\big) 
f^{a b e} \big( k^2 + (k-q)^2 + M^2 \big) \big( k^{\beta} + (k^{\beta}-q^{\beta})\big) 
\nn\\
&=& \frac{1}{d-1}
\left\{ 
\left( \frac{4 M^2}{q^2} + 1 \right)  I(M^2, M^2,q^2)
+  I(0,0,q^2) \right\} \left(q^2 \delta^{\alpha  \beta }-q^{\alpha } q^{\beta }\right)
f^{a b c} f^{a b e}
+ \cdots
\eea
where the ellipsis indicates the massive tadpoles left out of the expression, as asserted above (in fact, the crossed terms only yield massive tadpoles) and we have defined
\bea
I(\mu_1^2, \mu_2^2,q^2) &\equiv& \mu^{4-d} \int \frac{d^d k}{(2\pi)^d}
\frac{1}{k^2+\mu_1^2}\frac{1}{(k-q)^2+\mu_2^2}
\nn\\
&=& \frac1{16\pi^2}\left[  \frac1{\epsilon} + \gamma -\log 4\pi 
- \int_0^1 dx \log \left(  \frac{x \mu_1^2 + (1-x) \mu_2^2 - q^2x(1-x)}{\mu^2}\right)    \right] .
\eea
The logarithmic integral has the form
\bea
&& \int_0^1 dx \log \left(  \frac{x \mu_1^2 + (1-x) \mu_2^2 -q^2x(1-x)}{\mu^2}\right)  
= \log \left(-\frac{q^2}{\mu^2} \right)-2~,~~~~~~~~~~~~~ \mu_1=\mu_2 = 0 
\nonumber \\
 &&= \log \frac{M^2}{\mu^2} + \left( 1-\frac{M^2}{q^2} \right) \log \left( 1-\frac{q^2}{M^2} \right)-2~,~~~~~~~~~~~~~~~ \mu_1=0, ~\mu_2=M  \nonumber \\
 &&= \log \frac{M^2}{\mu^2}  + \sqrt{1-\frac{4M^2}{q^2}} \log \left( \frac{\sqrt{1-4M^2/q^2}+1}{\sqrt{1-4M^2/q^2}-1} \right)   -2~,~~\mu_1=\mu_2=M   \ \ .
 \eea
In the low energy region, only the term with $ \mu_1=\mu_2 = 0$ gives kinematic logarithms. However at high energy each of them involves equal factors of $\log(-q^2)$.

In turn, one also has that
\bea
\mu ^{4-d} \int \frac{d^d p}{(2\pi)^d} 
\Gamma^{c}_{\alpha}(-p) \Gamma^{e}_{\beta}(p) \Pi^{c,\alpha;e,\beta}(p)
&=& \mu^{2 d-8} \int d^{d} x \int d^{d} x' 
\mu^{4-d} \int \frac{d^d k}{(2\pi)^d}  \Gamma^{c}_{\alpha}(k) e^{- i k \cdot x}
\nn\\
&\times& \mu^{4-d} \int \frac{d^d k'}{(2\pi)^d}  \Gamma^{e}_{\beta}(k') e^{- i k' \cdot x'}
\mu^{4-d} \int \frac{d^d p}{(2\pi)^d} \Pi^{c,\alpha;e,\beta}(p) e^{- i p \cdot (x-x')}
\nn\\
&=& \mu^{2 d-8} \int d^{d} x \int d^{d} x' \Gamma^{c}_{\alpha}(x) \Pi^{c,\alpha;e,\beta}(x-x') 
\Gamma^{e}_{\beta}(x')
\eea
where
\bea
\Pi^{c,\alpha;e,\beta}(x) &=&
\mu^{4-d} \int \frac{d^d p}{(2\pi)^d} \Pi^{c,\alpha;e,\beta}(p) e^{- i p \cdot x}
\nn\\
&=& \frac{f^{a b c} f^{a b e}}{d-1}
\left\{ 
\left( - \frac{4 M^2}{\Box} + 1 \right) 
\left( -\Box \delta^{\alpha  \beta } + \partial^{\alpha } \partial^{\beta }\right) D^{2}_{F}(M^2,x)
+ \left( -\Box \delta^{\alpha  \beta } + \partial^{\alpha } \partial^{\beta }\right) D^{2}_{F}(x) \right\} 
\nn\\
\eea
where $D^{2}_{F}(M^2,x)$ ($D^{2}_{F}(x)$) is the inverse Fourier transform of $ I(M^2, M^2,q^2)$ 
($I(0,0,q^2)$).

To display the beta functions we have to look at the high-energy behavior $q^2 \gg M^2$ of the amplitude. In this limit, essentially we are left with only $D^{2}_{F}(x)$:
\beq
\Pi^{c,\alpha;e,\beta}(x) \bigg|_{q^2 \gg M^2} \to \frac{2}{d-1} f^{a b c} f^{a b e}
\left( -\Box \delta^{\alpha  \beta } + \partial^{\alpha } \partial^{\beta }\right) D^{2}_{F}(x) .
\eeq
Hence, to quadratic order in $\Gamma$ we find that the contribution of the field strength 
${\cal R}_{\mu\nu}$ to the one-loop effective action reads
\beq
S^{\Gamma}_{\textrm{eff}} = \frac{1}{4(d-1)}
 \mu^{2 d-8} \int d^{d} x \int d^{d} x'  
 \textrm{Tr} \bigl( {\cal R}_{\mu\nu}(x)  {\cal R}^{\mu\nu}(x') \bigr)
D^{2}_{F}(x-x') .
\eeq
With a bit more effort one can show that the contribution from the vertices containing three and four $\Gamma$s yields the same result, thereby completing the expression for $ {\cal R}_{\mu\nu}(x)  {\cal R}^{\mu\nu}(x')$. In $d=4$ dimensions we see that we are able to recover the result coming from the heat-kernel computation.

Now let us calculate the bubble associated with the four-point function for the $\Delta$s. From the expansion of the one-loop effective action, we find that
\bea
&-&\frac{1}{2}\textrm{Tr}\left[\frac{1}{\Box^{2} - M^2 \Box} \partial_X^{(k)} 
\frac{1}{\Box^{2} - M^2 \Box} \partial_X^{(k)} \right]
\nn\\
&=& -\frac{1}{2} \int d^{d}x  \int d^{d}x' \int d^{d}y \int d^{d}z
{}^a\left\langle x \left|\frac{1}{\Box^{2} - M^2 \Box} \right| x' \right\rangle^b
{}^b\left\langle x' \left| \partial_X^{(k)} \right| y \right\rangle^c
{}^c\left\langle y \left|\frac{1}{\Box^{2} - M^2 \Box} \right| z \right\rangle^d
{}^d\left\langle z \left| \partial_X^{(k)} \right| x \right\rangle^a
\nn\\
&=& - \frac{1}{2} \int d^{d}x  \int d^{d}x' \int d^{d}y \int d^{d}z
D^{ab}(x-x') \big( \partial_{X,x'}^{(k)} \big)^{bc} \delta(x'-y)
D^{cd}(y-z) \big( \partial_{X,x}^{(k)} \big)^{da} \delta(z-x) .
\eea
To calculate the four-point function for the $\Delta$ field we only need the contribution
$\widetilde{{\cal B}}^{ab}_{s\mu\nu}$, whose expression is the same as ${\cal B}_{(\mu\nu)}^{ab}$ given in Appendix B but without the term $\frac{1}{c_0} \delta^{ab} \delta_{\nu\mu}$. Hence we need to calculate
\bea
&& \frac{f^4}{2} \int d^{d}x  \int d^{d}x' \int d^{d}y \int d^{d}z
D^{ab}(x-x') \widetilde{{\cal B}}_{s\mu\nu}^{bc}(x') \partial_{x'}^{\mu} \partial_y^{\nu} \delta(x'-y)
D^{cd}(y-z) \widetilde{{\cal B}}_{s\lambda\kappa}^{da}(z) 
\partial_{x}^{\lambda} \partial_z^{\kappa} \delta(z-x)
\nn\\
&& = \frac{f^4}{2} \mu^{2d-8} \int d^{d}x  \int d^{d}x' \,
\widetilde{{\cal B}}_{s\mu\nu}^{ab}(x)  
 {\cal M}^{\mu\nu\lambda\kappa}(x-x') 
\widetilde{{\cal B}}_{s\lambda\kappa}^{ba}(x') 
\eea
where
\beq
 {\cal M}^{\mu\nu\lambda\kappa}(x-x') = \partial_{x}^{\mu} \partial_{x}^{\nu} D(x-x') 
\partial_{x}^{\lambda}\partial_{x}^{\kappa} D(x'-x)
\eeq
and
\beq
D(x-x') = \int \frac{d^4 k}{(2\pi)^4} \frac{e^{-ik \cdot (x-x')}}{k^4 + M^2 k^2}
\eeq
Now let us work on ${\cal M}_{\mu\nu\lambda\kappa}(x)$. Resorting to a simple Fourier transform, one gets
\beq
{\cal M}^{\mu\nu\lambda\kappa}(q) =  \mu^{4-d}\int \frac{d^d k}{(2\pi)^d} 
\frac{1}{k^4 + M^2 k^2}
\frac{1}{(k-q)^4 + M^2 (k-q)^2}
k^{\mu} k^{\nu} (k-q)^{\lambda} (k-q)^{\kappa}
\eeq
where ${\cal M}_{\mu\nu\lambda\kappa}(q)$ is the Fourier transform associated with the quantity
${\cal M}_{\mu\nu\lambda\kappa}(x)$. By using partial fraction decomposition and the usual Passarino-Veltman tensor reduction, we obtain that (neglecting tadpoles)
\beq
{\cal M}^{\mu\nu\lambda\kappa}(q) =  \mathbb{M}^{\mu\nu\lambda\kappa}(q) I(M^2, M^2,q^2)
+ \mathbb{N}^{\mu\nu\lambda\kappa}(q) I(0, 0,q^2) 
- \mathbb{P}^{\mu\nu\lambda\kappa}(q) I(M, 0,q^2) .
\eeq
The explicit form of the tensor functions $\mathbb{M}^{\mu\nu\lambda\kappa}(q)$, $\mathbb{N}^{\mu\nu\lambda\kappa}(q)$ and $\mathbb{P}^{\mu\nu\lambda\kappa}(q)$ can be found in Appendix F. Hence
\bea
{\cal M}^{\mu\nu\lambda\kappa}(x) &=& \mu ^{4-d} \int \frac{d^d q}{(2\pi)^d} e^{-i q \cdot x}
{\cal M}^{\mu\nu\lambda\kappa}(q)
\nn\\
&=& \mathbb{M}^{\mu\nu\lambda\kappa}(\partial) D^{2}_{F}(M^2,x)
+ \mathbb{N}^{\mu\nu\lambda\kappa}(\partial) D^{2}_{F}(x)
-  \mathbb{P}^{\mu\nu\lambda\kappa}(\partial) G^2_F(x) 
\eea
where $G^2_F(x)$ is the inverse Fourier transform of $I(M, 0,q^2)$. Hence we get our final result for the one-loop effective action:
\bea
S_{\textrm{eff}} &=& \bar{S} 
+ \frac{1}{8(d-1)}
 \mu^{2 d-8} \int d^{d} x \int d^{d} x'  
\nn\\
&\times& \bigg[ \textrm{Tr} \bigl( {\cal R}_{\mu\nu}(x)  {\cal R}^{\mu\nu}(x') \bigr)
D^{2}_{F}(x-x') 
+  \textrm{Tr} \bigl( {\cal R}_{\mu\nu}(x)  {\cal R}^{\mu\nu}(x') \bigr)
\left( - \frac{4 M^2}{\Box} + 1 \right) D^{2}_{F}(M^2,x-x') 
\bigg]
\nn\\
&+& \frac{f^4}{4} \mu^{2d-8} \int d^{d}x \int d^{d}x'\, 
\widetilde{{\cal B}}_{s\mu\nu}^{ab}(x)  
\widetilde{{\cal B}}_{s\lambda\kappa}^{ba}(x')
\nn\\
&\times& \Bigl[
 \mathbb{M}^{\mu\nu\lambda\kappa}(\partial_x) D^{2}_{F}(M^2,x-x')
+ \mathbb{N}^{\mu\nu\lambda\kappa}(\partial_x) D^{2}_{F}(x-x')
-  \mathbb{P}^{\mu\nu\lambda\kappa}(\partial_x) G^2_F(x-x')
\Bigr] .
\eea
As argued above, in order to display the beta functions we go to the high-energy limit, which amounts to setting $M \to 0$.  After tensor reduction, the term 
$\textrm{Tr} \bigl( {\cal R}_{\mu\nu}(x)  {\cal R}^{\mu\nu}(x') \bigr)$ gets generated by a bubble integral containing only normal propagators; however, the term 
$\widetilde{{\cal B}}_{s\mu\nu}^{ab}(x)  \widetilde{{\cal B}}_{s\lambda\kappa}^{ba}(x')$ is actually generated by three kinds of bubbles, namely one with normal propagators, one with quartic propagators and one with one quartic propagator and one normal propagator. The normal bubble is actually UV divergent, but the other ones are not -- they are, in fact, IR divergent. However, these IR divergent terms need to be considered, otherwise we will not be able to reproduce the result from the heat-kernel calculation. The purely quartic bubble that we need to consider now is
\beq
\widetilde{{\cal M}}^{\mu\nu\lambda\kappa}(q) =  \mu^{4-d}\int \frac{d^d k}{(2\pi)^d} 
\frac{1}{k^4} \frac{1}{(k-q)^4}
k^{\mu} k^{\nu} (k-q)^{\lambda} (k-q)^{\kappa}.
\eeq
After tensor reduction, this turns into:
\beq
\widetilde{{\cal M}}^{\mu\nu\lambda\kappa}(q) =  
\widetilde{\mathbb{M}}^{\mu\nu\lambda\kappa}(q) B_1(q^2)
+ \widetilde{\mathbb{N}}^{\mu\nu\lambda\kappa}(q) B_2(q^2) 
- \widetilde{\mathbb{P}}^{\mu\nu\lambda\kappa}(q) B_3(q^2)
\eeq
where explicit expressions for the functions $\widetilde{\mathbb{M}}^{\mu\nu\lambda\kappa}(q)$, $\widetilde{\mathbb{N}}^{\mu\nu\lambda\kappa}(q)$ and $\widetilde{\mathbb{P}}^{\mu\nu\lambda\kappa}(q)$ can be found in Appendix F and
\bea
B_1(q^2) &=& \mu^{4-d} \int \frac{d^d k}{(2\pi)^d} 
\frac{1}{k^4} \frac{1}{(k-q)^4}
= \frac{\mu^{4-d}}{(4\pi)^{d/2}} \Gamma(4-d/2)
\int_{0}^{1} dx \frac{x (1-x)}{[ -q^2 x(1-x) ]^{4-d/2}}
\nn\\
B_2(q^2) &=& \mu^{4-d}\int \frac{d^d k}{(2\pi)^d} 
\frac{1}{k^2} \frac{1}{(k-q)^2}
= \frac{\mu^{4-d}}{(4\pi)^{d/2}} \Gamma(2-d/2)
\int_{0}^{1} dx \frac{1}{[ -q^2 x(1-x) ]^{2-d/2}}
\nn\\
B_3(q^2) &=& \mu^{4-d}\int \frac{d^d k}{(2\pi)^d} 
\frac{1}{k^4} \frac{1}{(k-q)^2}
= - \frac{\mu^{4-d}}{(4\pi)^{d/2}} \Gamma(3-d/2)
\int_{0}^{1} dx \frac{(1-x)}{[ -q^2 x(1-x) ]^{3-d/2}}
\eea
Here the IR divergences generated by $B_1(q^2)$ and $B_3(q^2)$ always appear with powers of the external momentum $q$ in the denominator and give rise to apparently nonlocal terms. However, after performing the tensor reduction, we verify that the momenta in turn always appears in the combination $q^2$ in the numerator and cancel the inverse powers of $q$. Such logs of infrared origin should also appear as coefficients of local operators. 

Now the one-loop effective action reads
\bea
S_{\textrm{eff}} &=& \bar{S} 
+ \frac{1}{4(d-1)}
 \mu^{2 d-8} \int d^{d} x \int d^{d} x'  
 \textrm{Tr} \bigl( {\cal R}_{\mu\nu}(x)  {\cal R}^{\mu\nu}(x') \bigr)
D^{2}_{F}(x-x') 
\nn\\
&+& \frac{f^4}{4} \mu^{2d-8} 
( T_{\mu\nu}^{\alpha\beta} )^{ab,cd} ( T_{\lambda\kappa}^{\gamma\delta} )^{ba,ef}
\int d^{d}x \int d^{d}x'\, 
\Delta_{\alpha}^c(x) \Delta_{\beta}^d(x) \Delta_{\gamma}^e(x') \Delta_{\delta}^f(x')  
\nn\\
&\times& \Bigl[
 \widetilde{\mathbb{M}}^{\mu\nu\lambda\kappa}(\partial_x) G^{2}_{1}(x-x')
+ \widetilde{\mathbb{N}}^{\mu\nu\lambda\kappa}(\partial_x) D^{2}_{F}(x-x')
- \widetilde{\mathbb{P}}^{\mu\nu\lambda\kappa}(\partial_x) G^2_3(x-x')
\Bigr] 
\eea
where
\bea
G^{2}_{1}(x) &=&  \mu ^{4-d} \int \frac{d^d q}{(2\pi)^d} e^{-i q \cdot x} B_1(q^2)
\nn\\
G^2_3(x) &=&  \mu ^{4-d} \int \frac{d^d q}{(2\pi)^d} e^{-i q \cdot x} B_3(q^2)
\eea
and
\bea
( T_{\mu\nu}^{\alpha\beta} )^{ab,cd} &=& 
\left[ - \frac{2}{\alpha_3^2} \textrm{Tr}\left( \left\{ t^a, t^{(c} \right\}  \left\{ t^b, t^{d)} \right\} \right)
+ 2 \left( \frac{2}{\alpha_2^2} - \frac{1}{\alpha_3^2} \right) 
\textrm{Tr}\left( \left[ t^a, t^{(c} \right] \left[ t^b, t^{d)} \right] \right)
\right.
\nn\\
&+& \left. 2 \left( - \frac{3}{f^2} + \frac{1}{\alpha_2^2} - \frac{1}{\alpha_3^2}  \right)  
\textrm{Tr}\left( \left\{ t^a, t^{(c} \right\} \left[ t^b, t^{d)} \right] \right)
\right.
\nn\\
&+& \left. 2 \left( \frac{1}{f^2} - \frac{1}{\alpha_2^2} - \frac{1}{\alpha_3^2} \right) 
\textrm{Tr}\left( \left[ t^a, t^{(c} \right] \left\{ t^b, t^{d)} \right\} \right)
- \frac{4}{\alpha_3^2} \textrm{Tr}\Bigl( \{ t^c, t^d \} t^a t^b   \Bigr)
- \frac{4}{\alpha_4^2} \delta^{cd} \delta^{ab}
- \frac{4}{\alpha_5^2} \delta^{a (c} \delta^{d) b}
 \right] \delta^{\alpha\beta} \delta_{\nu\mu} 
\nn\\
&+& \left[ - 2 \left( \frac{1}{f^2} + \frac{1}{\alpha_2^2} \right)
\textrm{Tr}\left( 2 [t^a, t^{(d}] [t^b, t^{c)}] - \frac{1}{2} \{ t^a, t^{(d} \} [t^b, t^{c)}]
+ \frac{1}{2} [t^a, t^{(d}] \{ t^b, t^{c)} \} \right) 
\right.
\nn\\
&-& \left. \frac{1}{\alpha_3^2}
\textrm{Tr}\Bigl( 8 t^{(c} t^{d)}  t^a t^b + 4 ( t^{(d} t^{c)} t^b t^a + t^{(c} t^{d)}  t^a t^b )
+ 4 \left\{ t^a, t^{(c} \right\} \left\{ t^b, t^{d)} \right\} \Bigr)
\right.
\nn\\
&-& \left. \frac{8}{\alpha_4^2}  \delta^{a(c} \delta^{d)b}
- \frac{4}{\alpha_5^2} \Bigl(  \delta^{cd} \delta^{ab} + \delta^{a(d} \delta^{c)b} \Bigr) 
\right] \delta^{\alpha}_{\nu}\delta^{\beta}_{\mu}  .
\eea
One can show that the divergent part coincides with the one calculated from heat-kernel methods.

This procedure also reduces to the Gasser and Leutwyler's result of Ref. \cite{Gasser:1984gg} in the appropriate limit, as a similar calculation leads us to
\bea
S_{\textrm{eff}} &=& \bar{S} 
+ \frac{1}{8(d-1)}
 \mu^{2 d-8} \int d^{d} x \int d^{d} x' 
 \textrm{Tr} \bigl( {\cal R}_{\mu\nu}(x)  {\cal R}^{\mu\nu}(x') \bigr)
D^{2}_{F}(x-x') 
\nn\\
&+& \frac{1}{4} \mu^{2 d-8} \int d^{d} x \int d^{d} x'
\textrm{Tr} \bigl( \hat{\sigma}(x) \hat{\sigma}(x') \bigr)
D^{2}_{F}(x-x')  \ \ .
\eea
Again the divergent part coincides with the one calculated from heat-kernel approach.

\section{Discussion and conclusions}

 We have seen a variety of outcomes for the running couplings and for the comparison with other methods. The techniques that correctly identify physical running couplings in standard theories using a mass independent renormalization scheme are seen to often fail when used in a theory with an intrinsic mass scale such as the higher derivative theories explored in this paper. We have used new techniques, also described in \cite{Buccio:2024hys}, to pull out the dependence on the energy scale of the theory which goes in to the physical running coupling. 
 
The essential lesson demonstrated in these calculations is that in theories with operators of different dimensions, following the divergences through $\log \Lambda$ or $\log \mu$ does not always yield the correct behavior of running couplings in physical processes.

Perhaps the most interesting case was that of the fundamental coupling $f$ of the HD SU(N) nonlinear sigma model, which does not run at any energy  in physical processes, It is easy to understand why this is the case.  Because the interactions only involve an even number of fields, the renormalization of the coupling in the propagator uniquely involves the tadpole diagram of Fig. 1b. This does not contain any information on the momentum flowing in the propagator, and hence cannot generate any factor of $\log E^2$. However it is logarithmically divergent, and hence has been previously thought to lead to a running coupling. In particular,  Hasenfratz \cite{Hasenfratz:1988rf} argued that this coupling ran towards asymptotic freedom, by studying the UV cutoff of the theory. Percacci and Zanusso \cite{Percacci:2009fh} followed the IR cutoff using FRG techniques and reached the same conclusion. It follows from direct calculation that these conclusions are misleading for the behavior of physical amplitudes and that amplitudes formed in this theory do not have any physical running at all for the coupling $f$, at this loop order. The study of the UV or IR cutoff dependence of this coupling has been misleading. 
 
In other cases, the running of the couplings depends on whether the amplitude is studied at low energy or high energy. The basic coupling $g$ of the HD U(1) nonlinear sigma model does not run at low energy but does at high energy. In contrast, the coupling $\lambda$ of the HD linear sigma model runs at low energy but not at high energy. Various other couplings have patterns which differ from results reported in the literature. Effective field theory methods are useful in understanding these patterns. 

{Another way to describe our results involves the effective action. The running with the energy scale can be encoded in position space through the use of the operator $\log \Box$, which is the Fourier transform of $\log q^2$. This is a non-local operator acting on the fields in the effective action. With the usual two derivative kinetic energy, the non-local terms in the effective action have been extensively catalogued by Barvinsky and Vilkovisky and collaborators \cite{Barvinsky:1990up, Barvinsky:1993en}.  In theories without a mass scale, the coefficients of the $\log \Box$ terms are determined by the divergences in the local operators, as by dimensional analysis  loops generate $\log (\Box/\mu^2) $ and the coefficient of $\log  \mu^2 $ is tied to the divergence. This yields the usual connection to running couplings. However for higher derivative theories with an intrinsic mass scale the divergences in the local operators and the appearance of $\log \Box$ can  become disconnected in ways that we have documented above. The usual treatment of the heat kernel is used to identify the divergences in the  local operators, and we have provided techniques for identifying the physics which follows from the $\log \Box$ operators. }
 
 These results raise this issue of the usefulness of results following from methods tracing the cutoff dependence, such as the FRG, {in physical reactions}. If the running found in these methods is not reflected in the running of parameters in physical amplitudes, what can be said about the utility of the method? For example, Weinberg's original formulation of Asymptotic Safety was in terms of the scaling behavior of cross-sections~\cite{Weinberg}. These would involve the physical running constants. However, most of the present practice of Asymptotic Safety studies uses the Functional Renormalization Group , which can give running behavior which differs from the running in physical processes. If a coupling runs to a UV fixed point in the FRG, {or to asymptotic freedom} but does not run at all in physical amplitudes (such as the coupling $f$) what is the value of that fixed point determination? 
 
Presumably the correct behavior of amplitudes is contained in the FRG {effective action} if treated completely. However, kinematic logarithms appear as {\em nonlocal} contributions to the effective action.  {We have seen that focusing on the cutoff dependence of the local operators does not always reveal the nonlocal kinematic logs.  However, most current methods only study the local couplings.  At the least, our results tell us that the FRG running of the local couplings should not be used in physical applications unless care is taken to also identify the kinematic logarithms.}
 
We have also provided a roadmap for determining the physical beta functions in theories of this class. At low energy, one can integrate out the heavy degrees of freedom to form a low energy effective field theory. That EFT reveals the correct low energy running. At high energy, one uses the full theory, but needs to separate the kinematic running from the non-kinematic effects of $\log m^2$. This requires a direct calculation of amplitudes. Generalizing the results of Ref. \cite{Buccio:2023lzo, Buccio:2024hys} and identifying tadpoles and bubble diagrams in the heat kernel expansion, we show how to get the high energy physical beta functions in Section \ref{physical}.

\section*{Acknowledgements}

We thank Diego Buccio and Roberto Percacci for many discussions. JFD acknowledges partial support from the U.S. National Science Foundation under grant NSF-PHY-21-12800. GM acknowledges partial support from Conselho Nacional de Desenvolvimento Cient\'ifico e Tecnol\'ogico - CNPq under grant 317548/2021-2, Funda\c{c}\~ao Carlos Chagas Filho de Amparo \`a Pesquisa do Estado de S\~ao Paulo - FAPESP under grant n. 2023/06508-8 and Funda\c{c}\~ao Carlos Chagas Filho de Amparo \`a Pesquisa do Estado do Rio de Janeiro - FAPERJ under grants E- 26/202.725/2018 and E-26/201.142/2022.

\section*{Appendix A. The low energy limit of the linear sigma model}

The U(1) linear sigma model with a higher derivative interaction can be defined by the Lagrangian with a complex scalar field $\chi$
\beq
{\cal L}=  \partial_\mu \chi^* \partial^\mu \chi- \frac1{m^2}\Box \chi^* \Box\chi -\lambda\left(\chi^*\chi - \frac12 v^2\right)^2 \ \ .
\eeq
The U(1) symmetry is $\chi \to e^{i \theta}\chi$. The spectrum of this model can be identified using the parameterization $\chi = \frac1{\sqrt{2}}( v+\sigma) e^{i\phi/v}$. The U(1) symmetry here is 
now manifest as a shift symmetry of the $\phi$ field, $\phi \to \phi+\theta v$. Without any approximation this results in
\beq
{\cal L}= {\cal L}_\phi +{\cal L}_\sigma + {\cal L}_{int} \ \ .
\eeq
Here
\beq
{\cal L}_\phi = \frac12 \partial_\mu\phi  \partial^\mu \phi -\frac1{2m^2}\Box\phi \Box\phi 
\eeq
and 
\beq
{\cal L}_\sigma = \frac12\left(\partial_\mu \sigma \partial^\mu \sigma - m_\sigma^2 \sigma^2 \right) - \frac1{2m^2} \Box \sigma \Box  \sigma
\eeq
with $m_\sigma^2 =2\lambda v^2$. The interaction term has several components,
\bea\label{HDint}
{\cal L}_{int} &=&  \left(\frac{\sigma}{v} +\frac{\sigma^2}{2v^2} \right) \partial_\mu \phi \partial^\mu \phi   -\frac{\lambda}{4} \sigma^4 -\lambda v \sigma^3 
-\frac1{m^2}\left(   \frac{\sigma}{v} +\frac{\sigma^2}{2v^2} \right)\Box \phi \Box \phi -\frac{2}{m^2v^2}(\partial_\mu \phi \partial^\mu \sigma)^2  \nonumber \\
&~& + \frac2{m^2v}\left(1+\frac{\sigma}{v}\right) \Box \sigma \partial_\mu \phi \partial^\mu \phi  -\frac{2}{m^2v}\left(1+\frac{\sigma}{v}\right) \partial_\mu \sigma\partial^\mu \phi \Box \phi - \frac1{2m^2v^2}\left( 1+\frac{\sigma}{v} \right)^2 \left( 
\partial_\mu \phi \partial^\mu \phi \right)^2
\eea

Without the higher derivative term, the U(1) non-linear sigma model is formed by integrating out the $\sigma$ in the usual U(1) sigma model and keeping the leading interaction term involving the field $\phi$ at low energy \cite{Burgess:2020tbq,Donoghue:2022azh}. For the usual sigma model this involves the tree-level exchange of the $\sigma$, {which comes from the very first term in ${\cal L}_{int}$, such that the leading interaction carries four derivatives of $\phi$ . With the presence of the higher derivative term, there will also be new interactions which can be identified by inverse powers of $m^2$. In this case, it is the last term of ${\cal L}_{int}$ which gives the leading result. After some algebra, which is mostly converting factors of $v^2$ to $m_\sigma^2$ we find }
\beq
\frac{g}{M^4} = \frac{\lambda}{m_\sigma^4}+\frac{\lambda}{m^2 m_\sigma^2} \ \ .
\eeq 
We recover the usual U(1) nonlinear sigma model when $m\to \infty$, but the higher derivative term is important for $m<m_\sigma$. 

There is an interesting point here. If $m\ll m_\sigma$, then there is a hierarchy of scales, i.e.  $m\ll M \ll m_\sigma$, if $\lambda$ is not unusual in size. Then there is a weakly coupled EFT at energies below $m$, a different weakly coupled EFT from $m<E<M$, a strongly interacting region from $M \sim \sqrt{mm_\sigma}<E<m_\sigma$, then the full linear sigma model emerges above $m_\sigma$. The latter can again be weakly coupled if $\lambda$ is not too large.

\section*{Appendix B. Details of the traces for the SU(N) calculation}

In this appendix we collect all lengthy expressions concerning the one-loop renormalization of the HDNLSM put forward in the main text. We begin quoting the associated expressions for the matrices 
${\cal B}_{\nu\mu}^{ab}$, $Z_{\mu}^{ab (i)}$ and $Q^{ab (i)}$. One finds
\begingroup
\allowdisplaybreaks
\bea
\hspace{-5mm}
{\cal B}_{\mu\nu}^{ab} &=& 
\frac{1}{c_0} \delta^{ab} \delta_{\nu\mu}
\nn\\
&+&  \Biggl\{ 
\left[ - 2\alpha_3^2 \textrm{Tr}\left( \left\{ t^a, t^{(c} \right\}  \left\{ t^b, t^{d)} \right\} \right)
+ 2 \left( 2 \alpha_2^2 - \alpha_3^2 \right) 
\textrm{Tr}\left( \left[ t^a, t^{(c} \right] \left[ t^b, t^{d)} \right] \right)
+ 2 \left( - \frac{3}{f^2} + \alpha_2^2 - \alpha_3^2  \right)  
\textrm{Tr}\left( \left\{ t^a, t^{(c} \right\} \left[ t^b, t^{d)} \right] \right)
\right.
\nn\\
&+& \left. 2 \left( \frac{1}{f^2} - \alpha_2^2 - \alpha_3^2 \right) 
\textrm{Tr}\left( \left[ t^a, t^{(c} \right] \left\{ t^b, t^{d)} \right\} \right)
- 4\alpha_3^2 \textrm{Tr}\Bigl( \{ t^c, t^d \} t^a t^b   \Bigr)
- 4\alpha_4^2 \delta^{cd} \delta^{ab}
- 4\alpha_5^2 \delta^{a (c} \delta^{d) b}
 \right] \delta^{\alpha\beta} \delta_{\nu\mu} 
\nn\\
&+& \left[ - 2 \left( \frac{1}{f^2} + \alpha_2^2 \right)
\textrm{Tr}\left( 2 [t^a, t^d] [t^b, t^c] - \frac{1}{2} \{ t^a, t^d \} [t^b, t^c]
+ \frac{1}{2} [t^a, t^d] \{ t^b, t^c \} \right) 
\right.
\nn\\
&-& \left. \alpha_3^2
\textrm{Tr}\Bigl( 8 t^c t^d  t^a t^b + 4 ( t^d t^c t^b t^a + t^c t^d  t^a t^b )
+ 4 \left\{ t^a, t^c \right\} \left\{ t^b, t^d \right\} \Bigr)
- 8\alpha_4^2  \delta^{ac} \delta^{bd}
- 4\alpha_5^2 \Bigl(  \delta^{cd} \delta^{ab} + \delta^{ad} \delta^{bc} \Bigr) 
\right] \delta^{\alpha}_{\nu}\delta^{\beta}_{\mu} \Biggr\}
(\Delta_{\alpha})^c (\Delta_{\beta})^d  
\nn\\
\eea
\bea
Z_{\mu}^{ab (1)} &=& \textrm{Tr}\biggl[ \Bigl( 4  [ t^b, [ t^c, t^a] ] t^d 
- 4 t^a  t^b [t^c, t^d]  +  t^a [ [t^c, t^d], t^b ]  + 2 t^d [t^c , t^a] t^b - 2 t^c t^a [t^d, t^b ] \Bigr)
\delta^{\gamma\delta}\delta^{\alpha}_{\mu}  
\nn\\
&+& \Bigl[ 2  t^d \Bigl( 2  t^c  \{ t^b,  t^a \}  - 2 \{ t^c,  t^a \}  t^b - \left\{ t^c, \{ t^b,  t^a \}  \right\} 
+ 2 t^a \left\{ t^c, t^b  \right\} \Bigr)
+ 4 t^a t^b [t^c, t^d] - t^a [  [t^c, t^d] , t^b ] 
\nn\\
&+& 2 t^c [t^d, t^a] t^b  - 2 t^d  t^a [ t^c, t^b ] \Bigr] \delta^{\alpha\gamma} \delta^{\delta}_{\mu} 
\nn\\
&+& \Bigl[ 2 t^d \Bigl( - 2  \{ t^b,  t^a \} t^c + 2  t^b \{ t^c,  t^a \}  - \left\{ t^c, \{ t^b, t^a \} \right\} 
+ 2 t^a  \left\{ t^c,  t^b \right\} \Bigr) - 4 t^d t^a [t^b, t^c] - 4 t^c t^a [ t^b, t^d] \Bigr] 
 \delta^{\alpha\delta} \delta^{\gamma}_{\mu} \biggr] 
 \Delta^c_{\alpha} (d_{\gamma}\Delta_{\delta})^d
\nn\\
&+& \textrm{Tr}\biggl[ 2 [t^d, t^e] 
\Bigl( 2  t^c  \{ t^b,  t^a \}  - 2 \{ t^c,  t^a \}  t^b - \left\{ t^c, \{ t^b,  t^a \}  \right\} 
+ 2 t^a \left\{ t^c, t^b  \right\} \Bigr) 
\nn\\
&+& 2 [t^e, t^d] \Bigl( - 2  \{ t^b,  t^a \} t^c + 2  t^b \{ t^c,  t^a \}  - \left\{ t^c, \{ t^b, t^a \} \right\} 
+ 2 t^a  \left\{ t^c,  t^b \right\} \Bigr) -  2 [[ t^e,  t^d ],t^a] [t^c, t^b]
\nn\\
&-& 4 \Bigl(  \left\{ t^d, t^a \right\} t^e - t^d \left\{ t^e, t^a \right\} 
- t^e \left\{ t^d, t^a \right\} + \left\{ t^e, t^a \right\} t^d 
+ \frac12 \left\{ t^d, \left\{ t^e, t^a \right\} \right\} 
- \frac12 \left\{ t^e, \left\{ t^d, t^a \right\} \right\} 
\Bigr) [ t^c, t^b ] \biggr] \delta^{\alpha\delta} \delta^{\gamma}_{\mu}
\Delta^{c}_{\alpha} \Delta^d_{\gamma} \Delta^e_{\delta} 
\nn\\
&-& \textrm{Tr}\biggl[  4 \left\{ d_{[\mu} \Delta_{\nu]}, t^a \right\} [ \Delta^{\nu}, t^b ] 
+ 4 t^a  t^b d^{\nu}  d_{[\nu} \Delta_{\mu]} \biggr]
\nn\\
Z_{\mu}^{ab (2)} &=& - \Biggl\{ 
- \textrm{Tr}\biggl[ \Bigl( - 4  [ t^b, [ t^c, t^a] ] t^d 
- 4 t^a  t^b [t^c, t^d]  +  t^a [ [t^c, t^d], t^b ]  + 2 t^d [t^c , t^a] t^b - 2 t^c t^a [t^d, t^b ] \Bigr)
\delta^{\gamma\delta}\delta^{\alpha}_{\mu}  
\nn\\
&+& \Bigl[ 2  t^d \Bigl( 2  t^c  \{ t^b,  t^a \}  - 2 \{ t^c,  t^a \}  t^b - \left\{ t^c, \{ t^b,  t^a \}  \right\} 
+ 2 t^a \left\{ t^c, t^b  \right\} \Bigr)
+ 4 t^a t^b [t^c, t^d] - t^a [  [t^c, t^d] , t^b ] 
\nn\\
&+& 2 t^c [t^d, t^a] t^b  - 2 t^d  t^a [ t^c, t^b ] \Bigr] \delta^{\alpha\gamma} \delta^{\delta}_{\mu} 
\nn\\
&+& \Bigl[ 2 t^d \Bigl( - 2  \{ t^b,  t^a \} t^c + 2  t^b \{ t^c,  t^a \}  - \left\{ t^c, \{ t^b, t^a \} \right\} 
+ 2 t^a  \left\{ t^c,  t^b \right\} \Bigr) - 4 t^d t^a [t^b, t^c] - 4 t^c t^a [ t^b, t^d] \Bigr] 
 \delta^{\alpha\delta} \delta^{\gamma}_{\mu} \biggr] 
 \Delta^c_{\alpha} (d_{\gamma}\Delta_{\delta})^d
\nn\\
&-& \textrm{Tr}\biggl[ 2 [t^d, t^e] 
\Bigl( 2  t^c  \{ t^b,  t^a \}  - 2 \{ t^c,  t^a \}  t^b - \left\{ t^c, \{ t^b,  t^a \}  \right\} 
+ 2 t^a \left\{ t^c, t^b  \right\} \Bigr) 
\nn\\
&+& 2 [t^e, t^d] \Bigl( - 2  \{ t^b,  t^a \} t^c + 2  t^b \{ t^c,  t^a \}  - \left\{ t^c, \{ t^b, t^a \} \right\} 
+ 2 t^a  \left\{ t^c,  t^b \right\} \Bigr) -  2 [[ t^e,  t^d ],t^a] [t^c, t^b]
\nn\\
&-& 4 \Bigl(  \left\{ t^d, t^a \right\} t^e - t^d \left\{ t^e, t^a \right\} 
- t^e \left\{ t^d, t^a \right\} + \left\{ t^e, t^a \right\} t^d 
+ \frac12 \left\{ t^d, \left\{ t^e, t^a \right\} \right\} 
- \frac12 \left\{ t^e, \left\{ t^d, t^a \right\} \right\} 
\Bigr) [ t^c, t^b ] \biggr] \delta^{\alpha\delta} \delta^{\gamma}_{\mu}
\Delta^{c}_{\alpha} \Delta^d_{\gamma} \Delta^e_{\delta} 
\nn\\
&+& \textrm{Tr}\biggl[  4 \left\{ d_{[\mu} \Delta_{\nu]}, t^a \right\} [ \Delta^{\nu}, t^b ] 
+ 4 t^a  t^b d^{\nu}  d_{[\nu} \Delta_{\mu]} \biggr]
\Biggr\}
\nn\\
Z_{\mu}^{ab (3)} &=& 
- \Biggl\{ 8 \textrm{Tr}\biggl[ t^a t^b  \bigl\{ t^c, t^d \bigr\} \delta^{\alpha\delta} \delta^{\gamma}_{\mu}
+ t^a t^b t^d t^c \delta^{\gamma\delta} \delta^{\alpha}_{\mu} 
+ t^a t^b t^c t^d \delta^{\alpha\gamma} \delta^{\delta}_{\mu} \biggr]
\Delta^c_{\alpha} (d_{\gamma}\Delta_{\delta})^d  
\nn\\
&+& 4 \textrm{Tr}\biggl[ \left( t^a t^d  t^b t^c + t^a t^d t^c t^b  + t^d t^a t^b t^c + t^d t^a t^c t^b 
+ t^a t^c t^d  t^b + t^c t^a t^b t^d \right)  
\delta^{\gamma\delta} \delta^{\alpha}_{\mu}
\nn\\
&+& \left( t^a t^c  t^b t^d + t^a t^c t^d t^b + t^c t^a t^b t^d + t^c t^a t^d t^b 
+ t^a t^d t^c t^b + t^d t^a t^b t^c \right) 
\delta^{\alpha\gamma} \delta^{\delta}_{\mu}
\nn\\
&+& \bigl( t^a t^d  t^b t^c + t^a t^c t^b t^d + t^d t^a t^c t^b + t^c t^a t^d t^b \bigr) 
\delta^{\alpha\delta} \delta^{\gamma}_{\mu} \biggr]
\Delta^c_{\alpha} \partial_{\gamma} \Delta^d_{\delta}
\nn\\
&+& \textrm{Tr}\biggl[ 8 \bigl( - t^c t^e \bigl[ t^b, t^a \bigr] t^d - t^c t^e  t^d \bigl[ t^b, t^a \bigr]
+ t^c t^e \bigl\{ t^b , [ t^a, t^d ] \bigr\}  - [ t^c, t^a ] t^e t^b t^d 
- [ t^c, t^a ] t^e t^d t^b - t^c  [ t^e, t^a ]   t^b t^d
\nn\\
&-& t^c [ t^e, t^a ] t^d t^b - t^d  t^c \left[ t^b, t^a \right] t^e
+ t^d t^c t^b \left[ t^a, t^e \right] - t^c t^d t^e \left[ t^b, t^a \right] 
+ t^c t^d \left[ t^a, t^e \right]  t^b - t^b t^c \left[ t^d, t^a \right] t^e  
\nn\\
&-& t^b t^c t^d \left[ t^e, t^a \right] - t^c t^b  \left[ t^e, t^a \right] t^d
- t^c t^b t^e \left[ t^d, t^a \right] \bigr) \delta^{\alpha\delta} \delta^{\gamma}_{\mu} \biggr]
\Delta^{c}_{\alpha} \Delta^d_{\gamma} \Delta^e_{\delta} 
\Biggr\}
\nn\\
Z_{\mu}^{ab (4)} &=& - 8 \biggl[ 
 \Bigl( \delta^{cd} \delta^{ab} 
\delta^{\alpha\delta} \delta^{\gamma}_{\mu} 
+ \delta^{da} \delta^{cb}
\delta^{\gamma\delta} \delta^{\alpha}_{\mu} 
+ \delta^{ca} \delta^{db}
\delta^{\alpha\gamma} \delta^{\delta}_{\mu} \Bigr)
\Delta^c_{\alpha} (d_{\gamma}\Delta_{\delta})^d
+ 2 \textrm{Tr}\Bigl( t^d [t^e, t^c] \Bigr) \delta^{ab}
\delta^{\alpha\delta} \delta^{\gamma}_{\mu} 
\Delta^{c}_{\alpha} \Delta^d_{\gamma} \Delta^e_{\delta} 
\biggr]
\nn\\
Z_{\mu}^{ab (5)} &=& - 2 \biggl[ 
2 \Bigl( ( \delta^{cd} \delta^{ab} + \delta^{ac} \delta^{db} )
\delta^{\gamma\delta} \delta^{\alpha}_{\mu}
+ ( \delta^{cd} \delta^{ab} + \delta^{da} \delta^{cb} )
\delta^{\alpha\gamma} \delta^{\delta}_{\mu} 
+ ( \delta^{da} \delta^{cb} + \delta^{ca} \delta^{db} )
\delta^{\alpha\delta} \delta^{\gamma}_{\mu} \Bigr)
\Delta^c_{\alpha} (d_{\gamma}\Delta_{\delta})^d
\nn\\
&+& 4 \textrm{Tr}\Bigl( t^c [t^d, t^e] \Bigr) \delta^{ab}
\delta^{\alpha\delta} \delta^{\gamma}_{\mu} 
\Delta^{c}_{\alpha} \Delta^d_{\gamma} \Delta^e_{\delta} 
\biggr]
\eea
and
\bea
\hspace{-12mm}
Q^{ab (1)} &=&  \Biggl\{
 \biggl\{ \textrm{Tr}\biggl[ 2 [ t^a, t^c ] t^b t^d +  2 t^a \left\{ t^b, t^c \right\} t^d -  4 t^a t^b t^c t^d 
- \Bigl( \left\{ t^c, t^a \right\} \left\{ t^d, t^b \right\} - 2 t^a t^c \left\{ t^d, t^b \right\} + 4 t^a t^c t^b t^d 
- 2 \left\{ t^c , t^a \right\} t^b t^d \Bigr) \biggr]  
\delta^{\alpha\beta} \delta^{\gamma\delta} 
\nn\\
&+&\textrm{Tr}\biggl[ - 2 \left\{ t^d, t^a t^b \right\} t^c + 4 t^a \left\{ t^d,  t^b \right\} t^c - 4 t^a t^b t^d t^c  
- \Bigl( \left\{ t^c, t^a \right\}  \left\{ t^d, t^b \right\} 
- 2 \left\{ t^c , t^a \right\} t^b t^d 
- 2 t^a t^d \left\{ t^c, t^b \right\}  
+ 4 t^a t^d  t^b t^c \Bigr) \biggr] \delta^{\alpha\gamma} \delta^{\beta\delta} \biggr\}
\nn\\ 
&\times& (d_{\alpha} \Delta_{\beta})^c (d_{\gamma} \Delta_{\delta})^d
\nn\\
&+& \biggl\{ \textrm{Tr}\biggl[ 4 t^a [ \left\{ t^c, t^b \right\}, t^d ] t^e + 2 [ t^c,  [ t^d, t^a ] ] t^b t^e
+ 2 t^a \left\{ t^c, \left\{ t^d, t^b \right\} \right\} t^e  - 8 t^a t^b t^c t^d t^e 
- \Bigl( 2 \left\{ t^e, t^a \right\} [ \left\{ t^d, t^b \right\}, t^c ] 
\nn\\
&+& \left\{ t^e, t^a \right\} \left\{ \left\{ t^d, t^b \right\}, t^c  \right\}
- 4 \left\{ t^e , t^a \right\} t^b t^d t^c 
+  2 [ \left\{ t^d , t^a \right\}, t^c ] \left\{ t^e, t^b \right\}
- 4 [ \left\{ t^c, t^a \right\}, t^d ] t^b t^e 
\nn\\
&+& \left\{ \left\{ t^d, t^a \right\}, t^c \right\} \left\{ t^e, t^b \right\}
- 2 \left\{ \left\{ t^c, t^a \right\}, t^d \right\} t^b t^e  
- 4 t^a t^c t^d \left\{ t^e, t^b \right\} 
- 4 t^a t^e [ \left\{ t^d, t^b \right\}, t^c ] 
\nn\\
&-& 2 t^a t^e \left\{ \left\{t^d, t^b \right\}, t^c \right\} 
+ 8 t^a t^e t^b t^d t^c 
+ 8 t^a t^c t^d t^b t^e \Bigr) \biggr] 
\delta^{\alpha\beta} \delta^{\gamma\delta} 
\nn\\
&+& \textrm{Tr}\biggl[ - 2 t^a [[ t^d, t^c], t^b] t^e  
- 4 \left\{ t^d, t^a \right\} \left\{ t^c,  t^b \right\} t^e  
+ 4 t^a \left\{ t^d, \left\{ t^c,  t^b \right\} \right\} t^e  
+ 4 t^d \left\{ t^c, t^a t^b \right\} t^e  
+ 4 \left\{ t^c, t^a t^b \right\} t^d t^e 
\nn\\
&-& 2 \Bigl\{ t^d, \left\{ t^c, t^a t^b \right\} \Bigr\} t^e   
-  4 t^a t^b \left\{t^d,t^c \right\} t^e  
+ \Bigl( - 2\left\{ t^e, t^a t^b \right\} + 4 t^a \left\{ t^e,  t^b \right\} -  4 t^a t^b  t^e  \Bigr) [t^c, t^d]  
\nn\\
&-& \Bigl( 2 \left\{ t^e, t^a \right\}  
\bigl(  \left\{ t^d, t^b \right\} t^c  - t^d  \left\{ t^c, t^b \right\} \bigr)
+ \left\{ t^e, t^a \right\}  \left\{ t^c , \left\{ t^d, t^b \right\} \right\} 
 - 2 \left\{ t^e , t^a \right\} t^b \left\{ t^d, t^c  \right\}  
+ 2 \bigl(  \left\{ t^d, t^a \right\} t^c  - t^d \left\{ t^c, t^a \right\} \bigr) \left\{ t^e, t^b \right\} 
\nn\\
&-& 4 \bigl(  \left\{ t^c, t^a \right\} t^d - t^c \left\{ t^d, t^a \right\} \bigr) t^b t^e 
+ \left\{ t^d, \left\{ t^c, t^a \right\} \right\} \left\{ t^e, t^b \right\}  
- 2 \left\{ t^c, \left\{ t^d, t^a \right\} \right\} t^b t^e 
- 2 t^a \left\{t^d, t^c\right\}  \left\{ t^e, t^b \right\} 
\nn\\
&-& 4 t^a t^e  \bigl( \left\{ t^c, t^b \right\} t^d 
- t^c \left\{ t^d, t^b \right\} \bigr)
- 2 t^a t^e \left\{ t^c, \left\{ t^d, t^b \right\} \right\} 
+ 4 t^a t^e t^b \left\{ t^c, t^d \right\}
+ 4 t^a   \left\{t^d, t^c\right\}  t^b t^e \Bigr)
\biggr] \delta^{\alpha\gamma} \delta^{\beta\delta} \biggr\}
(\Delta_{\alpha})^c (\Delta_{\beta})^d (d_{\gamma} \Delta_{\delta})^e
 \nn\\
&+& \biggl\{ - \textrm{Tr}\biggl[ 4 [ \left\{ t^c , t^a \right\}, t^d ] [ \left\{ t^e, t^b \right\}, t^f  ]
+ 2 [ \left\{ t^c, t^a \right\}, t^d ] \left\{ \left\{ t^e, t^b \right\}, t^f \right\}
- 8 [ \left\{ t^c, t^a \right\}, t^d ] t^b  t^f t^e 
+ 2 \left\{ \left\{ t^c, t^a \right\}, t^d \right\} [ \left\{ t^e, t^b \right\}, t^f ]
\nn\\
&+& \left\{ \left\{ t^c, t^a \right\}, t^d \right\} \left\{ \left\{ t^e, t^b \right\}, t^f \right\}
- 4 \left\{ \left\{ t^c, t^a \right\}, t^d \right\} t^b  t^f t^e 
- 8 t^a t^c t^d [ \left\{ t^e, t^b \right\}, t^f ] 
\nn\\
&-& 4 t^a t^c t^d \left\{ \left\{t^e, t^b \right\}, t^f  \right\} 
+ 16 t^a  t^c t^d  t^b t^e t^f \biggr]
\delta^{\alpha\beta} \delta^{\gamma\delta} 
\nn\\
&+& \textrm{Tr}\biggl[ - 2 t^a [[ t^e, t^f], t^b] [t^c, t^d]
- 4 \left\{ t^f, t^a \right\} \left\{ t^e,  t^b \right\}  [t^c, t^d]
+ 4  t^a \left\{ t^f, \left\{ t^e,  t^b \right\} \right\} [t^c, t^d]
 + 4 t^f \left\{ t^e, t^a t^b \right\} [t^c, t^d]
\nn\\
&+& 4 \left\{ t^f, t^a t^b \right\} t^e  [t^c, t^d]
- 2 \Bigl\{ t^f, \left\{ t^e, t^a t^b \right\} \Bigr\} [t^c, t^d] 
-  4 t^a t^b  \left\{t^f,t^e \right\} [t^c, t^d]  
- [[ t^c, t^d],t^a] [ [ t^e, t^f], t^b ]
+  [ t^a, [ [ t^c, t^d], t^b] ] [ t^e, t^f] 
\nn\\
&-& \Bigl( 4 \bigl(  \left\{ t^c, t^a \right\} t^d  - t^c \left\{ t^d, t^a \right\} \bigr)
\bigl(  \left\{ t^e, t^b \right\} t^f  - t^e \left\{ t^f, t^b \right\} \bigr)
+ 2 \bigl(  \left\{ t^c, t^a \right\} t^d - t^c \left\{ t^d, t^a \right\} \bigr) 
\left\{ t^e, \left\{ t^f, t^b \right\} \right\} 
\nn\\
&-& 4 \bigl(  \left\{ t^e, t^a \right\} t^f - t^e \left\{ t^f, t^a \right\} \bigr) t^b \left\{ t^d, t^c \right\}
+ 2 \left\{ t^c, \left\{ t^d, t^a \right\} \right\} 
\bigl( \left\{ t^e, t^b \right\} t^f  - t^e \left\{ t^f, t^b \right\} \bigr)
+ \left\{ t^c, \left\{ t^d , t^a \right\} \right\} \left\{ t^e, \left\{ t^f, t^b \right\} \right\} 
\nn\\
&-& 2 \left\{ t^e, \left\{ t^f, t^a \right\} \right\} t^b \left\{ t^d, t^c \right\} 
- 4 t^a \left\{ t^d, t^c \right\} \bigl(  \left\{ t^e, t^b \right\} t^f - t^e \left\{ t^f, t^b \right\} \bigr)
\nn\\
&-& 2 t^a \left\{t^d, t^c\right\} \left\{ t^e, \left\{ t^f, t^b \right\} \right\} 
+ 4 t^a  \left\{t^d, t^c\right\} t^b \left\{ t^f, t^e \right\} \Bigr)
\biggr] \delta^{\alpha\gamma} \delta^{\beta\delta} \biggr\}
(\Delta_{\alpha})^c (\Delta_{\beta})^d (\Delta_{\gamma})^e (\Delta_{\delta})^f
 \Biggr\}
\eea
\bea
\hspace{-10mm}
Q^{ab (2)} &=& -  \Biggl\{
 \biggl\{ \textrm{Tr}\biggl[ 2 [ t^a, t^c ] t^b t^d +  2 t^a \left\{ t^b, t^c \right\} t^d -  4 t^a t^b t^c t^d 
- \Bigl( \left\{ t^c, t^a \right\} \left\{ t^d, t^b \right\} - 2 t^a t^c \left\{ t^d, t^b \right\} + 4 t^a t^c t^b t^d 
- 2 \left\{ t^c , t^a \right\} t^b t^d \Bigr) \biggr]  
\delta^{\alpha\beta} \delta^{\gamma\delta} 
\nn\\
&-&\textrm{Tr}\biggl[ - 2 \left\{ t^d, t^a t^b \right\} t^c + 4 t^a \left\{ t^d,  t^b \right\} t^c - 4 t^a t^b t^d t^c  
- \Bigl( \left\{ t^c, t^a \right\}  \left\{ t^d, t^b \right\} 
- 2 \left\{ t^c , t^a \right\} t^b t^d 
- 2 t^a t^d \left\{ t^c, t^b \right\}  
+ 4 t^a t^d  t^b t^c \Bigr) \biggr] \delta^{\alpha\gamma} \delta^{\beta\delta} \biggr\}
\nn\\ 
&\times& (d_{\alpha} \Delta_{\beta})^c (d_{\gamma} \Delta_{\delta})^d
\nn\\
&+& \biggl\{ \textrm{Tr}\biggl[ 4 t^a [ \left\{ t^c, t^b \right\}, t^d ] t^e + 2 [ t^c,  [ t^d, t^a ] ] t^b t^e
+ 2 t^a \left\{ t^c, \left\{ t^d, t^b \right\} \right\} t^e  - 8 t^a t^b t^c t^d t^e 
- \Bigl( 2 \left\{ t^e, t^a \right\} [ \left\{ t^d, t^b \right\}, t^c ] 
\nn\\
&+& \left\{ t^e, t^a \right\} \left\{ \left\{ t^d, t^b \right\}, t^c  \right\}
- 4 \left\{ t^e , t^a \right\} t^b t^d t^c 
+  2 [ \left\{ t^d , t^a \right\}, t^c ] \left\{ t^e, t^b \right\}
- 4 [ \left\{ t^c, t^a \right\}, t^d ] t^b t^e 
\nn\\
&+& \left\{ \left\{ t^d, t^a \right\}, t^c \right\} \left\{ t^e, t^b \right\}
- 2 \left\{ \left\{ t^c, t^a \right\}, t^d \right\} t^b t^e  
- 4 t^a t^c t^d \left\{ t^e, t^b \right\} 
- 4 t^a t^e [ \left\{ t^d, t^b \right\}, t^c ] 
\nn\\
&-& 2 t^a t^e \left\{ \left\{t^d, t^b \right\}, t^c \right\} 
+ 8 t^a t^e t^b t^d t^c 
+ 8 t^a t^c t^d t^b t^e \Bigr) \biggr] 
\delta^{\alpha\beta} \delta^{\gamma\delta} 
\nn\\
&-& \textrm{Tr}\biggl[ - 2 t^a [[ t^d, t^c], t^b] t^e  
- 4 \left\{ t^d, t^a \right\} \left\{ t^c,  t^b \right\} t^e  
+ 4 t^a \left\{ t^d, \left\{ t^c,  t^b \right\} \right\} t^e  
+ 4 t^d \left\{ t^c, t^a t^b \right\} t^e  
+ 4 \left\{ t^c, t^a t^b \right\} t^d t^e 
\nn\\
&-& 2 \Bigl\{ t^d, \left\{ t^c, t^a t^b \right\} \Bigr\} t^e   
-  4 t^a t^b \left\{t^d,t^c \right\} t^e  
+ \Bigl( - 2\left\{ t^e, t^a t^b \right\} + 4 t^a \left\{ t^e,  t^b \right\} -  4 t^a t^b  t^e  \Bigr) [t^c, t^d]  
\nn\\
&-& \Bigl( 2 \left\{ t^e, t^a \right\}  
\bigl(  \left\{ t^d, t^b \right\} t^c  - t^d  \left\{ t^c, t^b \right\} \bigr)
+ \left\{ t^e, t^a \right\}  \left\{ t^c , \left\{ t^d, t^b \right\} \right\} 
 - 2 \left\{ t^e , t^a \right\} t^b \left\{ t^d, t^c  \right\}  
+ 2 \bigl(  \left\{ t^d, t^a \right\} t^c  - t^d \left\{ t^c, t^a \right\} \bigr) \left\{ t^e, t^b \right\} 
\nn\\
&-& 4 \bigl(  \left\{ t^c, t^a \right\} t^d - t^c \left\{ t^d, t^a \right\} \bigr) t^b t^e 
+ \left\{ t^d, \left\{ t^c, t^a \right\} \right\} \left\{ t^e, t^b \right\}  
- 2 \left\{ t^c, \left\{ t^d, t^a \right\} \right\} t^b t^e 
- 2 t^a \left\{t^d, t^c\right\}  \left\{ t^e, t^b \right\} 
\nn\\
&-& 4 t^a t^e  \bigl( \left\{ t^c, t^b \right\} t^d 
- t^c \left\{ t^d, t^b \right\} \bigr)
- 2 t^a t^e \left\{ t^c, \left\{ t^d, t^b \right\} \right\} 
+ 4 t^a t^e t^b \left\{ t^c, t^d \right\}
+ 4 t^a   \left\{t^d, t^c\right\}  t^b t^e \Bigr)
\biggr] \delta^{\alpha\gamma} \delta^{\beta\delta} \biggr\}
(\Delta_{\alpha})^c (\Delta_{\beta})^d (d_{\gamma} \Delta_{\delta})^e
 \nn\\
&+& \biggl\{ - \textrm{Tr}\biggl[ 4 [ \left\{ t^c , t^a \right\}, t^d ] [ \left\{ t^e, t^b \right\}, t^f  ]
+ 2 [ \left\{ t^c, t^a \right\}, t^d ] \left\{ \left\{ t^e, t^b \right\}, t^f \right\}
- 8 [ \left\{ t^c, t^a \right\}, t^d ] t^b  t^f t^e 
+ 2 \left\{ \left\{ t^c, t^a \right\}, t^d \right\} [ \left\{ t^e, t^b \right\}, t^f ]
\nn\\
&+& \left\{ \left\{ t^c, t^a \right\}, t^d \right\} \left\{ \left\{ t^e, t^b \right\}, t^f \right\}
- 4 \left\{ \left\{ t^c, t^a \right\}, t^d \right\} t^b  t^f t^e 
- 8 t^a t^c t^d [ \left\{ t^e, t^b \right\}, t^f ] 
\nn\\
&-& 4 t^a t^c t^d \left\{ \left\{t^e, t^b \right\}, t^f  \right\} 
+ 16 t^a  t^c t^d  t^b t^e t^f \biggr]
\delta^{\alpha\beta} \delta^{\gamma\delta} 
\nn\\
&-& \textrm{Tr}\biggl[ - 2 t^a [[ t^e, t^f], t^b] [t^c, t^d]
- 4 \left\{ t^f, t^a \right\} \left\{ t^e,  t^b \right\}  [t^c, t^d]
+ 4  t^a \left\{ t^f, \left\{ t^e,  t^b \right\} \right\} [t^c, t^d]
 + 4 t^f \left\{ t^e, t^a t^b \right\} [t^c, t^d]
\nn\\
&+& 4 \left\{ t^f, t^a t^b \right\} t^e  [t^c, t^d]
- 2 \Bigl\{ t^f, \left\{ t^e, t^a t^b \right\} \Bigr\} [t^c, t^d] 
-  4 t^a t^b  \left\{t^f,t^e \right\} [t^c, t^d]  
-  [[ t^c, t^d],t^a] [ [ t^e, t^f], t^b ]
+  [ t^a, [ [ t^c, t^d], t^b] ] [ t^e, t^f] 
\nn\\
&-& \Bigl( 4 \bigl(  \left\{ t^c, t^a \right\} t^d  - t^c \left\{ t^d, t^a \right\} \bigr)
\bigl(  \left\{ t^e, t^b \right\} t^f  - t^e \left\{ t^f, t^b \right\} \bigr)
+ 2 \bigl(  \left\{ t^c, t^a \right\} t^d - t^c \left\{ t^d, t^a \right\} \bigr) 
\left\{ t^e, \left\{ t^f, t^b \right\} \right\} 
\nn\\
&-& 4 \bigl(  \left\{ t^e, t^a \right\} t^f - t^e \left\{ t^f, t^a \right\} \bigr) t^b \left\{ t^d, t^c \right\}
+ 2 \left\{ t^c, \left\{ t^d, t^a \right\} \right\} 
\bigl( \left\{ t^e, t^b \right\} t^f  - t^e \left\{ t^f, t^b \right\} \bigr)
+ \left\{ t^c, \left\{ t^d , t^a \right\} \right\} \left\{ t^e, \left\{ t^f, t^b \right\} \right\} 
\nn\\
&-& 2 \left\{ t^e, \left\{ t^f, t^a \right\} \right\} t^b \left\{ t^d, t^c \right\} 
- 4 t^a \left\{ t^d, t^c \right\} \bigl(  \left\{ t^e, t^b \right\} t^f - t^e \left\{ t^f, t^b \right\} \bigr)
\nn\\
&-& 2 t^a \left\{t^d, t^c\right\} \left\{ t^e, \left\{ t^f, t^b \right\} \right\} 
+ 4 t^a  \left\{t^d, t^c\right\} t^b \left\{ t^f, t^e \right\} \Bigr)
\biggr] \delta^{\alpha\gamma} \delta^{\beta\delta} \biggr\}
(\Delta_{\alpha})^c (\Delta_{\beta})^d (\Delta_{\gamma})^e (\Delta_{\delta})^f
 \Biggr\}
\eea
\endgroup
\bea
Q^{ab (3)} &=& \Biggl\{
 \textrm{Tr}\biggl[8 t^e t^f \Bigr( \left\{ \bigl[ [t^c, t^b ], t^a \bigr], t^d \right\} + \left[ t^c, t^a \right] \left[ t^d, t^b \right]  \Bigl) + 4  \left\{ [ t^c, t^a ], t^d \right\} \left\{ [ t^e, t^b ], t^f \right\}
\biggr] \delta^{\alpha\beta} \delta^{\gamma\delta}
\nn\\
&+&  \textrm{Tr}\biggl[8 t^e t^f \Bigr(  \left[ \left[ t^c, t^a \right], t^b \right] t^d 
+ \left[ \left[ t^c, t^a \right], t^b \right] t^d 
+  [ t^c, t^a ] [ t^d, t^b ] \Bigl)
+ 4 \left[ t^c, t^a \right] t^d \left[ t^e t^f, t^b \right]   
+ 4 t^d  \left[ t^c, t^a \right]  \left[ t^f t^e, t^b \right] 
\biggr] \delta^{\alpha\gamma} \delta^{\beta\delta} \Biggr\}
\nn\\
&\times& (\Delta_{\alpha})^c (\Delta_{\beta})^d (\Delta_{\gamma})^e (\Delta_{\delta})^f
 \eea
\bea
Q^{ab (4)} &=& 4 \delta^{cd} f^{eak} f^{fbk}
(\Delta_{\nu})^c (\Delta^{\nu})^d (\Delta_{\mu})^e (\Delta^{\mu})^f
\eea
\bea
Q^{ab (5)} &=& 4 \biggl( 
 \delta^{cd} f^{eak} f^{fbk}
+ 2 \textrm{Tr}\Bigl( t^d \left\{ t^c, t^a \right\} 
- t^c \left\{ t^d, t^a \right\} \Bigr)
\textrm{Tr}\Bigl( t^f \left\{ t^e,  t^b \right\} 
- t^e \left\{ t^f, t^b \right\} \Bigr)
\biggr) (\Delta_{\nu})^c  (\Delta_{\mu})^d (\Delta^{\nu})^e (\Delta^{\mu})^f .
\nn\\
\eea
On the other hand, the function ${\cal B} = \delta_{\mu\nu} {\cal B}^{\mu\nu}$  is given by
\bea
{\cal B}^{ab} &=& \frac{4}{c_0} \delta^{ab}
+  \Biggl\{ - 12 \alpha_3^2 \textrm{Tr}\left( \left\{ t^a, t^{(c} \right\} \left\{ t^b, t^{d)} \right\} \right)
+ 4 \left( - \frac{1}{f^2} + 3\alpha_2^2 - 2 \alpha_3^2 \right) 
\textrm{Tr}\left( \left[ t^a, t^{(c} \right] \left[ t^b, t^{d)} \right] \right)
\nn\\
&+& \left( - \frac{23}{f^2} + 9\alpha_2^2 - 8\alpha_3^2 \right) 
\textrm{Tr}\left( \left\{ t^a, t^{(c} \right\} \left[ t^b, t^{d)} \right] \right)
+ \left( \frac{7}{f^2} - 9\alpha_2^2 - 8\alpha_3^2  \right) 
\textrm{Tr}\left( \left[ t^a, t^{(c} \right] \left\{ t^b, t^{d)} \right\} \right)
\nn\\
&-& 2 \alpha_3^2
\textrm{Tr}\Bigl( 10 \{ t^c, t^d \}  t^a t^b + \{ t^c, t^d \} \{ t^a, t^b \} \Bigr)
- 4 \left( 4\alpha_4^2 + \alpha_5^2 \right) \delta^{cd} \delta^{ab}  
- 4 \left( 2\alpha_4^2 + 5\alpha_5^2 \right) \delta^{a (c} \delta^{d) b}
 \Biggr\}
\delta^{\alpha\beta} (\Delta_{\alpha})^c (\Delta_{\beta})^d  .
\nn\\
\eea
Let us present here all traces necessary for the above computations. We find that
\beq
\textrm{Tr} \bigl( {\cal B} \bigr) = \frac{4 \left(N^2-1\right)}{c_0}
+ 2 \left( \frac{N}{f^2} - 3 N \alpha _2^2 
+ \left(\frac{12}{N}-7 N\right) \alpha _3^2
+  \left(4 - 8 N^2\right) \alpha _4^2
-  \left(8+2N^2\right) \alpha _5^2 \right)  
\delta^{\alpha\beta} \delta^{cd} (\Delta_{\alpha})^c (\Delta_{\beta})^d
\eeq
\bea
\textrm{Tr} \bigl( {\cal B}_{(\mu\nu)} {\cal B}^{\mu\nu} \bigr) &=& \frac{4 (N^2-1)}{c_0^2}
+ \frac{4}{c_0}
\left( \frac{N}{f^2} - {3 N}{\alpha _2^2} + {\left(\frac{12}{N}-7 N\right)}{\alpha _3^2} 
+ {\left(4 - 8 N^2\right)}{\alpha _4^2} - {(8+2 N^2)}{\alpha _5^2} \right)
  \delta^{\alpha\beta} \delta^{cd}
 (\Delta_{\alpha})^c (\Delta_{\beta})^d
\nn\\
&+& \biggl[ \Bigl( b_1 \delta ^{e f} \delta ^{c d} + b_2 \text{Tr}(t^e t^f t^c t^d ) \Bigr)
\delta^{\alpha\beta} \delta^{\gamma\delta}
+ \Bigl( b_3 \delta ^{e f} \delta ^{c d} + b_4 \text{Tr}(t^e t^f t^c t^d) \Bigr)
\delta^{\alpha\gamma} \delta^{\beta\delta}
 \biggr]
(\Delta_{\alpha})^c (\Delta_{\beta})^d (\Delta_{\gamma})^e (\Delta_{\delta})^f
\nn\\
\eea
where
\bea
b_1 &=& - {4}{\alpha _2^4}+{40}{\alpha _5^4}
+ {2 \left(\frac{64}{N^2}+5\right)}{\alpha _3^4}
+{\left({50}{\alpha _3^2}+{48 N}{\alpha _4^2}+{16 N}{\alpha _5^2}\right)}{\alpha _2^2}
+ {2 \left(32 N^2+16\right)}{\alpha _4^4}
+ {2 \left(16 N^2+80\right)}{\alpha _4^2 \alpha _5^2}
\nn\\
&+& \frac{\left({34}{\alpha _2^2}
-{38}{\alpha _3^2}
-{16 N}{\alpha _4^2}\right)}{f^2}
+  {\left[{2 \left(32 \left( \frac{N^2-1}{2N} \right)-8 N\right)}{\alpha _5^2}
+ {2 \left(192 \left( \frac{N^2-1}{2N} \right)-40 N\right)}{\alpha _4^2}\right]}{\alpha _3^2}
-\frac{26}{f^4}
\nn\\
b_2 &=& 
{\left({672}{\alpha _5^2}+{448}{\alpha _4^2}\right)}{\alpha _3^2}
+{61 N}{\alpha _2^4}
+{\left({192}{\alpha _4^2}+{288}{\alpha _5^2}-{20 N}{\alpha _3^2}\right)}{\alpha _2^2}
+\frac{117 N}{f^4}+\frac{{64}{\alpha _4^2}-{32}{\alpha _5^2}-{142 N}{\alpha _2^2}
+{108 N}{\alpha _3^2}}{f^2}
\nn\\
&+& {\left[2560 \left( \frac{N^2-1}{2N} \right)-1112 N\right]}{\alpha _3^4}
\nn\\
b_3 &=& \frac{23}{2 } \alpha _2^4
+ {32}{\alpha _4^4}
+{160}{\alpha _4^2 \alpha _5^2}
+ {\left(\frac{64}{N^2}+42\right)}{\alpha _3^4}
+ {\left(-{38}{\alpha _3^2}-{16 N}{\alpha _5^2}\right)}{\alpha _2^2}
+ {\left(16 N^2+120\right)}{\alpha _5^4}
\nn\\
&+& \frac{-{\alpha _2^2}
+{2}{\alpha _3^2}-{16 N}{\alpha _5^2}}{f^2}
+ {\left[128 \left( \frac{N^2-1}{2N} \right)-16 N\right]}{\alpha _3^2 \alpha _5^2}
+\frac{7}{2 f^4}
\nn\\
b_4 &=& {\left({480}{\alpha _5^2}+{320}{\alpha _4^2}\right)}{\alpha _3^2}
+ {5 N}{\alpha _2^4}
+ {\left( - {192}{\alpha _4^2} - {288}{\alpha _5^2}
- {28 N}{\alpha _3^2}\right)}{\alpha _2^2}+\frac{5 N}{f^4}
\nn\\
&+& \frac{- {64}{\alpha _4^2} + {32}{\alpha _5^2}
+ {10 N}{\alpha _2^2} - {28 N}{\alpha _3^2}}{f^2}
+ {\left[512 \left( \frac{N^2-1}{2N} \right)-216 N\right]}{\alpha _3^4}.
\eea
Likewise
\bea
\textrm{Tr} \bigl( {\cal B}^2 \bigr) &=& \frac{16 (N^2-1)}{c_0^2}
+ \frac{16}{c_0}
\left( \frac{N}{f^2} - {3 N}{\alpha _2^2} 
+ {\left(\frac{12}{N}-7 N\right)}{\alpha _3^2} 
+ {\left(4-8 N^2\right)}{\alpha _4^2} 
- {(8+2 N^2)}{\alpha _5^2}  \right)
\delta^{\alpha\beta} \delta^{c d} (\Delta_{\alpha})^c (\Delta_{\beta})^d
\nn\\
&+&  \biggl[ \Bigl( b_5 \delta ^{e f} \delta ^{c d} + b_6 \text{Tr}(t^e t^f t^c t^d)  \Bigr)
\delta^{\alpha\beta} \delta^{\gamma\delta}
+ \Bigl( b_7 \delta ^{e f} \delta ^{c d} + b_8 \text{Tr}(t^e t^f t^c t^d)  \Bigr)
\delta^{\alpha\gamma} \delta^{\beta\delta} \biggr]
(\Delta_{\alpha})^c (\Delta_{\beta})^d (\Delta_{\gamma})^e (\Delta_{\delta})^f
\nn\\ 
\eea
where
\bea
b_5 &=& -\frac{45}{2 } \alpha _2^4
+ {256 N^2}{\alpha _4^4}
+ {\left({174}{\alpha _3^2}
+ {192 N}{\alpha _4^2} + {48 N}{\alpha _5^2}\right)}{\alpha _2^2}
+ {\left(\frac{576}{N^2}+48\right)}{\alpha _3^4}
+ {\left(128 N^2+576\right)}{\alpha _4^2 \alpha _5^2}
+ {\left(16 N^2+144\right)}{\alpha _5^4}
\nn\\
&+& \frac{{123}{\alpha _2^2} - {178}{\alpha _3^2} - {64 N}{\alpha _4^2}
- {16 N}{\alpha _5^2}}{f^2}
+ {\left\{{\left[384 \left( \frac{N^2-1}{2 N}  \right)-80 N\right]}{\alpha _5^2}
- {64 \left[5 N-24 \left( \frac{N^2-1}{2 N}  \right)\right]}{\alpha _4^2}\right\}}{\alpha _3^2}-\frac{221}{2 f^4}
\nn\\
b_6 &=& {\left({2240}{\alpha _5^2} + {896}{\alpha _4^2}\right)}{\alpha _3^2}
+ {234 N}{\alpha _2^4}
+ {\left({384}{\alpha _4^2} + {960}{\alpha _5^2} - {24 N}{\alpha _3^2}\right)}{\alpha _2^2}
+\frac{458 N}{f^4}
\nn\\
&+& \frac{- {128}{\alpha _4^2} - {320}{\alpha _5^2} - {588 N}{\alpha _2^2}
+ {488 N}{\alpha _3^2}}{f^2}
+ {\left[9216 \left( \frac{N^2-1}{2 N}  \right)-4016 N\right]}{\alpha _3^4}
\nn\\
b_7 &=& {36}{\alpha _2^4}
- {120}{\alpha _2^2 \alpha _3^2}
+ {100}{\alpha _3^4}
+ {64}{\alpha _4^4}
+ {320}{\alpha _4^2 \alpha _5^2}
+ {400}{\alpha _5^4}+\frac{4}{f^4}
+ \frac{{40}{\alpha _3^2} - {24}{\alpha _2^2}}{f^2}
\nn\\
b_8 &=& {\left({1600}{\alpha _5^2} + {640}{\alpha _4^2}\right)}{\alpha _3^2}
+ {\left(-{960}{\alpha _5^2}-{384}{\alpha _4^2}\right)}{\alpha _2^2}
+\frac{{320}{\alpha _5^2} + {128}{\alpha _4^2}}{f^2} .
\eea
The first contribution in the trace of the ${\cal B}^2$ terms is associated with the presence of a~``cosmological constant"~term, that in principle should be included in the bare action due to the renormalization procedure. On the other hand, the second contribution gives us the hidden tadpole terms as mentioned above. Only the third contribution is actually a bubble.

As for ${\cal E}^{ab}$, we find that
\bea
\textrm{Tr} \bigl( {\cal E} \bigr) &=& - \frac{N}{c_0}  \delta ^{c d} \delta^{\alpha\beta}  
(\Delta_{\alpha})^c (\Delta_{\beta})^d
+ \frac{1}{f^2} \Biggl[ -\frac{N}{2}  \delta^{c d} 
\Bigl( \delta^{\alpha\beta} \delta^{\gamma\delta} + \delta^{\alpha\gamma} \delta^{\beta\delta} \Bigr) 
(d_{\alpha} \Delta_{\beta})^c (d_{\gamma} \Delta_{\delta})^d 
\nn\\
&-& \Biggl( \frac{1}{2} \Bigl(  \delta ^{e f} \delta ^{c d} + 4 N \text{Tr}(t^e t^f t^c t^d) \Bigr) \delta^{\alpha\beta} \delta^{\gamma\delta} 
+ \delta ^{e f} \delta ^{c d}  \delta^{\alpha\gamma} \delta^{\beta\delta}
\Biggr)
(\Delta_{\alpha})^c (\Delta_{\beta})^d (\Delta_{\gamma})^e (\Delta_{\delta})^f
\Biggr]
\nn\\
&+& \alpha_2^2  \frac{N}{2}  \delta^{c d} 
\Bigl( \delta^{\alpha\beta} \delta^{\gamma\delta} - \delta^{\alpha\gamma} \delta^{\beta\delta} \Bigr) 
(d_{\alpha} \Delta_{\beta})^c (d_{\gamma} \Delta_{\delta})^d 
\nn\\
&+& {2N}{\alpha_3^2} \text{Tr}(t^e t^f t^c t^d) 
\Bigl( 5 \delta^{\alpha\beta} \delta^{\gamma\delta} 
+ 3 \delta^{\alpha\gamma} \delta^{\beta\delta} \Bigr)  
(\Delta_{\alpha})^c (\Delta_{\beta})^d (\Delta_{\gamma})^e (\Delta_{\delta})^f
\nn\\
&+& {\alpha_4^2} \left[ 4 N \delta ^{e f} \delta^{cd}  \delta^{\alpha\beta} \delta^{\gamma\delta}
- 16 \text{Tr}(t^e t^f t^c t^d)
\Bigr( \delta^{ \alpha\beta}\delta^{\gamma \delta} - \delta^{\alpha \gamma}\delta^{\beta \delta}  \Bigl)
\right] (\Delta_{\alpha})^c (\Delta_{\beta})^d (\Delta_{\gamma})^e (\Delta_{\delta})^f 
\nn\\
&+& {\alpha_5^2} \left[ 4 N \delta ^{e f} \delta^{cd} \delta^{\alpha\gamma} \delta^{\beta\delta}
+ 8 \text{Tr}(t^e t^f t^c t^d)
\Bigr( \delta^{ \alpha\beta}\delta^{\gamma \delta} - \delta^{\alpha \gamma}\delta^{\beta \delta}  \Bigl)
\right] (\Delta_{\alpha})^c (\Delta_{\beta})^d (\Delta_{\gamma})^e (\Delta_{\delta})^f .
\eea
In turn
\bea
\textrm{Tr} \bigl( {\cal R}_{\mu\nu}  {\cal R}^{\mu\nu} \bigr)
&=& 2 N \textrm{Tr}\bigl( [t^e, t^f] [t^c, t^d] \bigr) \delta^{\alpha\gamma} \delta^{\beta\delta}
(\Delta_{\alpha})^c (\Delta_{\beta})^d (\Delta_{\gamma})^e (\Delta_{\delta})^f 
\eea
and
\bea
\textrm{Tr} \bigl( \hat{\sigma}^2 \bigr)
&=& 4 \textrm{Tr}\left( [ \Delta_{\mu},t^a ] [ \Delta^{\mu},t^b ] \right)
\textrm{Tr}\left( [ \Delta_{\nu},t^b ] [ \Delta^{\nu},t^a ] \right)
\nn\\
&=& \left( \frac{1}{2} \delta^{e f} \delta^{c d}  + 2 N \text{Tr}(t^e t^f t^c t^d) \right)
\delta^{\alpha\beta} \delta^{\gamma\delta}
\Delta^c_{\alpha} \Delta^{d}_{\beta} \Delta^e_{\gamma} \Delta^{f}_{\delta}
+ \delta^{e f} \delta^{c d} 
\delta^{\alpha\gamma} \delta^{\beta\delta}
\Delta^c_{\alpha} \Delta^{d}_{\beta} \Delta^e_{\gamma} \Delta^{f}_{\delta}.
\eea
All traces and SU(N) algebraic manipulations were carried out with the help of computer symbolic operations, performed by means of {\it Wolfram Mathematica} and packages such as {\it FeynCalc}~\cite{Shtabovenko:2020gxv,Shtabovenko:2016sxi,Mertig:1990an} and {\it FeynArts}~\cite{Hahn:2000kx}.

\section*{Appendix C. Brief explanation of the calculation for the coefficients $a_0$, $a_1$ and $a_2$}

As discussed above, we used heat-kernel techniques in order to evaluate one-loop divergences. In order to derive the expansion of the heat kernel $\langle x|  e^{-\tau{\cal D}} |x\rangle$ in terms of the $a_n$ coefficients,
\beq
\langle x|  e^{-\tau{\cal D}} |x\rangle = \frac{i}{(4\pi)^{d/2}} \frac{e^{-\tau^{1/2} m^2}}{\tau^{d/4}}
\sum_{n=0}^{\infty} \tau^{n/2} a_n(x),
\eeq
one usually starts by inserting a complete set of momentum eigenstates. We obtain that
\beq
\langle x|  e^{-\tau{\cal D}} |x\rangle = \int \frac{d^{d} p}{(2\pi)^d} 
e^{- i p \cdot x} e^{-\tau{\cal D}} e^{ i p \cdot x} .
\eeq
The next steps are the use of the identities
\bea
D_{\mu} e^{ i p \cdot x} &=& e^{ i p \cdot x} ( i p_{\mu} + D_{\mu} )
\nn\\
D^{\mu} D_{\mu} e^{ i p \cdot x} &=& e^{ i p \cdot x} ( i p_{\mu} + D_{\mu} ) ( i p^{\mu} + D^{\mu} )
\nn\\
D^{\nu} D_{\nu} D^{\mu} D_{\mu} e^{ i p \cdot x} &=& 
e^{ i p \cdot x} ( i p_{\nu} + D_{\nu} ) ( i p^{\nu} + D^{\nu} )
( i p_{\mu} + D_{\mu} ) ( i p^{\mu} + D^{\mu} )
\eea
and the Taylor expansion of the exponential containing the interesting physics in powers of $\tau$, keeping terms which contribute up to order $\tau$ after the integration over momentum is performed. After a straightforward calculation, one finds that
\bea
\langle x|  e^{-\tau{\cal D}} |x\rangle &=& \frac{i}{(4\pi)^{d/2}}  \frac{e^{-\tau^{1/2} m^2}}{\tau^{d/4}}
\left\{ \frac{\Gamma(d/4)}{2 \Gamma(d/2)}
+ \tau^{1/2} \frac{\Gamma \left(\frac{d/2-1}{2}\right)}{2 \Gamma \left(\frac{d}{2}-1\right)}
\frac{(-f^2 {\cal B})}{2d}
\right.
\nn\\
&+& \left. \tau \frac{\Gamma(d/4)}{4\Gamma(d/2)}
\left[ \frac{(d-2)}{6} {\cal R}_{\mu\nu}  {\cal R}^{\mu\nu}
+ \frac{1}{(d + 2)} \left(  \frac{1}{2} (-f^2 {\cal B}_{(\mu\nu)} ) (-f^2 {\cal B}^{\mu\nu} )
+ \frac{1}{4} (- f^2 {\cal B})^2  \right)
- 2 ( - f^2 {\cal E} )
\right]
+ {\cal O}(\tau^{3/2}) \right\}
\nn\\
\eea
or, identifying the coefficients in the expansion:
\beq
a_0 = \frac{\Gamma(d/4)}{2 \Gamma(d/2)} {\bf 1}
\eeq
\beq
a_1 = - \frac{\Gamma \left(\frac{d/2-1}{2}\right)}{2 \Gamma \left(\frac{d}{2}-1\right)}
\frac{f^2 {\cal B}}{2d}
\eeq 
and
\beq
a_2 = \frac{\Gamma(d/4)}{4\Gamma(d/2)}
\left[ \frac{(d-2)}{6} {\cal R}_{\mu\nu}  {\cal R}^{\mu\nu}
+ \frac{1}{(d + 2)} \left(  \frac{f^4}{2} {\cal B}_{(\mu\nu)}  {\cal B}^{\mu\nu} 
+ \frac{f^4}{4}  {\cal B}^2  \right) + 2 f^2 {\cal E} 
\right]
\eeq
which are precisely the expressions quoted in the main text. Apart from total derivative terms, our expressions fully coincide with the results of Ref.~\cite{Barvinsky:2021ijq}.

\section*{Appendix D. Vertices needed for the computation of the one-loop effective action}

We are going to present some interaction vertices that are necessary in the course of the calculation of the one-loop effective action. Such vertices can be found below.

Two pions and one $\Gamma$:
\beq
{\cal V}_{\Gamma}^{abc,\alpha} = - \int d^4 x_1 \int d^4 x_2 \int d^4 x_3\,
e^{i(k \cdot x_{1} + p \cdot x_{2} + q \cdot x_3)}
\frac{\delta}{\delta \pi^a (x_1)} \frac{\delta}{\delta \pi^b(x_2)} \frac{\delta}{\delta \Gamma^c_{\alpha}(x_3)}	S_{\pi\pi\Gamma}
\eeq
where 
\beq
S_{\pi\pi\Gamma} = -i f^{abc}  \int d^4 x \, \Gamma^c_{\mu}
\Bigl[ (\partial_{\nu} \partial^{\nu} \partial^{\mu} \pi^b)  \pi^a
+ (\partial_{\nu}  \partial^{\nu} \pi^a) \partial^{\mu} \pi^b 
+ M^2 \partial^{\mu} \pi^a  \pi^b  \Bigr].
\eeq
We get
\beq
{\cal V}_{\Gamma}^{abc,\alpha} = 
(2\pi)^4 \delta(k+p+q)  f^{a b c} ( k^2 + p^2 + M^2 ) ( k^{\alpha} - p^{\alpha}) .
\eeq

Two pions and two $\Delta$s:
\beq
{\cal V}_{\Delta}^{abcd,\alpha \beta} = - \int d^4 x_1 \int d^4 x_2 \int d^4 x_3 \int d^4 x_4\,
e^{i(k \cdot x_{1} + p \cdot x_{2} + q \cdot x_3 + r \cdot x_4 )}
\frac{\delta}{\delta \pi^a (x_1)} \frac{\delta}{\delta \pi^b(x_2)} \frac{\delta}{\delta \Delta^c_{\alpha}(x_3)}
\frac{\delta}{\delta \Delta^d_{\beta}(x_4)} S_{\pi \pi \Delta \Delta} 
\eeq
where
\bea
S_{\pi \pi \Delta \Delta} &=& -\frac{f^2}{2} \int d^4 x \, \pi^a 
\bigl( \widetilde{{\cal B}}_{s\mu\nu}^{ab} \partial^{\nu} \partial^{\mu} 
+ {\cal C}_{\mu}^{ab} \partial^{\mu} + {\cal E}^{ab} \bigr) \pi^b
\eea
where it is to be understood that in the above equation we keep only terms quadratic in $\Delta$. We find that
\bea
{\cal V}_{\Delta}^{abcd,\alpha \beta} &=& - \frac{f^2}{2}  (2\pi)^4 \delta(k+p+q+r) 
\Bigl[ (B^{\alpha\beta}_{\mu\nu})^{ab,cd} p^{\nu} p^{\mu} 
+ (B^{\alpha\beta}_{\mu\nu})^{ba,cd} k^{\nu} k^{\mu} 
+ i  (C^{\alpha\beta}_{\mu})^{ab,cd} p^{\mu} + i (C^{\alpha\beta}_{\mu})^{ba,cd} k^{\mu} 
\nn\\
&-& \bigl( (E^{\alpha\beta})^{ab,cd} + (E^{\alpha\beta})^{ba,cd} \bigr) \Bigr]
\eea
where
\bea
(B^{\alpha\beta}_{\mu\nu})^{ab,cd} &=& 
 \Biggl\{ 
\left[ - \frac{2}{\alpha_3^2} \textrm{Tr}\left( \left\{ t^a, t^{(e} \right\}  \left\{ t^b, t^{f)} \right\} \right)
+ 2 \left( \frac{2}{\alpha_2^2} - \frac{1}{\alpha_3^2} \right) 
\textrm{Tr}\left( \left[ t^a, t^{(e} \right] \left[ t^b, t^{f)} \right] \right)
\right.
\nn\\
&+& \left. 2 \left( - \frac{3}{f^2} + \frac{1}{\alpha_2^2} - \frac{1}{\alpha_3^2}  \right)  
\textrm{Tr}\left( \left\{ t^a, t^{(e} \right\} \left[ t^b, t^{f)} \right] \right)
+ 2 \left( \frac{1}{f^2} - \frac{1}{\alpha_2^2} - \frac{1}{\alpha_3^2} \right) 
\textrm{Tr}\left( \left[ t^a, t^{(e} \right] \left\{ t^b, t^{f)} \right\} \right)
\right.
\nn\\
&-& \left. \frac{4}{\alpha_3^2} \textrm{Tr}\Bigl( \{ t^e, t^f \} t^a t^b   \Bigr)
- \frac{4}{\alpha_4^2} \delta^{ef} \delta^{ab}
- \frac{4}{\alpha_5^2} \delta^{a (e} \delta^{f) b}
 \right] \delta^{\gamma\delta} \delta_{\nu\mu} 
\nn\\
&+& \left[ - 2 \left( \frac{1}{f^2} + \frac{1}{\alpha_2^2} \right)
\textrm{Tr}\left( 2 [t^a, t^{(f}] [t^b, t^{e)}] - \frac{1}{2} \{ t^a, t^{(f} \} [t^b, t^{e)}]
+ \frac{1}{2} [t^a, t^{(f}] \{ t^b, t^{e)} \} \right) 
\right.
\nn\\
&-& \left. \frac{1}{\alpha_3^2}
\textrm{Tr}\Bigl( 8 t^{(e} t^{f)}  t^a t^b + 4 ( t^{(f} t^{e)} t^b t^a + t^{(e} t^{f)}  t^a t^b )
+ 4 \left\{ t^a, t^{(e} \right\} \left\{ t^b, t^{f)} \right\} \Bigr)
\right.
\nn\\
&-& \left. \frac{8}{\alpha_4^2}  \delta^{a(e} \delta^{f)b}
- \frac{4}{\alpha_5^2} \Bigl(  \delta^{ef} \delta^{ab} + \delta^{a(f} \delta^{e)b} \Bigr) 
\right] \delta^{\gamma}_{\nu}\delta^{\delta}_{\mu} \Biggr\}
\bigl( \delta^{ec}  \delta^{df} \delta^{\alpha}_{\gamma} \delta^{\beta}_{\delta}
+ \delta^{ed}  \delta^{cf} \delta^{\beta}_{\gamma} \delta^{\alpha}_{\delta} \bigr),
\eea
or its non-symmetrized version,
\bea
(\bar{B}^{\lambda\rho}_{\ \ \mu\nu})^{ab,lm} &=& 
 \Biggl\{ 
\left[ - \frac{2}{\alpha_3^2} \textrm{Tr}\left( \left\{ t^a, t^{(c} \right\}  \left\{ t^b, t^{d)} \right\} \right)
+ 2 \left( \frac{2}{\alpha_2^2} - \frac{1}{\alpha_3^2} \right) 
\textrm{Tr}\left( \left[ t^a, t^{(c} \right] \left[ t^b, t^{d)} \right] \right)
\right.
\nn\\
&+& \left. 2 \left( - \frac{3}{f^2} + \frac{1}{\alpha_2^2} - \frac{1}{\alpha_3^2}  \right)  
\textrm{Tr}\left( \left\{ t^a, t^{(c} \right\} \left[ t^b, t^{d)} \right] \right)
+ 2 \left( \frac{1}{f^2} - \frac{1}{\alpha_2^2} - \frac{1}{\alpha_3^2} \right) 
\textrm{Tr}\left( \left[ t^a, t^{(c} \right] \left\{ t^b, t^{d)} \right\} \right)
\right.
\nn\\
&-& \left. \frac{4}{\alpha_3^2} \textrm{Tr}\Bigl( \{ t^c, t^d \} t^a t^b   \Bigr)
- \frac{4}{\alpha_4^2} \delta^{cd} \delta^{ab}
- \frac{4}{\alpha_5^2} \delta^{a (c} \delta^{d) b}
 \right] \delta^{\alpha\beta} \delta_{\nu\mu} 
\nn\\
&+& \left[ - 2 \left( \frac{1}{f^2} + \frac{1}{\alpha_2^2} \right)
\textrm{Tr}\left( 2 [t^a, t^d] [t^b, t^c] - \frac{1}{2} \{ t^a, t^d \} [t^b, t^c]
+ \frac{1}{2} [t^a, t^d] \{ t^b, t^c \} \right) 
\right.
\nn\\
&-& \left. \frac{1}{\alpha_3^2}
\textrm{Tr}\Bigl( 8 t^c t^d  t^a t^b + 4 ( t^d t^c t^b t^a + t^c t^d  t^a t^b )
+ 4 \left\{ t^a, t^c \right\} \left\{ t^b, t^d \right\} \Bigr)
\right.
\nn\\
&-& \left. \frac{8}{\alpha_4^2}  \delta^{ac} \delta^{bd}
- \frac{4}{\alpha_5^2} \Bigl(  \delta^{cd} \delta^{ab} + \delta^{ad} \delta^{bc} \Bigr) 
\right] \delta^{\alpha}_{\nu}\delta^{\beta}_{\mu} \Biggr\}
( \delta^{cl} \delta^{dm} \delta^{\lambda}_{\alpha} \delta^{\rho}_{\beta} 
+ \delta^{cm} \delta^{dl} \delta^{\rho}_{\alpha} \delta^{\lambda}_{\beta} ),
\eea
and
\begingroup
\allowdisplaybreaks
\bea
(C^{\alpha\beta}_{\mu})^{ab,cd} &\equiv& (C^{\alpha\beta}_{\mu})^{ab,cd}(q,r) =
- i \bigl( r_{\delta} \delta^{ce}  \delta^{df} \delta^{\alpha}_{\gamma} \delta^{\beta}_{\nu} 
+ q_{\delta} \delta^{cf}  \delta^{de} \delta^{\alpha}_{\nu} \delta^{\beta}_{\gamma} \bigr)
\Biggl\{ \frac{1}{f^2} \biggl\{ 
\textrm{Tr}\Bigl[ 4  [ t^b, [ t^e, t^a] ] t^f - 4 t^a  t^b [t^e, t^f]  +  t^a [  [t^e, t^f] , t^b ] 
\nn\\
&+& 2  t^f [t^e, t^a] t^b - 2 t^e t^a [ t^f, t^b ]  \Bigr]  \delta^{\delta\nu} \delta^{\gamma}_{\mu}
\nn\\
&+& \textrm{Tr}\Bigl[ 2 t^f \Bigl( 2  t^e  \{ t^b,  t^a \}  - 2 \{ t^e,  t^a \}  t^b - \left\{ t^e, \{ t^b,  t^a \}  \right\} 
+ 2 t^a \left\{ t^e, t^b  \right\} \Bigr) - 4 t^a  t^b [t^f, t^e] +  t^a [  [t^f, t^e] , t^b ] 
\nn\\
&+& 2 t^e [ t^f, t^a] t^b - 2 t^f  t^a [t^e, t^b ]  \Bigr]  \delta^{\gamma\delta} \delta^{\nu}_{\mu}
\nn\\
&+& \textrm{Tr}\Bigl[ 2  t^f \Bigl( - 2  \{ t^b,  t^a \} t^e + 2  t^b \{ t^e,  t^a \}  - \left\{ t^e, \{ t^b, t^a \} \right\} 
+ 2 t^a  \left\{ t^e,  t^b \right\} \Bigr)
- 4 t^f  t^a [t^b, t^e] - 4 t^e t^a [ t^b, t^f] \Bigr] \delta^{\gamma\nu} \delta^{\delta}_{\mu} \biggr\} 
\nn\\
&-& \frac{1}{\alpha_2^2} \biggl\{ 
- \textrm{Tr}\Bigl[ - 4  [ t^b, [ t^e, t^a] ] t^f - 4 t^a  t^b [t^e, t^f]  +  t^a [  [t^e, t^f] , t^b ]  + 2  t^f [t^e, t^a] t^b 
- 2 t^e t^a [ t^f, t^b ]  \Bigr] \delta^{\delta\nu}  \delta^{\gamma}_{\mu}
\nn\\
&-& \textrm{Tr}\Bigl[ 2 t^f \Bigl( 2  t^e  \{ t^b,  t^a \}  - 2 \{ t^e,  t^a \}  t^b - \left\{ t^e, \{ t^b,  t^a \}  \right\} 
+ 2 t^a \left\{ t^e, t^b  \right\} \Bigr) - 4 t^a  t^b [t^f, t^e] +  t^a [  [t^f, t^e] , t^b ] 
\nn\\
&+& 2 t^e [ t^f, t^a] t^b - 2 t^f  t^a [t^e, t^b ]  \Bigr] \delta^{\gamma\delta} \delta^{\nu}_{\mu}
\nn\\
&-& \textrm{Tr}\Bigl[ 2  t^f \Bigl( - 2  \{ t^b,  t^a \} t^e + 2  t^b \{ t^e,  t^a \}  - \left\{ t^e, \{ t^b, t^a \} \right\} 
+ 2 t^a  \left\{ t^e,  t^b \right\} \Bigr)
- 4 t^f  t^a [t^b, t^e] - 4 t^e t^a [ t^b, t^f] \Bigr] \delta^{\gamma\nu} \delta^{\delta}_{\mu}  \biggr\}
\nn\\
&-& \frac{4}{\alpha_3^2} \textrm{Tr}\biggl[ ( t^a t^f t^b t^e + t^b t^f t^a t^e  + t^a t^f t^e t^b + t^a t^e t^f t^b 
+ t^b t^e t^f t^a + t^b t^f t^e t^a + 2 t^a t^b t^f t^e )
\delta^{\gamma}_{\mu} \delta^{\delta\nu}
\nn\\
&+& ( t^a t^e t^b t^f + t^b t^e t^a t^f + t^a t^e t^f t^b  + t^a t^f t^e t^b + t^b t^f t^e t^a 
+ t^b t^e t^f t^a + 2 t^a t^b t^e t^f )
\delta^{\gamma\delta} \delta^{\nu}_{\mu}
 \nn\\
&+& 2 ( t^a t^f t^b t^e + t^a t^e t^b t^f + t^a t^b t^f t^e + t^a t^b t^e t^f) 
\delta^{\gamma\nu} \delta^{\delta}_{\mu} 
\biggr] 
\nn\\
&-& \frac{8}{\alpha_4^2} \biggl( \delta^{fe}\delta^{ab} \delta^{\gamma\nu} \delta^{\delta}_{\mu} 
+ \delta^{fa} \delta^{eb}\delta^{\gamma}_{\mu} \delta^{\delta\nu} 
+ \delta^{ea} \delta^{fb} \delta^{\gamma\delta} \delta^{\nu}_{\mu} \biggr) 
\nn\\
&-& \frac{4}{\alpha_5^2} \biggl[ 
\delta^{fe}\delta^{ab} ( \delta^{\gamma}_{\mu} \delta^{\delta\nu} 
+  \delta^{\gamma\delta} \delta^{\nu}_{\mu} ) 
+  \delta^{fa} \delta^{eb} ( \delta^{\gamma\delta} \delta^{\nu}_{\mu} 
+  \delta^{\gamma\nu} \delta^{\delta}_{\mu} )
+ \delta^{ea} \delta^{fb} ( \delta^{\gamma\nu} \delta^{\delta}_{\mu} 
+ \delta^{\gamma}_{\mu} \delta^{\delta\nu} )
\biggr] \Biggr\} 
\eea
and also
\bea
(E^{\alpha\beta})^{ab,cd} &\equiv& (E^{\alpha\beta})^{ab,cd}(q,r) 
= m^2 \textrm{Tr}\left( [ t^e ,t^a ] [ t^f,t^b ] \right) 
( \delta^{ce}  \delta^{df} + \delta^{cf} \delta^{de} ) \delta^{\alpha\beta}
\nn\\
&-& \bigl( q_{\gamma} r_{\mu} \delta^{ce} \delta^{df} \delta^{\alpha}_{\delta} \delta^{\beta}_{\nu}
+ r_{\gamma} q_{\mu} \delta^{cf}  \delta^{de} \delta^{\alpha}_{\nu} \delta^{\beta}_{\delta} \bigr)
\Biggl\{ \frac{1}{f^2} \biggl\{ \textrm{Tr}\biggl[ 2 [ t^a, t^e ] t^b t^f +  2 t^a \left\{ t^b, t^e \right\} t^f -  4 t^a t^b t^e t^f 
\nn\\
&-& \Bigl( \left\{ t^e, t^a \right\} \left\{ t^f, t^b \right\} - 2 t^a t^e \left\{ t^f, t^b \right\} + 4 t^a t^e t^b t^f 
- 2 \left\{ t^e , t^a \right\} t^b t^f \Bigr) \biggr]  
\delta^{\gamma\delta} \delta^{\mu\nu} 
\nn\\
&+&\textrm{Tr}\biggl[ - 2 \left\{ t^f, t^a t^b \right\} t^e + 4 t^a \left\{ t^f,  t^b \right\} t^e - 4 t^a t^b t^f t^e  
\nn\\
&-& \Bigl( \left\{ t^e, t^a \right\}  \left\{ t^f, t^b \right\} 
- 2 \left\{ t^e , t^a \right\} t^b t^f 
- 2 t^a t^f \left\{ t^e, t^b \right\}  
+ 4 t^a t^f  t^b t^e \Bigr) \biggr] \delta^{\gamma\mu} \delta^{\delta\nu} \biggr\}
\nn\\
&-& \frac{1}{\alpha_2^2}
 \biggl\{ \textrm{Tr}\biggl[ 2 [ t^a, t^e ] t^b t^f +  2 t^a \left\{ t^b, t^e \right\} t^f -  4 t^a t^b t^e t^f 
\nn\\
&-& \Bigl( \left\{ t^e, t^a \right\} \left\{ t^f, t^b \right\} - 2 t^a t^e \left\{ t^f, t^b \right\} + 4 t^a t^e t^b t^f 
- 2 \left\{ t^e , t^a \right\} t^b t^f \Bigr) \biggr]  
\delta^{\gamma\delta} \delta^{\mu\nu} 
\nn\\
&-&\textrm{Tr}\biggl[ - 2 \left\{ t^f, t^a t^b \right\} t^e + 4 t^a \left\{ t^f,  t^b \right\} t^e - 4 t^a t^b t^f t^e  
\nn\\
&-& \Bigl( \left\{ t^e, t^a \right\}  \left\{ t^f, t^b \right\} 
- 2 \left\{ t^e , t^a \right\} t^b t^f 
- 2 t^a t^f \left\{ t^e, t^b \right\}  
+ 4 t^a t^f  t^b t^e \Bigr) \biggr] \delta^{\gamma\mu} \delta^{\delta\nu} \biggr\} \Biggr\}
 \eea
\endgroup
where repeated use was made of the following results:
\bea
u A_{\mu}(x)u^{-1} &=& 2 \Delta_{\mu} 
\nn\\
u \partial_{\nu} A_{\mu} u^{-1} &=& 2 ( d_{\nu} \Delta_{\mu} + 2 \Delta_{[\mu} \Delta_{\nu]} ) .
\eea

Two pions and four $\Delta$s:
\bea
{\cal V}_{\Delta}^{abcdef,\alpha \beta\gamma\delta} &=& - \int d^4 x_1 \int d^4 x_2 \int d^4 x_3 
\int d^4 x_4\, \int d^4 x_5\, \int d^4 x_6\, 
e^{i(k \cdot x_{1} + p \cdot x_{2} + q \cdot x_3 + r \cdot x_4 + s \cdot x_5 + v \cdot x_6)}
\nn\\
&\times& 
\frac{\delta}{\delta \pi^a (x_1)} \frac{\delta}{\delta \pi^b(x_2)} \frac{\delta}{\delta \Delta^c_{\alpha}(x_3)}
\frac{\delta}{\delta \Delta^d_{\beta}(x_4)} \frac{\delta}{\delta \Delta^e_{\gamma}(x_5)} 
\frac{\delta}{\delta \Delta^f_{\delta}(x_6)} S_{\pi \pi \Delta \Delta \Delta \Delta} .
\eea
where
\beq
S_{\pi \pi \Delta \Delta \Delta \Delta} = - \frac{f^2}{2} \int d^4 x \, \pi^a {\cal E}^{ab} \pi^b
\eeq
where only terms quartic in $\Delta$ enter the latter interaction action. We obtain
\bea
{\cal V}_{\Delta}^{ijmnkl, \mu\nu\kappa\lambda} &=& \frac{f^2}{2}
(2\pi)^4 \delta(k+p+q+r+s+v)
\nn\\
&\times&
(E^{\alpha\beta\gamma\delta})^{ab,cdef}
( \delta^{ai} \delta^{bj} + \delta^{bi} \delta^{aj} ) 
\Bigl[ 
 ( \delta^{mc} \delta^{nd} \delta^{\mu}_{\alpha} \delta^{\nu}_{\beta} 
  + \delta^{md} \delta^{nc} \delta^{\mu}_{\beta} \delta^{\nu}_{\alpha}  ) 
 ( \delta^{ke} \delta^{lf} \delta^{\kappa}_{\gamma} \delta^{\lambda}_{\delta}
  + \delta^{kf} \delta^{le} \delta^{\kappa}_{\delta} \delta^{\lambda}_{\gamma} )
\nn\\
&+& 
( \delta^{mc} \delta^{ne} \delta^{\mu}_{\alpha} \delta^{\nu}_{\gamma}  
+  \delta^{me} \delta^{nc} \delta^{\mu}_{\gamma} \delta^{\nu}_{\alpha} ) 
( \delta^{kd} \delta^{lf} \delta^{\kappa}_{\beta} \delta^{\lambda}_{\delta}
+ \delta^{kf} \delta^{ld} \delta^{\kappa}_{\delta} \delta^{\lambda}_{\beta} )
\nn\\
&+&  ( \delta^{mc} \delta^{nf} \delta^{\mu}_{\alpha} \delta^{\nu}_{\delta} 
 +  \delta^{mf} \delta^{nc} \delta^{\mu}_{\delta} \delta^{\nu}_{\alpha} ) 
 ( \delta^{ke} \delta^{ld} \delta^{\kappa}_{\gamma} \delta^{\lambda}_{\beta} 
 + \delta^{kd} \delta^{le} \delta^{\kappa}_{\beta} \delta^{\lambda}_{\gamma} )
 \Bigr]
\eea
where
\bea
&& (E^{\alpha\beta\gamma\delta})^{ab,cdef} \equiv 
\frac{1}{f^2}
\Biggl\{ - \textrm{Tr}\biggl[ 4 [ \left\{ t^c , t^a \right\}, t^d ] [ \left\{ t^e, t^b \right\}, t^f  ]
+ 2 [ \left\{ t^c, t^a \right\}, t^d ] \left\{ \left\{ t^e, t^b \right\}, t^f \right\}
\nn\\
&& - 8 [ \left\{ t^c, t^a \right\}, t^d ] t^b  t^f t^e 
+ 2 \left\{ \left\{ t^c, t^a \right\}, t^d \right\} [ \left\{ t^e, t^b \right\}, t^f ]
+ \left\{ \left\{ t^c, t^a \right\}, t^d \right\} \left\{ \left\{ t^e, t^b \right\}, t^f \right\}
- 4 \left\{ \left\{ t^c, t^a \right\}, t^d \right\} t^b  t^f t^e 
\nn\\
&& - 8 t^a t^c t^d [ \left\{ t^e, t^b \right\}, t^f ] 
- 4 t^a t^c t^d \left\{ \left\{t^e, t^b \right\}, t^f  \right\} 
+ 16 t^a  t^c t^d  t^b t^e t^f \biggr]
\delta^{\alpha\beta} \delta^{\gamma\delta} 
\nn\\
&&+ \textrm{Tr}\biggl[ - 2 t^a [[ t^e, t^f], t^b] [t^c, t^d]
- 4 \left\{ t^f, t^a \right\} \left\{ t^e,  t^b \right\}  [t^c, t^d]
+ 4  t^a \left\{ t^f, \left\{ t^e,  t^b \right\} \right\} [t^c, t^d]
 + 4 t^f \left\{ t^e, t^a t^b \right\} [t^c, t^d]
\nn\\
&&+ 4 \left\{ t^f, t^a t^b \right\} t^e  [t^c, t^d]
- 2 \Bigl\{ t^f, \left\{ t^e, t^a t^b \right\} \Bigr\} [t^c, t^d] 
-  4 t^a t^b  \left\{t^f,t^e \right\} [t^c, t^d]  
- [[ t^c, t^d],t^a] [ [ t^e, t^f], t^b ]
+  [ t^a, [ [ t^c, t^d], t^b] ] [ t^e, t^f] 
\nn\\
&&- \Bigl( 4 \bigl(  \left\{ t^c, t^a \right\} t^d  - t^c \left\{ t^d, t^a \right\} \bigr)
\bigl(  \left\{ t^e, t^b \right\} t^f  - t^e \left\{ t^f, t^b \right\} \bigr)
+ 2 \bigl(  \left\{ t^c, t^a \right\} t^d - t^c \left\{ t^d, t^a \right\} \bigr) 
\left\{ t^e, \left\{ t^f, t^b \right\} \right\} 
\nn\\
&&- 4 \bigl(  \left\{ t^e, t^a \right\} t^f - t^e \left\{ t^f, t^a \right\} \bigr) t^b \left\{ t^d, t^c \right\}
+ 2 \left\{ t^c, \left\{ t^d, t^a \right\} \right\} 
\bigl( \left\{ t^e, t^b \right\} t^f  - t^e \left\{ t^f, t^b \right\} \bigr)
+ \left\{ t^c, \left\{ t^d , t^a \right\} \right\} \left\{ t^e, \left\{ t^f, t^b \right\} \right\} 
\nn\\
&&- 2 \left\{ t^e, \left\{ t^f, t^a \right\} \right\} t^b \left\{ t^d, t^c \right\} 
- 4 t^a \left\{ t^d, t^c \right\} \bigl(  \left\{ t^e, t^b \right\} t^f - t^e \left\{ t^f, t^b \right\} \bigr)
\nn\\
&& - 2 t^a \left\{t^d, t^c\right\} \left\{ t^e, \left\{ t^f, t^b \right\} \right\} 
+ 4 t^a  \left\{t^d, t^c\right\} t^b \left\{ t^f, t^e \right\} \Bigr)
\biggr] \delta^{\alpha\gamma} \delta^{\beta\delta}  \Biggr\}
\nn\\
&&- \frac{1}{\alpha_2^2}
\Biggl\{ - \textrm{Tr}\biggl[ 4 [ \left\{ t^c , t^a \right\}, t^d ] [ \left\{ t^e, t^b \right\}, t^f  ]
+ 2 [ \left\{ t^c, t^a \right\}, t^d ] \left\{ \left\{ t^e, t^b \right\}, t^f \right\}
- 8 [ \left\{ t^c, t^a \right\}, t^d ] t^b  t^f t^e 
+ 2 \left\{ \left\{ t^c, t^a \right\}, t^d \right\} [ \left\{ t^e, t^b \right\}, t^f ]
\nn\\
&&+ \left\{ \left\{ t^c, t^a \right\}, t^d \right\} \left\{ \left\{ t^e, t^b \right\}, t^f \right\}
- 4 \left\{ \left\{ t^c, t^a \right\}, t^d \right\} t^b  t^f t^e 
- 8 t^a t^c t^d [ \left\{ t^e, t^b \right\}, t^f ] 
\nn\\
&& - 4 t^a t^c t^d \left\{ \left\{t^e, t^b \right\}, t^f  \right\} 
+ 16 t^a  t^c t^d  t^b t^e t^f \biggr]
\delta^{\alpha\beta} \delta^{\gamma\delta} 
\nn\\
&&- \textrm{Tr}\biggl[ - 2 t^a [[ t^e, t^f], t^b] [t^c, t^d]
- 4 \left\{ t^f, t^a \right\} \left\{ t^e,  t^b \right\}  [t^c, t^d]
+ 4  t^a \left\{ t^f, \left\{ t^e,  t^b \right\} \right\} [t^c, t^d]
 + 4 t^f \left\{ t^e, t^a t^b \right\} [t^c, t^d]
\nn\\
&&+ 4 \left\{ t^f, t^a t^b \right\} t^e  [t^c, t^d]
- 2 \Bigl\{ t^f, \left\{ t^e, t^a t^b \right\} \Bigr\} [t^c, t^d] 
-  4 t^a t^b  \left\{t^f,t^e \right\} [t^c, t^d]  
-  [[ t^c, t^d],t^a] [ [ t^e, t^f], t^b ]
+  [ t^a, [ [ t^c, t^d], t^b] ] [ t^e, t^f] 
\nn\\
&&- \Bigl( 4 \bigl(  \left\{ t^c, t^a \right\} t^d  - t^c \left\{ t^d, t^a \right\} \bigr)
\bigl(  \left\{ t^e, t^b \right\} t^f  - t^e \left\{ t^f, t^b \right\} \bigr)
+ 2 \bigl(  \left\{ t^c, t^a \right\} t^d - t^c \left\{ t^d, t^a \right\} \bigr) 
\left\{ t^e, \left\{ t^f, t^b \right\} \right\} 
\nn\\
&&- 4 \bigl(  \left\{ t^e, t^a \right\} t^f - t^e \left\{ t^f, t^a \right\} \bigr) t^b \left\{ t^d, t^c \right\}
+ 2 \left\{ t^c, \left\{ t^d, t^a \right\} \right\} 
\bigl( \left\{ t^e, t^b \right\} t^f  - t^e \left\{ t^f, t^b \right\} \bigr)
+ \left\{ t^c, \left\{ t^d , t^a \right\} \right\} \left\{ t^e, \left\{ t^f, t^b \right\} \right\} 
\nn\\
&&- 2 \left\{ t^e, \left\{ t^f, t^a \right\} \right\} t^b \left\{ t^d, t^c \right\} 
- 4 t^a \left\{ t^d, t^c \right\} \bigl(  \left\{ t^e, t^b \right\} t^f - t^e \left\{ t^f, t^b \right\} \bigr)
\nn\\
&& - 2 t^a \left\{t^d, t^c\right\} \left\{ t^e, \left\{ t^f, t^b \right\} \right\} 
+ 4 t^a  \left\{t^d, t^c\right\} t^b \left\{ t^f, t^e \right\} \Bigr)
\biggr] \delta^{\alpha\gamma} \delta^{\beta\delta}  \Biggr\}
\nn\\
&&+ \frac{1}{\alpha_3^2}
\Biggl\{
 \textrm{Tr}\biggl[8 t^e t^f \Bigr( \left\{ \bigl[ [t^c, t^b ], t^a \bigr], t^d \right\} + \left[ t^c, t^a \right] \left[ t^d, t^b \right]  \Bigl) + 4  \left\{ [ t^c, t^a ], t^d \right\} \left\{ [ t^e, t^b ], t^f \right\}
\biggr] \delta^{\alpha\beta} \delta^{\gamma\delta}
\nn\\
&&+  \textrm{Tr}\biggl[8 t^e t^f \Bigr(  \left[ \left[ t^c, t^a \right], t^b \right] t^d 
+ \left[ \left[ t^c, t^a \right], t^b \right] t^d 
+  [ t^c, t^a ] [ t^d, t^b ] \Bigl)
+ 4 \left[ t^c, t^a \right] t^d \left[ t^e t^f, t^b \right]   
+ 4 t^d  \left[ t^c, t^a \right]  \left[ t^f t^e, t^b \right] 
\biggr] \delta^{\alpha\gamma} \delta^{\beta\delta} \Biggr\}
\nn\\
&&+ \frac{4}{\alpha_4^2} \delta^{cd} f^{eak} f^{fbk} \delta^{\alpha\beta} \delta^{\gamma\delta}
+ \frac{4}{\alpha_5^2} \biggl( 
 \delta^{cd} f^{eak} f^{fbk}
+ 2 \textrm{Tr}\Bigl( t^d \left\{ t^c, t^a \right\} 
- t^c \left\{ t^d, t^a \right\} \Bigr)
\textrm{Tr}\Bigl( t^f \left\{ t^e,  t^b \right\} 
- t^e \left\{ t^f, t^b \right\} \Bigr)
\biggr) \delta^{\alpha\gamma} \delta^{\beta\delta} .
\eea

A mixed vertex with two pions, one $\Gamma$ and two $\Delta$s:
\bea
{\cal V}_{\Gamma\Delta}^{abcde,\alpha\beta\gamma} &=& 
- \int d^4 x_1 \int d^4 x_2 \int d^4 x_3\, \int d^4 x_4\,
\int d^4 x_5\, e^{i(k \cdot x_{1} + p \cdot x_{2} + q \cdot x_3 + r \cdot x_4 + s \cdot x_5)}
\nn\\
&\times&
\frac{\delta}{\delta \pi^a (x_1)} \frac{\delta}{\delta \pi^b(x_2)} \frac{\delta}{\delta \Gamma^c_{\alpha}(x_3)}\frac{\delta}{\delta \Delta^d_{\beta}(x_4)} \frac{\delta}{\delta \Delta^e_{\gamma}(x_5)} 
S_{\pi \pi \Gamma \Delta \Delta}
\eea
where
\bea
S_{\pi \pi \Gamma \Delta \Delta} &=& -\frac{f^2}{2} \int d^4 x \, \pi^a 
\bigl(  \widetilde{{\cal B}}_{s}^{ab, \mu\nu} \partial_{\nu} \hat{\Gamma}^{bc}_{\mu} \pi^c
+ 2 \widetilde{{\cal B}}_{s}^{ab, \mu\nu}  \hat{\Gamma}^{bc}_{\mu} \partial_{\nu} \pi^c 
+ {\cal C}_{\mu}^{ab} \partial_{\mu} \pi^b  
+ {\cal C}_{\mu}^{ab} \hat{\Gamma}^{bc}_{\mu} \pi^c + {\cal E}^{ab} \pi^b \bigr) 
\nn\\
\widetilde{{\cal B}}_{s\mu\nu}^{ab} &=&  \frac{1}{2} \bigl( \widetilde{{\cal B}}_{\mu\nu}^{ab} 
+ \widetilde{{\cal B}}_{\nu\mu}^{ab} \bigr) 
\nn\\
\widetilde{{\cal B}}_{\mu\nu}^{ab} &=& \frac{1}{f^2} Y_{\mu\nu}^{ab (1)} 
+ \sum_{i=2}^{5} \frac{1}{\alpha_i^2} Y_{\mu\nu}^{ab (i)} 
\eea
where it is to be understood that in the above equations we keep only terms quadratic in $\Delta$ and linear in $\Gamma$. We get
\bea
{\cal V}_{\Gamma\Delta}^{ijklm,\kappa\lambda\rho} &=& 
\frac{f^2}{2} (2\pi)^4 \delta(k+p+q+r+s)
 \nn\\
&\times&
\bigl(  - f^{bcd} \delta^{dk} \delta^{ia} \delta^{jc} 
(B^{\lambda\rho\mu\nu})^{ab,lm} \delta^{\kappa}_{\mu}  q_{\nu} 
- f^{bcd}  \delta^{dk} \delta^{ic} \delta^{ja} 
(B^{\lambda\rho\mu\nu})^{ab,lm} \delta^{\kappa}_{\mu}  q_{\nu} 
\nn\\
&-& 2 f^{bcd} \delta^{dk}  \delta^{ia} \delta^{jc} 
(B^{\lambda\rho\mu\nu})^{ab,lm} \delta^{\kappa}_{\mu} p_{\nu} 
- 2 f^{bcd} \delta^{dk}  \delta^{ic} \delta^{ja} 
(B^{\lambda\rho\mu\nu})^{ab,lm} \delta^{\kappa}_{\mu} k_{\nu}
\nn\\
&-& i \delta^{ia} \delta^{jb} (Z^{\kappa\lambda\rho}_{\mu})^{ab,klm} p^{\mu}  
- i \delta^{ib} \delta^{ja} (Z^{\kappa\lambda\rho}_{\mu})^{ab,klm} k^{\mu}  
\nn\\
&-& i  f^{bcd} \delta^{dk} \delta^{\kappa\mu} \delta^{ia} \delta^{jc} (C^{\lambda\rho}_{\mu})^{ab,lm}(r,s) 
- i f^{bcd} \delta^{dk} \delta^{\kappa\mu} \delta^{ic} \delta^{ja} (C^{\lambda\rho}_{\mu})^{ab,lm}(r,s)
\nn\\
&+& \delta^{ia} \delta^{jb} (\mathbb{E}^{\kappa\lambda\rho})^{ab,klm} 
+ \delta^{ib} \delta^{ja} (\mathbb{E}^{\kappa\lambda\rho})^{ab,klm} \bigr) 
\eea
where
\bea
(Z^{\kappa\lambda\rho}_{\mu})^{ab,klm} &=& - i f^{def} ( \delta^{fk} \delta^{cl} \delta^{em}
\delta^{\kappa}_{\gamma} \delta^{\lambda}_{\alpha} \delta^{\rho}_{\delta} 
+ \delta^{fk} \delta^{cm} \delta^{el}
\delta^{\kappa}_{\gamma}  \delta^{\rho}_{\alpha} \delta^{\lambda}_{\delta}  )
\nn\\
&\times& \Biggl\{
\frac{1}{f^2} \textrm{Tr}\biggl[ \Bigl( 4  [ t^b, [ t^c, t^a] ] t^d 
- 4 t^a  t^b [t^c, t^d]  +  t^a [ [t^c, t^d], t^b ]  + 2 t^d [t^c , t^a] t^b - 2 t^c t^a [t^d, t^b ] \Bigr)
\delta^{\gamma\delta}\delta^{\alpha}_{\mu}  
\nn\\
&+& \Bigl[ 2  t^d \Bigl( 2  t^c  \{ t^b,  t^a \}  - 2 \{ t^c,  t^a \}  t^b - \left\{ t^c, \{ t^b,  t^a \}  \right\} 
+ 2 t^a \left\{ t^c, t^b  \right\} \Bigr)
+ 4 t^a t^b [t^c, t^d] - t^a [  [t^c, t^d] , t^b ] 
\nn\\
&+& 2 t^c [t^d, t^a] t^b  - 2 t^d  t^a [ t^c, t^b ] \Bigr] \delta^{\alpha\gamma} \delta^{\delta}_{\mu} 
\nn\\
&+& \Bigl[ 2 t^d \Bigl( - 2  \{ t^b,  t^a \} t^c + 2  t^b \{ t^c,  t^a \}  - \left\{ t^c, \{ t^b, t^a \} \right\} 
+ 2 t^a  \left\{ t^c,  t^b \right\} \Bigr) - 4 t^d t^a [t^b, t^c] - 4 t^c t^a [ t^b, t^d] \Bigr] 
 \delta^{\alpha\delta} \delta^{\gamma}_{\mu} \biggr] 
\nn\\
&-& \frac{1}{\alpha_2^2} \Biggl\{ 
- \textrm{Tr}\biggl[ \Bigl( - 4  [ t^b, [ t^c, t^a] ] t^d 
- 4 t^a  t^b [t^c, t^d]  +  t^a [ [t^c, t^d], t^b ]  + 2 t^d [t^c , t^a] t^b - 2 t^c t^a [t^d, t^b ] \Bigr)
\delta^{\gamma\delta}\delta^{\alpha}_{\mu}  
\nn\\
&+& \Bigl[ 2  t^d \Bigl( 2  t^c  \{ t^b,  t^a \}  - 2 \{ t^c,  t^a \}  t^b - \left\{ t^c, \{ t^b,  t^a \}  \right\} 
+ 2 t^a \left\{ t^c, t^b  \right\} \Bigr)
+ 4 t^a t^b [t^c, t^d] - t^a [  [t^c, t^d] , t^b ] 
\nn\\
&+& 2 t^c [t^d, t^a] t^b  - 2 t^d  t^a [ t^c, t^b ] \Bigr] \delta^{\alpha\gamma} \delta^{\delta}_{\mu} 
\nn\\
&+& \Bigl[ 2 t^d \Bigl( - 2  \{ t^b,  t^a \} t^c + 2  t^b \{ t^c,  t^a \}  - \left\{ t^c, \{ t^b, t^a \} \right\} 
+ 2 t^a  \left\{ t^c,  t^b \right\} \Bigr) - 4 t^d t^a [t^b, t^c] - 4 t^c t^a [ t^b, t^d] \Bigr] 
 \delta^{\alpha\delta} \delta^{\gamma}_{\mu} \biggr] 
\Biggr\}
\nn\\
&-& \frac{1}{\alpha_3^2} \Biggl\{ 8 \textrm{Tr}\biggl[ t^a t^b  \bigl\{ t^c, t^d \bigr\} \delta^{\alpha\delta} \delta^{\gamma}_{\mu}
+ t^a t^b t^d t^c \delta^{\gamma\delta} \delta^{\alpha}_{\mu} 
+ t^a t^b t^c t^d \delta^{\alpha\gamma} \delta^{\delta}_{\mu} \biggr]
\Biggr\}
\nn\\
&-& \frac{8}{\alpha_4^2}  \Bigl( \delta^{cd} \delta^{ab} 
\delta^{\alpha\delta} \delta^{\gamma}_{\mu} 
+ \delta^{da} \delta^{cb}
\delta^{\gamma\delta} \delta^{\alpha}_{\mu} 
+ \delta^{ca} \delta^{db}
\delta^{\alpha\gamma} \delta^{\delta}_{\mu} \Bigr)
\nn\\
&-& \frac{4}{\alpha_5^2} \Bigl( ( \delta^{cd} \delta^{ab} + \delta^{ac} \delta^{db} )
\delta^{\gamma\delta} \delta^{\alpha}_{\mu}
+ ( \delta^{cd} \delta^{ab} + \delta^{da} \delta^{cb} )
\delta^{\alpha\gamma} \delta^{\delta}_{\mu} 
+ ( \delta^{da} \delta^{cb} + \delta^{ca} \delta^{db} )
\delta^{\alpha\delta} \delta^{\gamma}_{\mu} \Bigr) \Biggr\}
\eea
and
\bea
&& (\mathbb{E}^{\kappa\lambda\rho})^{ab,klm} \equiv (\mathbb{E}^{\kappa\lambda\rho})^{ab,klm}(r,s)
= - \frac{1}{2} (\bar{B}^{\lambda\rho}_{\ \ \nu\mu})^{ac,lm}
( q^{\mu} \delta^{\kappa\nu} - q^{\nu} \delta^{\kappa\mu} )  f^{cbf} \delta^{kf}
\nn\\
&& - \Biggl\{ \frac{1}{f^2} 
 \biggl\{ \textrm{Tr}\biggl[ 2 [ t^a, t^c ] t^b t^d +  2 t^a \left\{ t^b, t^c \right\} t^d -  4 t^a t^b t^c t^d 
- \Bigl( \left\{ t^c, t^a \right\} \left\{ t^d, t^b \right\} - 2 t^a t^c \left\{ t^d, t^b \right\} + 4 t^a t^c t^b t^d 
- 2 \left\{ t^c , t^a \right\} t^b t^d \Bigr) \biggr]  
\delta^{\alpha\beta} \delta^{\gamma\delta} 
\nn\\
&& + \textrm{Tr}\biggl[ - 2 \left\{ t^d, t^a t^b \right\} t^c + 4 t^a \left\{ t^d,  t^b \right\} t^c - 4 t^a t^b t^d t^c  
- \Bigl( \left\{ t^c, t^a \right\}  \left\{ t^d, t^b \right\} 
- 2 \left\{ t^c , t^a \right\} t^b t^d 
- 2 t^a t^d \left\{ t^c, t^b \right\}  
+ 4 t^a t^d  t^b t^c \Bigr) \biggr] \delta^{\alpha\gamma} \delta^{\beta\delta} \biggr\}
\nn\\
&& - \frac{1}{\alpha_2^2}
\biggl\{ \textrm{Tr}\biggl[ 2 [ t^a, t^c ] t^b t^d +  2 t^a \left\{ t^b, t^c \right\} t^d -  4 t^a t^b t^c t^d 
- \Bigl( \left\{ t^c, t^a \right\} \left\{ t^d, t^b \right\} - 2 t^a t^c \left\{ t^d, t^b \right\} + 4 t^a t^c t^b t^d 
- 2 \left\{ t^c , t^a \right\} t^b t^d \Bigr) \biggr]  
\delta^{\alpha\beta} \delta^{\gamma\delta} 
\nn\\
&& - \textrm{Tr}\biggl[ - 2 \left\{ t^d, t^a t^b \right\} t^c + 4 t^a \left\{ t^d,  t^b \right\} t^c - 4 t^a t^b t^d t^c  
- \Bigl( \left\{ t^c, t^a \right\}  \left\{ t^d, t^b \right\} 
- 2 \left\{ t^c , t^a \right\} t^b t^d 
- 2 t^a t^d \left\{ t^c, t^b \right\}  
+ 4 t^a t^d  t^b t^c \Bigr) \biggr] \delta^{\alpha\gamma} \delta^{\beta\delta} \biggr\} \Biggr\}
\nn\\ 
&& \times 
\Bigl[ f^{dfg} \delta^{kg} \delta^{\kappa}_{\gamma} 
( r_{\alpha} \delta^{lc}  \delta^{mf} \delta^{\lambda}_{\beta} \delta^{\rho}_{\delta}
+ s_{\alpha} \delta^{mc}  \delta^{lf} \delta^{\rho}_{\beta} \delta^{\lambda}_{\delta} )
+ f^{cfg} \delta^{kg} \delta^{\kappa}_{\alpha}   
( r_{\gamma} \delta^{ld}  \delta^{mf} \delta^{\lambda}_{\delta}  \delta^{\rho}_{\beta} 
+ s_{\gamma} \delta^{md}  \delta^{lf} \delta^{\rho}_{\delta} \delta^{\lambda}_{\beta} )
\Bigr].
\eea

A mixed vertex with two pions, two $\Gamma$s and two $\Delta$s:
\bea
{\cal V}_{\Gamma\Delta}^{abcdef,\alpha\beta\gamma\delta} &=& - \int d^4 x_1 \int d^4 x_2 \int d^4 x_3\, 
\int d^4 x_4\, \int d^4 x_5\, \int d^4 x_6\, 
e^{i(k \cdot x_{1} + p \cdot x_{2} + q \cdot x_3 + r \cdot x_4 + s \cdot x_5 + v \cdot x_6)}
\nn\\
&\times&
\frac{\delta}{\delta \pi^a (x_1)} \frac{\delta}{\delta \pi^b(x_2)} \frac{\delta}{\delta \Gamma^c_{\alpha}(x_3)}\frac{\delta}{\delta \Gamma^d_{\beta}(x_4)} \frac{\delta}{\delta \Delta^e_{\gamma}(x_5)} 
\frac{\delta}{\delta \Delta^f_{\delta}(x_6)}  S_{\pi \pi \Gamma \Gamma \Delta \Delta} 
\eea
where
\beq
S_{\pi \pi \Gamma \Gamma \Delta \Delta} = -\frac{f^2}{2} \int d^4 x \, \pi^a 
\bigl( \widetilde{{\cal B}}_{s}^{ab,\mu\nu} \ \hat{\Gamma}^{bc}_{\nu}  \hat{\Gamma}^{cd}_{\mu} \pi^d  
+ {\cal C}_{\mu}^{ab} \hat{\Gamma}^{bc}_{\mu} \pi^c + {\cal E}^{ab} \pi^b \bigr) 
\eeq
where it is to be understood that we are keeping here only terms quadratic in $\Delta$ and quadratic in $\Gamma$. We find
\bea
{\cal V}_{\Gamma\Delta}^{ijknlm,\kappa\sigma\lambda\rho} &=& 
\frac{f^2}{2}(2\pi)^4 \delta(k+p+q+r+s+v)
\nn\\ 
&\times&  \Bigl[ 
- f^{bcf} f^{cdg} ( \delta^{ia} \delta^{jd} + \delta^{ja} \delta^{id} ) (B^{\lambda\rho\mu\nu})^{ab,lm}
( \delta^{kf} \delta^{ng} \delta^{\kappa}_{\nu}  \delta^{\sigma}_{\mu} 
+  \delta^{nf} \delta^{kg}  \delta^{\sigma}_{\nu} \delta^{\kappa}_{\mu} )
\nn\\
&-& i f^{bcf} ( \delta^{ia} \delta^{jc} + \delta^{ja} \delta^{ic} )
[ (Z^{\kappa\lambda\rho}_{\mu})^{ab,klm} \delta^{nf} \delta^{\sigma}_{\mu}
+ (Z^{\sigma\lambda\rho}_{\mu})^{ab,nlm} \delta^{kf} \delta^{\kappa}_{\mu} ]
\nn\\
&+& ( \delta^{ia} \delta^{jb} + \delta^{ja} \delta^{ib} ) 
(\mathbb{E}^{\kappa\sigma\lambda\rho})^{ab,knlm} \Bigr] 
\eea
where
\bea
&& (\mathbb{E}^{\kappa\sigma\lambda\rho})^{ab,knlm} = 
\frac{1}{2} f^{cdf} f^{dbg} (\bar{B}^{\lambda\rho\nu\mu})^{ac,lm} 
( - \delta^{kf}   \delta^{ng} \delta^{\kappa}_{\mu} \delta^{\sigma}_{\nu}
- \delta^{nf}  \delta^{kg}  \delta^{\sigma}_{\mu} \delta^{\kappa}_{\nu}
+ \delta^{kf}  \delta^{ng} \delta^{\kappa}_{\nu} \delta^{\sigma}_{\mu}
+ \delta^{nf}  \delta^{kg} \delta^{\sigma}_{\nu} \delta^{\kappa}_{\mu} )
\nn\\
&& - f^{chg} f^{dft} \Biggl\{ \frac{1}{f^2} 
 \biggl\{ \textrm{Tr}\biggl[ 2 [ t^a, t^c ] t^b t^d +  2 t^a \left\{ t^b, t^c \right\} t^d -  4 t^a t^b t^c t^d 
\nn\\
&& - \Bigl( \left\{ t^c, t^a \right\} \left\{ t^d, t^b \right\} - 2 t^a t^c \left\{ t^d, t^b \right\} + 4 t^a t^c t^b t^d 
- 2 \left\{ t^c , t^a \right\} t^b t^d \Bigr) \biggr]  
\delta^{\alpha\beta} \delta^{\gamma\delta} 
\nn\\
&& + \textrm{Tr}\biggl[ - 2 \left\{ t^d, t^a t^b \right\} t^c + 4 t^a \left\{ t^d,  t^b \right\} t^c - 4 t^a t^b t^d t^c  
- \Bigl( \left\{ t^c, t^a \right\}  \left\{ t^d, t^b \right\} 
- 2 \left\{ t^c , t^a \right\} t^b t^d 
- 2 t^a t^d \left\{ t^c, t^b \right\}  
+ 4 t^a t^d  t^b t^c \Bigr) \biggr] \delta^{\alpha\gamma} \delta^{\beta\delta} \biggr\}
\nn\\
&& - \frac{1}{\alpha_2^2}
\biggl\{ \textrm{Tr}\biggl[ 2 [ t^a, t^c ] t^b t^d +  2 t^a \left\{ t^b, t^c \right\} t^d -  4 t^a t^b t^c t^d 
- \Bigl( \left\{ t^c, t^a \right\} \left\{ t^d, t^b \right\} - 2 t^a t^c \left\{ t^d, t^b \right\} + 4 t^a t^c t^b t^d 
- 2 \left\{ t^c , t^a \right\} t^b t^d \Bigr) \biggr]  
\delta^{\alpha\beta} \delta^{\gamma\delta} 
\nn\\
&&- \textrm{Tr}\biggl[ - 2 \left\{ t^d, t^a t^b \right\} t^c + 4 t^a \left\{ t^d,  t^b \right\} t^c - 4 t^a t^b t^d t^c  
- \Bigl( \left\{ t^c, t^a \right\}  \left\{ t^d, t^b \right\} 
- 2 \left\{ t^c , t^a \right\} t^b t^d 
- 2 t^a t^d \left\{ t^c, t^b \right\}  
+ 4 t^a t^d  t^b t^c \Bigr) \biggr] \delta^{\alpha\gamma} \delta^{\beta\delta} \biggr\}
\Biggr\}
\nn\\
&& \times ( \delta^{kg} \delta^{nt} \delta^{\kappa}_{\alpha}  \delta^{\sigma}_{\gamma}
+ \delta^{ng}  \delta^{kt} \delta^{\sigma}_{\alpha}  \delta^{\kappa}_{\gamma} )
( \delta^{lh}  \delta^{mf} \delta^{\lambda}_{\beta} \delta^{\rho}_{\delta} 
+ \delta^{mh}  \delta^{lf} \delta^{\rho}_{\beta}\delta^{\lambda}_{\delta} ) .
 \eea
We repeatedly used the results
$$
\hat{\Gamma}^{ab}_{\mu} = - 2  \textrm{Tr}\left( [ t^a,t^b ] \Gamma_{\mu} \right),
$$
and
$$
i f^{abc} = 2 \textrm{Tr}\left( [ t^a,t^b ] t^c  \right) .
$$

\section*{Appendix E. The tadpole diagrams of the effective action}

The tadpole integrals we are going to calculate are given by (in momentum space)
\bea
{\cal D}^{cd,\alpha\beta} &=&  \mu^{4-d} \frac12 \int \frac{d^d k}{(2\pi)^d}
D^{ab}(k) {\cal V}_{\Delta}^{abcd,\alpha\beta}(k,q,r) 
\nn\\
{\cal D}^{cdef,\alpha\beta\gamma\delta} &=&  \mu^{4-d} \frac12 \int \frac{d^d k}{(2\pi)^d}
D^{ab}(k){\cal V}_{\Delta}^{abcdef,\alpha\beta\gamma\delta}(k,q,r,s,v)
\eea
with two and four $\Delta$s in the interaction vertex, and
\bea
{\cal M}^{cde,\alpha\beta\gamma} &=&  \mu^{4-d} \frac12 \int \frac{d^d k}{(2\pi)^d}
D^{ab}(k){\cal V}_{\Gamma\Delta}^{abcde,\alpha\beta\gamma}(k,q,r,s)
\nn\\
{\cal M}^{cdef,\alpha\beta\gamma\delta} &=& \mu^{4-d} \frac12 \int \frac{d^d k}{(2\pi)^d}
D^{ab}(k){\cal V}_{\Gamma\Delta}^{abcdef,\alpha\beta\gamma\delta}(k,q,r,s,v)
\eea
which contain mixed vertices with pions, $\Gamma$s and $\Delta$s. The overall factor of 1/2 accounts for the same factor that appears in the expansion of the effective action displayed above. As mentioned above, explicit expressions for the interaction vertices ${\cal V}$ can be found in the Appendix D. The pion propagator reads
\beq
D^{ab}(k) = \frac{\delta^{ab}}{k^4 + M^2 k^2}
= \frac{\delta^{ab}}{M^2} \left( \frac{1}{k^2}-\frac{1}{k^2+M^2} \right).
\eeq
Using dimensional regularization, we find that 
\bea
{\cal D}^{cd,\alpha\beta} &=& f^2  \frac{1}{2M^2} \left[
 \frac{\delta^{\mu\nu}}{2} \textrm{Tr}[(B^{\alpha\beta}_{\mu\nu})^{cd}]
 +  \frac{d/2}{M^2} \textrm{Tr}[(E^{\alpha\beta})^{cd}(q,-q)] \right] 
\mu^{4-d} \left( \frac{M^2}{4\pi} \right)^{d/2}  \Gamma(-d/2) 
  \nn\\
&=&  \frac{f^2}{32 \pi ^2}
\left[ \textrm{Tr}[(B^{\alpha\beta}_{\mu\nu})^{cd}] \frac{M^2 \delta^{\mu\nu}}{4} \left( \frac{1}{ \epsilon } 
- \log \left(\frac{M^2}{\mu^2}\right) + \log 4\pi  - \gamma_E + \frac32 \right)
 \right.
 \nn\\
 &+& \left. \textrm{Tr}[(E^{\alpha\beta})^{cd}(q,-q)] 
 \left( \frac{1}{ \epsilon }  - \log \left(\frac{M^2}{\mu^2}\right) + \log 4 \pi - \gamma_E +1 \right)
 \right] 
\eea
\bea
{\cal D}^{mnkl, \mu\nu\kappa\lambda} &=&   \frac{f^2}{2M^2}
\textrm{Tr}[(E^{\alpha\beta\gamma\delta})^{cdef}]
\Bigl[ 
 ( \delta^{mc} \delta^{nd} \delta^{\mu}_{\alpha} \delta^{\nu}_{\beta} 
  + \delta^{md} \delta^{nc} \delta^{\mu}_{\beta} \delta^{\nu}_{\alpha}  ) 
 ( \delta^{ke} \delta^{lf} \delta^{\kappa}_{\gamma} \delta^{\lambda}_{\delta}
  + \delta^{kf} \delta^{le} \delta^{\kappa}_{\delta} \delta^{\lambda}_{\gamma} )
\nn\\
&+& 
( \delta^{mc} \delta^{ne} \delta^{\mu}_{\alpha} \delta^{\nu}_{\gamma}  
+  \delta^{me} \delta^{nc} \delta^{\mu}_{\gamma} \delta^{\nu}_{\alpha} ) 
( \delta^{kd} \delta^{lf} \delta^{\kappa}_{\beta} \delta^{\lambda}_{\delta}
+ \delta^{kf} \delta^{ld} \delta^{\kappa}_{\delta} \delta^{\lambda}_{\beta} )
\nn\\
&+&  ( \delta^{mc} \delta^{nf} \delta^{\mu}_{\alpha} \delta^{\nu}_{\delta} 
 +  \delta^{mf} \delta^{nc} \delta^{\mu}_{\delta} \delta^{\nu}_{\alpha} ) 
 ( \delta^{ke} \delta^{ld} \delta^{\kappa}_{\gamma} \delta^{\lambda}_{\beta} 
 + \delta^{kd} \delta^{le} \delta^{\kappa}_{\beta} \delta^{\lambda}_{\gamma} )
 \Bigr]
  \frac{d/2}{M^2} \mu^{4-d} \left( \frac{M^2}{4\pi} \right)^{d/2}  \Gamma(-d/2) 
  \nn\\
&=&   \frac{f^2}{32 \pi ^2} 
\textrm{Tr}[(E^{\alpha\beta\gamma\delta})^{cdef}]
\Bigl[ 
 ( \delta^{mc} \delta^{nd} \delta^{\mu}_{\alpha} \delta^{\nu}_{\beta} 
  + \delta^{md} \delta^{nc} \delta^{\mu}_{\beta} \delta^{\nu}_{\alpha}  ) 
 ( \delta^{ke} \delta^{lf} \delta^{\kappa}_{\gamma} \delta^{\lambda}_{\delta}
  + \delta^{kf} \delta^{le} \delta^{\kappa}_{\delta} \delta^{\lambda}_{\gamma} )
\nn\\
&+& 
( \delta^{mc} \delta^{ne} \delta^{\mu}_{\alpha} \delta^{\nu}_{\gamma}  
+  \delta^{me} \delta^{nc} \delta^{\mu}_{\gamma} \delta^{\nu}_{\alpha} ) 
( \delta^{kd} \delta^{lf} \delta^{\kappa}_{\beta} \delta^{\lambda}_{\delta}
+ \delta^{kf} \delta^{ld} \delta^{\kappa}_{\delta} \delta^{\lambda}_{\beta} )
\nn\\
&+&  ( \delta^{mc} \delta^{nf} \delta^{\mu}_{\alpha} \delta^{\nu}_{\delta} 
 +  \delta^{mf} \delta^{nc} \delta^{\mu}_{\delta} \delta^{\nu}_{\alpha} ) 
 ( \delta^{ke} \delta^{ld} \delta^{\kappa}_{\gamma} \delta^{\lambda}_{\beta} 
 + \delta^{kd} \delta^{le} \delta^{\kappa}_{\beta} \delta^{\lambda}_{\gamma} )
 \Bigr]
  \left( \frac{1}{ \epsilon }  - \log \left(\frac{M^2}{\mu^2}\right) + \log 4 \pi - \gamma_E +1 \right)
  \nn\\
\eea
\bea
{\cal M}^{klm,\kappa\lambda\rho} &=&   \frac{f^2}{2 M^2}
\Bigl(  f^{abd} \delta^{dk} (B^{\lambda\rho\mu\nu})^{ab,lm} \delta^{\kappa}_{\mu}  q_{\nu} 
+ i  f^{abd} \delta^{dk} \delta^{\kappa\mu}  (C^{\lambda\rho}_{\mu})^{ab,lm}(q,r) 
\nn\\
&+&  \textrm{Tr}[(\mathbb{E}^{\kappa\lambda\rho})^{klm}(r,-q-r)] 
\Bigr) 
  \frac{d/2}{M^2} \mu^{4-d} \left( \frac{M^2}{4\pi} \right)^{d/2}  \Gamma(-d/2) 
\nn\\
&=& \frac{f^2}{32 \pi ^2}  
\Bigl(  f^{abd} \delta^{dk} (B^{\lambda\rho\mu\nu})^{ab,lm} \delta^{\kappa}_{\mu}  q_{\nu} 
+ i  f^{abd} \delta^{dk} \delta^{\kappa\mu}  (C^{\lambda\rho}_{\mu})^{ab,lm}(r,-q-r) 
\nn\\
&+&  \textrm{Tr}[(\mathbb{E}^{\kappa\lambda\rho})^{klm}(r,-q-r)] 
\Bigr)  
 \left( \frac{1}{ \epsilon }  - \log \left(\frac{M^2}{\mu^2}\right) + \log 4 \pi - \gamma_E +1 \right)
\eea
and finally
\bea
{\cal M}^{knlm,\kappa\sigma\lambda\rho} &=&   \frac{f^2}{2 M^2}
\Bigl[ 
 f^{bcf} f^{acg} (B^{\lambda\rho\mu\nu})^{ab,lm}
( \delta^{kf} \delta^{ng} \delta^{\kappa}_{\nu}  \delta^{\sigma}_{\mu} 
+  \delta^{nf} \delta^{kg}  \delta^{\sigma}_{\nu} \delta^{\kappa}_{\mu} )
\nn\\
&+& i f^{abf}
[ (Z^{\kappa\lambda\rho}_{\mu})^{ab,klm} \delta^{nf} \delta^{\sigma}_{\mu}
+ (Z^{\sigma\lambda\rho}_{\mu})^{ab,nlm} \delta^{kf} \delta^{\kappa}_{\mu} ]
\nn\\
&+& \textrm{Tr}[(\mathbb{E}^{\kappa\sigma\lambda\rho})^{knlm}] \Bigr] 
  \frac{d/2}{M^2} \mu^{4-d} \left( \frac{M^2}{4\pi} \right)^{d/2}  \Gamma(-d/2) 
  \nn\\
&=&   \frac{f^2}{32 \pi ^2} 
\Bigl[ 
 f^{bcf} f^{acg} (B^{\lambda\rho\mu\nu})^{ab,lm}
( \delta^{kf} \delta^{ng} \delta^{\kappa}_{\nu}  \delta^{\sigma}_{\mu} 
+  \delta^{nf} \delta^{kg}  \delta^{\sigma}_{\nu} \delta^{\kappa}_{\mu} )
\nn\\
&+& i f^{abf}
[ (Z^{\kappa\lambda\rho}_{\mu})^{ab,klm} \delta^{nf} \delta^{\sigma}_{\mu}
+ (Z^{\sigma\lambda\rho}_{\mu})^{ab,nlm} \delta^{kf} \delta^{\kappa}_{\mu} ]
\nn\\
&+& \textrm{Tr}[(\mathbb{E}^{\kappa\sigma\lambda\rho})^{knlm}] \Bigr] 
  \left( \frac{1}{ \epsilon }  - \log \left(\frac{M^2}{\mu^2}\right) + \log 4 \pi - \gamma_E +1 \right)
\eea
where $\textrm{Tr}[(B^{\alpha\beta}_{\mu\nu})^{cd}] = \delta^{ab} (B^{\alpha\beta}_{\mu\nu})^{ab,cd}$, etc., and the second equality is the result after setting $d=4-2\epsilon$ and Taylor expanding the expression. 

The renormalization of the mass term comes from the term ${\cal D}^{cd,\alpha\beta}$:

\bea
\frac{b_0}{c_0}
\frac{1}{16 \pi ^2}  \left( \frac{1}{ \epsilon } - \log \left(\frac{M^2}{\mu^2}\right) + \cdots \right)
 \delta^{cd} \delta^{\gamma\delta}
\eea
where the ellipsis indicates unimportant numerical factors. This agrees with the previous result using the heat kernel method.

On the other hand, the renormalization of $f^2$, associated with the kinetic term 
${\cal P}^{\alpha\beta\gamma\delta} \delta^{cd}
( d_{\alpha} \Delta_{\beta} )^c ( d_{\gamma} \Delta_{\delta} )^d$, comes from all tadpole terms considered above except the term ${\cal D}^{mnkl, \mu\nu\kappa\lambda}$:
\bea
- \frac{N}{2} 
{\cal P}^{\alpha\gamma\beta\mu} q_{\gamma} q_{\mu}
\frac{1}{16 \pi ^2}  \left( \frac{1}{ \epsilon } - \log \left(\frac{M^2}{\mu^2}\right) + \cdots \right),
\eea
associated with the term ${\cal P}^{\alpha\gamma\beta\mu} \delta^{cd}
( \partial_{\gamma} \Delta_{\alpha} )^c ( \partial_{\mu} \Delta_{\beta} )^d$
\bea
&& - \frac{N}{2} 
{\cal P}^{\alpha\beta\delta\gamma} f^{cde} q_{\delta} 
\frac{1}{16 \pi ^2}  \left( \frac{1}{ \epsilon } - \log \left(\frac{M^2}{\mu^2}\right) + \cdots \right) 
\nn\\
&& - \frac{N}{2} 
{\cal P}^{\delta\beta\alpha\gamma} f^{cde} (- q_{\delta} )
\frac{1}{16 \pi ^2}  \left( \frac{1}{ \epsilon } - \log \left(\frac{M^2}{\mu^2}\right) + \cdots \right), 
\eea
associated with the terms ${\cal P}^{\alpha\beta\delta\gamma} \delta^{ce}
( \hat{\Gamma}^{cd}_{\alpha} \Delta_{\beta}^d ) ( \partial_{\delta} \Delta_{\gamma}^e )$ and 
${\cal P}^{\delta\beta\alpha\gamma} \delta^{ce}
 ( \partial_{\delta} \Delta_{\beta}^e ) ( \hat{\Gamma}^{cd}_{\alpha} \Delta_{\gamma}^d )$, and finally
\beq
- \frac{N}{2} {\cal P}^{\alpha\beta\gamma\delta}
f^{k m e} f^{l n e}
 \left( \delta^{\alpha  \kappa } \delta^{\beta  \rho } \delta^{\gamma  \sigma } 
\delta^{\delta  \lambda } + \delta^{\alpha  \sigma }\delta^{\beta  \lambda } \delta^{\gamma  \kappa } \delta^{\delta  \rho }\right)
\frac{1}{16 \pi ^2}  \left( \frac{1}{ \epsilon } - \log \left(\frac{M^2}{\mu^2}\right) + \cdots \right)
\eeq
which accounts for the term $(\Gamma \Delta)^2$. Comparing the corresponding expressions, it is also easy to see that we can recover the contribution to $\Delta^4$ coming from 
$\textrm{Tr} \bigl( {\cal E} \bigr)$ in the heat-kernel approach, also accompanied by 
$\log M^2/\mu^2$. Notice that, as expected, $\mu$-dependence does not correspond to a logarithmic momentum dependence and hence the $\mu$-dependence of the renormalization can be reabsorbed without producing any large logs with the physical energy scale.

\section*{Appendix F. Tensor functions arising from Passarino-Veltman reduction of bubble Feynman integrals}

We here collect the tensor functions that appear in the tensor reduction of the bubble diagram function ${\cal M}^{\mu\nu\lambda\kappa}(q)$ computed in section 6. These are given by
\bea
\mathbb{M}^{\mu\nu\lambda\kappa}(q) &=&
\frac{1}{16 \left(d^2-1\right) q^4} 
\nn\\
&\times& \Big[ ( - 4 (d+3) M^2 q^4 - (d+2) q^6 - 16 M^4 q^2 )
\big( q^{\mu } q^{\nu } \delta^{\kappa  \lambda }
 + q^{\lambda } q^{\nu } \delta^{\kappa  \mu }
 + q^{\kappa } q^{\nu } \delta^{\lambda  \mu }
 + q^{\lambda } q^{\mu } \delta^{\kappa  \nu }
 + q^{\kappa } q^{\mu } \delta^{\lambda  \nu }
 + q^{\kappa } q^{\lambda } \delta^{\mu  \nu } \big)
 \nn\\
 &+& ( 16 M^4 q^4 + 8 M^2 q^6 + q^8 ) \big( \delta^{\kappa  \nu } \delta^{\lambda  \mu }
 + \delta^{\kappa  \mu } \delta^{\lambda  \nu }
 + \delta^{\kappa  \lambda } \delta^{\mu  \nu } \big)
 + ( 48 M^4 + 24 (d+2) M^2 q^2 + (d+2) (d+4) q^4 ) q^{\kappa } q^{\lambda } q^{\mu } q^{\nu } \Big]
 \nn\\
&-& \frac{1}{8 (1-d) q^2}  
 \Big[ ( 12 M^2 - (d+2) q^2 )  q^{\lambda } q^{\mu } q^{\nu }
+ q^2 ( q^2 - 4 M^2 ) \big( q^{\nu } \delta^{\lambda  \mu }
+ q^{\mu } \delta^{\lambda  \nu } + q^{\lambda } \delta^{\mu  \nu } \big) \Big] q^{\kappa}
\nn\\
&-& \frac{1}{8 (1-d) q^2}  
\Big[ ( 12 M^2 - (d+2) q^2 )  q^{\kappa} q^{\mu } q^{\nu }
+ q^2 ( q^2 - 4 M^2 ) \big( q^{\nu } \delta^{\kappa  \mu }
+ q^{\mu } \delta^{\kappa  \nu } + q^{\kappa } \delta^{\mu  \nu } \big) \Big] q^{\lambda}
\nn\\
&+& \frac{1}{4 (1-d) q^2} 
\big[ - ( 4 M^2 + d q^2 ) q^{\mu } q^{\nu } + q^2 ( q^2 + 4 M^2 ) \delta^{\mu  \nu } \big] 
q^{\lambda} q^{\kappa}
\eea
\bea
\mathbb{N}^{\mu\nu\lambda\kappa}(q) &=& 
\frac{1}{16 \left(d^2-1\right)} 
\Big[ - (d+2) q^2 \big( q^{\mu } q^{\nu } \delta^{\kappa  \lambda }
 + q^{\lambda } q^{\nu } \delta^{\kappa  \mu }
 + q^{\kappa } q^{\nu } \delta^{\lambda  \mu }
 + q^{\lambda } q^{\mu } \delta^{\kappa  \nu }
 + q^{\kappa } q^{\mu } \delta^{\lambda  \nu }
 + q^{\kappa } q^{\lambda } \delta^{\mu  \nu } \big)
 \nn\\
 &+& q^4 \big( \delta^{\kappa  \nu } \delta^{\lambda  \mu }
 + \delta^{\kappa  \mu } \delta^{\lambda  \nu }
 + \delta^{\kappa  \lambda } \delta^{\mu  \nu } \big)
 + (d+2) (d+4) q^{\kappa } q^{\lambda } q^{\mu } q^{\nu } \Big]
 \nn\\
&-& \frac{1}{8 (d-1)} \big[  (d+2) q^{\lambda } q^{\mu } q^{\nu } 
- q^2 ( q^{\lambda } \delta^{\mu  \nu }
+ q^{\nu } \delta^{\lambda  \mu } + q^{\mu } \delta^{\lambda  \nu } )  \big] q^{\kappa}
\nn\\
&-& \frac{1}{8 (d-1)} \big[  (d+2) q^{\kappa } q^{\mu } q^{\nu } 
- q^2 ( q^{\kappa } \delta^{\mu  \nu }
+ q^{\nu } \delta^{\kappa  \mu } + q^{\mu } \delta^{\kappa  \nu } )  \big] q^{\lambda}
\nn\\
&+& \frac{1}{4 (d-1)} \big( d q^{\mu } q^{\nu } - q^2 \delta^{\mu  \nu } \big) q^{\lambda} q^{\kappa}
\eea
and
\bea
 \mathbb{P}^{\mu\nu\lambda\kappa}(q) &=& \frac{1}{8 (1-d) \left(d^2-1\right) q^8} 
  \Big[   \big( 2 (d^2-1) (M^2+q^2)^2 q^6 - 2 (d^2-1) (M^2+q^2)^3 q^4 -(1-d) (d+2) (M^2+q^2)^4 q^2 \big) 
\nn\\
&\times& \Big( \delta^{\mu  \nu } q^{\kappa } q^{\lambda } 
+ \delta^{\lambda  \nu } q^{\kappa } q^{\mu } 
+ \delta^{\kappa  \nu } q^{\lambda } q^{\mu } 
+ \delta^{\lambda  \mu } q^{\kappa } q^{\nu } 
+ \delta^{\kappa  \mu } q^{\lambda } q^{\nu } 
+ \delta^{\kappa  \lambda } q^{\mu } q^{\nu } \Big)
\nn\\
&+& (1-d) (M^2+q^2)^4 q^4 \Big( \delta^{\lambda  \mu } \delta^{\kappa  \nu } 
+ \delta^{\kappa  \mu } \delta^{\lambda  \nu } 
+ \delta^{\kappa  \lambda } \delta^{\mu  \nu } \Big)
\nn\\
&+& \Big( 4 (d^2-1) (d+2) (M^2+q^2)^3 q^2 + 8 (d^2-1) (1-d) q^8 
-16 (1-d) (d^2-1)  (M^2+q^2) q^6 
\nn\\
&-& 12 d (d^2-1)  (M^2+q^2)^2 q^4 
+(d+2) (d+4) (1-d) (M^2+q^2)^4 \Big) q^{\kappa } q^{\lambda } q^{\mu } q^{\nu } \Big]
 \nn\\
&-& \frac{1}{4 (d-1) q^4}  
\Big[ \Big( 3 d \left(M^2+q^2\right)^2 + 6 (1-d) q^2 \left(M^2+q^2\right) - 4 (1-d) q^4 \Big) 
q^{\lambda } q^{\mu } q^{\nu }
\nn\\
&-& q^2 \left(M^2+q^2\right)^2 \Big( q^{\nu } \delta^{\lambda  \mu }
+ q^{\mu } \delta^{\lambda  \nu }
+ q^{\lambda } \delta^{\mu  \nu } \Big) \Big] q^{\kappa}
\nn\\
&-& \frac{1}{4 (d-1) q^4}  
\Big[ \Big( 3 d \left(M^2+q^2\right)^2 + 6 (1-d) q^2 \left(M^2+q^2\right) - 4 (1-d) q^4 \Big) 
q^{\kappa } q^{\mu } q^{\nu }
\nn\\
&-& q^2 \left(M^2+q^2\right)^2 \Big( q^{\nu } \delta^{\kappa \mu }
+ q^{\mu } \delta^{\kappa  \nu }
+ q^{\kappa} \delta^{\mu  \nu } \Big) \Big] q^{\lambda}
\nn\\
&+& \frac{1}{2 (d-1) q^4} 
 \Big[ \Big( d (M^2+q^2)^2 + 2 (1-d) q^2 \left(M^2+q^2\right) - 2 (1-d) q^4 \Big) q^{\mu } q^{\nu }
- q^2 \left(M^2+q^2\right)^2 \delta^{\mu  \nu } \Big] q^{\lambda} q^{\kappa}.
\eea
On the other hand, for the purely quartic bubble $\widetilde{{\cal M}}^{\mu\nu\lambda\kappa}(q)$, the tensor functions are given by
\bea
\widetilde{\mathbb{M}}^{\mu\nu\lambda\kappa}(q) &=&
\frac{1}{16 (1-d) \left(d^2-1\right)}
\Big[ - 2 \left(d^2-1\right) q^2 \Big( q^{\lambda } q^{\nu } \delta^{\kappa  \mu }
+ q^{\kappa } q^{\nu } \delta^{\lambda  \mu }
+ q^{\lambda } q^{\mu } \delta^{\kappa  \nu }
+ q^{\kappa } q^{\mu } \delta^{\lambda  \nu } \Big)
\nn\\
&-& (d+2) (1-d) q^2 \Big( q^{\mu } q^{\nu } \delta^{\kappa  \lambda }
+ q^{\lambda } q^{\nu } \delta^{\kappa  \mu }
+  q^{\lambda } q^{\mu } \delta^{\kappa  \nu } 
+ q^{\kappa } q^{\nu } \delta^{\lambda  \mu }
+ q^{\kappa } q^{\mu } \delta^{\lambda  \nu }
+ q^{\kappa } q^{\lambda } \delta^{\mu  \nu } \Big)
\nn\\
&+& (1-d) q^4 \Big( \delta^{\kappa  \nu } \delta^{\lambda  \mu }
+ \delta^{\kappa  \mu } \delta^{\lambda  \nu }
+ \delta^{\kappa  \lambda } \delta^{\mu  \nu } \Big)
+ \Big( 3 d \left(d^2+d-2\right)-4 d\left(d^2-1\right)  \Big) 
q^{\kappa } q^{\lambda } q^{\mu } q^{\nu }
\Big]
\nn\\
\eea
\bea
\widetilde{\mathbb{N}}^{\mu\nu\lambda\kappa}(q) &=&
 \frac{1}{4 (1-d) \left(d^2-1\right) q^4}
 \Big[   
 \left(d^2-1\right) q^2 \Big( q^{\lambda } q^{\nu } \delta^{\kappa  \mu }
+  q^{\kappa } q^{\nu } \delta^{\lambda  \mu } 
+  q^{\lambda } q^{\mu } \delta^{\kappa  \nu }
+  q^{\kappa } q^{\mu } \delta^{\lambda  \nu } \Big)
\nn\\
&+& (1-d) d \, q^2 \Big( q^{\mu } q^{\nu } \delta^{\kappa  \lambda }
+ q^{\lambda } q^{\nu } \delta^{\kappa  \mu }
+  q^{\kappa } q^{\nu } \delta^{\lambda  \mu }
+  q^{\lambda } q^{\mu } \delta^{\kappa  \nu }
+ q^{\kappa } q^{\mu } \delta^{\lambda  \nu }
+ q^{\kappa } q^{\lambda } \delta^{\mu  \nu } \Big)
\nn\\
&+& (1-d) q^4 \Big( \delta^{\kappa  \nu } \delta^{\lambda  \mu }
+ \delta^{\kappa  \mu } \delta^{\lambda  \nu }
+  \delta^{\kappa  \lambda } \delta^{\mu  \nu } \Big)
- (d-2) (d-1) d \, q^{\kappa } q^{\lambda } q^{\mu } q^{\nu }
 \Big]
\eea
and finally
\bea
\widetilde{\mathbb{P}}^{\mu\nu\lambda\kappa}(q) &=& 
\frac{1}{4 (1-d) \left(d^2-1\right) q^2}
\Big[ - 2 \left(d^2-1\right) q^2 \Big( q^{\lambda } q^{\nu } \delta^{\kappa  \mu }
+ q^{\kappa } q^{\nu } \delta^{\lambda  \mu }
+ q^{\lambda } q^{\mu } \delta^{\kappa  \nu }
+ q^{\kappa } q^{\mu } \delta^{\lambda  \nu } \Big)
\nn\\
&-& (1-d) (d+3) q^2 \Big( q^{\mu } q^{\nu } \delta^{\kappa  \lambda }
+ q^{\lambda } q^{\nu } \delta^{\kappa  \mu }
+ q^{\kappa } q^{\nu } \delta^{\lambda  \mu }
+ q^{\lambda } q^{\mu } \delta^{\kappa  \nu }
+ q^{\kappa } q^{\mu } \delta^{\lambda  \nu }
+ q^{\kappa } q^{\lambda } \delta^{\mu  \nu } \Big)
\nn\\
&+& 2 (1-d) q^4 \Big( \delta^{\kappa  \nu } \delta^{\lambda  \mu }
+ \delta^{\kappa  \mu } \delta^{\lambda  \nu }
+  \delta^{\kappa  \lambda } \delta^{\mu  \nu } \Big)
+ 2 (d-2) (d-1) q^{\kappa } q^{\lambda } q^{\mu } q^{\nu }
\Big] .
\eea

 \end{document}